\documentclass[manuscript]{aastex63}
\usepackage{rotating}
\shorttitle{Low-Redshift FeLoBALQs}
\shortauthors{Leighly et al.}

\begin{document}

\title{The Physical Properties of Low Redshift FeLoBAL Quasars.
  IV. Optical-Near IR Spectral Energy Distributions and Near-IR
  Variability Properties}

\author[0000-0002-3809-0051]{Karen M.\ Leighly}
\affiliation{Homer L.\ Dodge Department of Physics and Astronomy, The
  University of Oklahoma, 440 W.\ Brooks St., Norman, OK 73019, USA}

\author[0000-0002-3173-1098]{Hyunseop Choi}
\affiliation{D\'{e}partement de Physique, Universit\'{e} de
  Montr\'{e}al, Succ.\ Centre-Ville, Montr\'{e}al, Qu\'{e}bec, H3C 3J7,
  Canada}

\author{Michael Eracleous} 
\affiliation{Department of Astronomy \& Astrophysics and Institute for
  Gravitation and the Cosmos, The Pennsylvania State University, 525
  Davey Lab, University Park, PA 16802, USA}

\author{Donald M.\ Terndrup}
\affiliation{Department of Astronomy, The Ohio State University, 140
  W. 18th Ave., Columbus, OH 43210}
\affiliation{Homer L.\ Dodge Department of Physics and Astronomy,
  The
  University of Oklahoma, 440 W.\ Brooks St., Norman, OK 73019, USA}

\author{Sarah C.\ Gallagher}
\affiliation{Department of Physics \& Astronomy, The University of
  Western Ontario, London, ON, N6A 3K7, Canada}
\affiliation{Institute for Earth and Space Exploration, The
  University of Western Ontario, London, ON, N6A 3K7, Canada}
\affiliation{The Rotman Institute of Philosophy, The University of
  Western Ontario, London, ON, N6A 3K7, Canada}

\author{Gordon T.\ Richards}
\affiliation{Department of Physics, Drexel University, 32 S. 32nd St.,
  Philadelphia, PA 19104}



\begin{abstract}
{ We present the optical-near infrared spectral energy
  distributions (SED) and near infrared variability properties of 30
  low-redshift iron low-ionization 
Broad Absorption Line quasars (FeLoBALQs) and matched samples of
LoBALQs and unabsorbed  quasars. Significant correlations between the
SED properties and accretion rate indicators found among the
unabsorbed comparison sample objects suggest an intrinsic origin for
SED differences.  A range of reddening likely mutes these correlations
among the FeLoBAL quasars.   The restframe optical-band reddening is
correlated with the location of the outflow, suggesting a link between
the outflows and the presence of dust.  We analyzed {\it WISE} 
variability and provide a correction for photometry uncertainties in
an appendix.  We found an anticorrelation between the variability
amplitude and inferred continuum emission region size, and suggest
that as the origin of the anticorrelation between variability
amplitude and luminosity typically observed in quasars.  We found that
the LoBALQ optical emission line and other parameters are more similar
to those of the unabsorbed continuum sample objects than the
FeLoBALQs. Thus, FeLoBAL quasars are a special population of objects. 

We interpret the results using an accretion-rate scenario for FeLoBAL
quasars.  The high accretion rate  FeLoBAL quasars are radiating
powerfully enough to drive a thick, high-velocity outflow.  Quasars
with intermediate accretion rates may have an outflow, but it is not
sufficiently thick to include \ion{Fe}{2} absorption.  Low accretion
rate FeLoBAL outflows originate in absorption in a failing torus, no
longer optically thick enough to reprocess radiation into the near-IR.  }

\end{abstract}



\section{Introduction} \label{intro}

Broad absorption lines observed in \ion{C}{4} are found  in about
10-26\% of optically selected quasars \citep{tolea02, hf03,
  reichard03, trump06, knigge08,   gibson09}.  These blueshifted lines
reveal
an unambiguous signature of outflow.  Therefore, Broad Absorption Line
Quasars (BALQs) may be
important sources of quasar feedback in galaxy evolution.  About 1.3\%
of quasars have broad \ion{Mg}{2} absorption \citep{trump06}; these
are called low-ionization broad absorption line quasars
(LoBALQs). About 0.3\% of quasars also have absorption from
\ion{Fe}{2}, and these are called iron low-ionization broad absorption
line quasars \citep[FeLoBALQs,][]{trump06}.  FeLoBAL quasars are rare,
and only a few were known before the Sloan Digital Sky Survey.
For example, \citet{hall02}  published spectra and discussed the wide
range of features observed in 23 unusual objects discovered in the
SDSS;  many of those objects are FeLoBAL quasars.  

How do BAL quasars, and FeLoBAL quasars specifically, fit in among
quasars in general?  Are FeLoBAL quasars fundamentally the same as all
other BAL quasars, and their magnificent spectra are observed because
of a select range of viewing angles?  Or do FeLoBAL quasars mark a
special stage in quasar evolution?  Or are  both factors important? 

The spectral energy distribution provides a potentially powerful probe
of quasar  physics.  Three components are generally observed: the
accretion disk emission, typified by a blue optical-UV continuum
spectrum; the torus, reprocessed continuum radiating in the infrared
and identified by a break at 1 micron; and the X-ray
emission, thought to originate in  inverse Compton scattering of the
accretion disk emission by energetic electrons.  Different relative
strengths of these components are observed among active galactic
nuclei (AGN) and quasars.  For example, the prominence of the torus
component relative to the optical power  
law has been observed to differ from object to object.  While detailed
models of the torus may be fit to extensive photometry data extending
to the far infrared \citep[e.g.,][]{lyu17}, fainter and less
well-studied objects are limited to detection of the presence of hot
dust that results in the upturn of the spectrum toward wavelengths
just longward of 1 micron.  Those that do not show this upturn are termed
Hot-Dust Deficient (HDD), and \citet{lyu17} found that HDD quasars
have relatively lower accretion rates than other PG quasars.    There may also be a
connection between the presence of 
a hot dust component and BAL outflows; a strong hot-dust component has
been found to be correlated with BAL strength \citep{zhang14},
\ion{C}{4} emission blue shift \citep{temple21}, and [\ion{O}{3}]
blueshift \citep{calistrorivera21}.  

Reddening from dust along the line of sight can also affect what we see,
and may also affect quasar selection.  Quasars are most easily
differentiated from stars by their blue optical band colors
\citep[e.g.,][]{green86}; red quasars may be confused with low-mass
stars. Modern data provides efficient selection of red quasars via
their radio emission \citep[e.g.,][]{glikman04} or {\it WISE} colors
\citep[e.g.,][]{stern12}.  They may also be detected spectroscopically
\citep[e.g.,][]{klindt19}.  Reddened quasars show a large range of
interesting differences from the more typical blue quasars.  For
example, a larger fraction of LoBAL and FeLoBAL quasars are found
among  reddened quasars \citep{urrutia09, fynbo13, krogager15,
  krogager16}.   Reddened quasars have also been found to show excess
near-infrared emission \citep{calistrorivera21}, and blue shifted
[\ion{O}{3}] line emission       \citep{dipompeo18}.  

Reddening in quasars and BAL outflows have also been associated with
quasar evolution \citep[e.g.,][]{urrutia09}.  A popular scenario holds
that quasar activity is initiated by mergers, with the early stages
shrouded by dust \citep{sanders88b,hopkins05}.  An object with a high
accretion rate may be initially observed as a red quasar
\citep[e.g.,][]{glikman12}.  Such an object produces 
copious radiation that can destroy or eject the dust and 
reveal the brilliant quasar. Eventually, the fuel source decreases,
turning off the quasar and the star formation, and 
producing a dormant early-type galaxy \citep[e.g.,][]{klindt19}.  Some
surveys of dust-reddened quasars have found a high fraction of LoBAL
quasars, prompting the hypothesis that the LoBAL phenomenon is associated with
young quasars \citep[e.g.,][]{urrutia09}.  The discovery of an enhanced
merger fraction in reddened quasars \citep[e.g.,][]{glikman15} and
LoBAL quasars \citep{lazarova23}  supports this idea.  

Variability offers another way to probe the properties of
quasars. It has long been known that the variability time
scale is correlated with luminosity, or, equivalently, that the
variability amplitude is anticorrelated with luminosity.  First shown
in X-ray data \citep[e.g.,][]{barr86,lp93,green93,nandra97}, this
trend is now well established in the optical band pass
\citep[e.g.,][]{vandenberk04, wilhite08, kelly09, macleod10, zuo12,
  gallastegui14, simm16, caplar17, sun18, laurenti20, suberlak21,
  decicco22, yu22}.  Various authors assert a further dependence on
Eddington ratio \citep[e.g.,][]{zuo12,simm16,yu22} and /  or black hole mass 
\citep[e.g.,][]{wilhite08, kelly09, macleod10, zuo12,suberlak21}. 

So far, our detailed studies have focused solely on FeLoBAL quasars, a
fact that provokes the question: are the profound differences between
FeLoBAL quasars and the unabsorbed matched comparison sample unique to
FeLoBALQS, or are they shared by other BAL quasars?  Confirmation that
they are unique would support an evolutionary origin of the FeLoBAL
quasar phenomenon.

Our paper is organized as follows.  Because this paper is the fourth
is a series, \S\ref{review} gives a review of the principal results
from the first three papers \citep{choi22, leighly22, choi22b}.
\S\ref{data} describes the data used in this paper.
\S\ref{preliminary} presents the extraction of characteristic
properties used to describe the spectral energy distributions and the
{\it   WISE} variability.  \S\ref{dist_comp} describes the
distributions of the extracted SED and {\it
  WISE} variability parameters. \S\ref{correlations} presents the
results of correlations of the optical emission line \citep{leighly22}
and {\it SimBAL} \citep{choi22} parameters with the SED and {\it WISE}
variability parameters. \S\ref{comp_lobal} reports the results of the
comparison of the optical emission line properties of a sample of
low-redshift LoBAL quasars with those of the FeLoBAL quasars and unabsorbed comparison
sample presented in \citet{leighly22}.  \S\ref{discussion} discusses
the results of the analysis and presents a speculative scenario to
explain the two groups of FeLoBAL quasars found in the first three
papers of this series.  Finally, \S\ref{future} gives a brief summmary
of results and conclusions.  

Finally, we note special terminology used throughout the paper.  We
plot our spectra as a function of wavelength exclusively.  We use the
terms ``steeper'' or ``bluer'' when referring to spectrum or spectral
energy distribution that is relatively brighter at short wavelengths,
and ``flatter'' or ``redder'' for one that is relatively brighter at
long wavelengths.  

\section{Brief Review of Papers I, II, and III}\label{review}

\citet{choi22} [Paper I] reported the results of analysis of a 
sample of 50 low-redshift ($0.66 < z < 1.63$) FeLoBAL quasars using
{\it SimBAL}.  The forward-modeling spectral synthesis code {\it
  SimBAL} was introduced in \citet{leighly18}; additional discussion
and features are discussed in \citet{leighly19} and \citet{choi20}.

{ The near-UV spectra of FeLoBAL quasars contain several to
  thousands of iron absorption lines, each characterized by its own
  critical density, excitation energy, and oscillator strength
  \citep[e.g.,][]{hall02}.  Many  investigators have used these lines
  and photoionization physics to   analyze the properties of their
  outflows \citep{dekool01, arav01,     everett02, dekool02a,
    dekool02b,arav08, korista08, moe09, dunn10,     bautista10,
    aoki11,     shi16, hamann19b}.  High densities ($n$) are
  recognized by the presence of lines from transitions with high
  critical density, while their absence signals a low density
  outflow; in fact, an estimate of the density can be 
  obtained by visual examination of the spectrum \citep{lucy14}.  High
  ionization parameters $U$ are recognized by the presence of
  transitions from relatively rare ions (e.g., from  high energy
  levels that are difficult to populate in a photoionized  gas), or
  low oscillator strengths \citep[e.g.,][]{lucy14}. Such outflows
  require a large column density to build up a detectable absorption
  line that is obtained from a large ionization parameter, i.e., a large
  Str\"omgren sphere 
  \citep[e.g.,][]{leighly09}.  The robust measurements 
  of $U$ and $n$ possible in FeLoBAL quasar spectra yield equally
  robust estimates  of the location of  the outflow, given by the
  standard formula  ($R=\sqrt{Q/4\pi U n     c}$).  The difference
  between {\it SimBAL} and previous analyses is that {\it SimBAL}
  uses forward modeling,  a venerable   technique ubiquitously
  used in X-ray spectral  modeling for decades.  The advantage of
  forward modeling over inference is that it can handle line blending
  and makes use of constraints obtained by lines that are not present
  in the spectrum.}  As shown in \citet{choi22} Appendix C,
  {\it SimBAL}  produces values of the location of outflow that are
  commensurate  with those obtained through inference methods.
  
Because of the rich information available in the rest near-UV spectra,
the {\it SimBAL} analysis resulted in fully characterized outflows.
The physical conditions of the outflow, including the ionization
parameter, density, column density, and covering fraction, were
measured for 60 outflow components (with some objects having more than
one component).  A principal result was that the
FeLoBAL outflows are present at a large range of distances from the
central engine ($0 \lesssim \log R \lesssim 4.4$ [pc]), with no
evidence that disk winds (with $R \ll 0.01$ pc) are manifest as UV
outflows.  For most of the objects, the 
outflow velocity was inversely related to the distance from the
central engine, as might be expected from radiative line driving.
However, eleven objects had FeLoBAL gas that lay less than 10 pc from
the central engine with low ($v_\mathrm{off} > -2000\rm \, km\,
s^{-1}$) outflow velocities.  These objects were
identified as ``loitering outflow''
objects that represent a newly discovered class of FeLoBALQs. This 
paper discussed the potential origins and
acceleration mechanisms of FeLoBAL outflows as a function of distance
from the active nucleus.

Among the 50 objects analyzed by \citet{choi22}, 30 had redshifts low
enough ($z<1$) that the H$\beta$ and [\ion{O}{3}] lines were visible
in the spectra.  \citet{leighly22} [Paper II] analyzed the rest-frame
optical emission-line properties of these spectra, focusing on the rich
diagnostic information obtainable from the H$\beta$ / [\ion{O}{3}] /
\ion{Fe}{2} region.  A
comparison sample of  132 unabsorbed quasars matched in
redshift, signal-to-noise ratio, and luminosity was analyzed in
parallel.  The description of the construction of the comparison
sample may be found in \citet[][\S 2.1]{leighly22}.  They found the  
anticorrelation between [\ion{O}{3}] equivalent width and
\ion{Fe}{2}/H$\beta$ that is ubiquitous in quasars.  They developed a
summary statistic called $E1$ (Fig.\ 3 and \S 3.1 in \citet{leighly22})
that characterizes this anticorrelation.  They found that while the
unabsorbed objects showed a single-peak distribution of $E1$, the
FeLoBALQs showed a broader and double-peaked distribution. 
$E1$ is strongly correlated with accretion rate
\citep[][Fig.\ 12]{leighly22}, implying that low-redshift FeLoBALQs
are characterized by either a low accretion rate ($E1<0$) or a high
accretion rate ($E1>0$).  { In addition, \citet{leighly22} reported that
the H$\beta$ FWHM width is systematically larger in
FeLoBAL quasars than in unabsorbed objects \citep[][\S 3.4.1,
  Fig.\ 13]{leighly22}, implying a larger-inclination viewing angle
{\it or} a consequence of diminished emission in the core of the line
profile.}

Finally, \citet{choi22b} [Paper III] compared the results of the {\it
  SimBAL} analysis presented in Paper I with the results of the
optical emission-line analysis in Paper II.  They found that the
outflow properties were also correlated with the accretion rates.  The
high accretion rate objects showed faster outflows closer to the central
engine, while the low accretion rate objects (the loitering outflow
objects among them) showed near-zero outflow velocities close to the
central engine, and higher velocities at large distances
\citep[][Fig.\ 4]{choi22b}.  High accretion rate objects showed
significantly lower volume filling factor outflows than the low
accretion rate objects.
These results are especially profound among the objects with
smaller-scale ($\log R < 2$ [pc]) outflows, and  provide further
support that, at least for the smaller-scale outflows, there are two
populations of FeLoBAL quasars. 

\section{Data} \label{data}

Most of the data and analysis results used in this paper were taken from
\citet{choi22}, \citet{leighly22}, and \citet{choi22b}.
Additionally, in this paper, we include broad-band photometry from the
30 low-redshift FeLoBALQs and the comparison sample identified in
Paper 2  (\S\ref{sed_modeling}), the {\it WISE} and {\it   NEOWISE}
photometry (\S\ref{wise_variability}),  measurements of the 
optical emission lines from a sample of LoBAL quasars
(\S\ref{lobals}), and an IRTF observation of SDSS~J144800.15$+$404311.7
(\S\ref{irtf}). 

\subsection{Spectral Energy
  Distributions}\label{sed_modeling}

We compiled the photometry
information for the FeLoBALQs and comparison sample from SDSS
\citep{blanton17}, 2MASS \citep{skrutskie06}, 
UKIDSS \citep{lawrence07}, and {\it WISE} \citep{wright10}.  In a very few cases
we also included GALEX \citep{martin05}.  These data were corrected for
redshift using the values presented in \citet{leighly22} and
Milky Way reddening \citep{ccm88}.  The results are shown 
in Fig.~\ref{phot1}.  If both  UKIDSS and 2MASS data were available,
the higher-sensitivity UKIDSS data are shown.   

\begin{figure*}[!t]
\epsscale{1.0}
\begin{center}
\includegraphics[width=5.5truein]{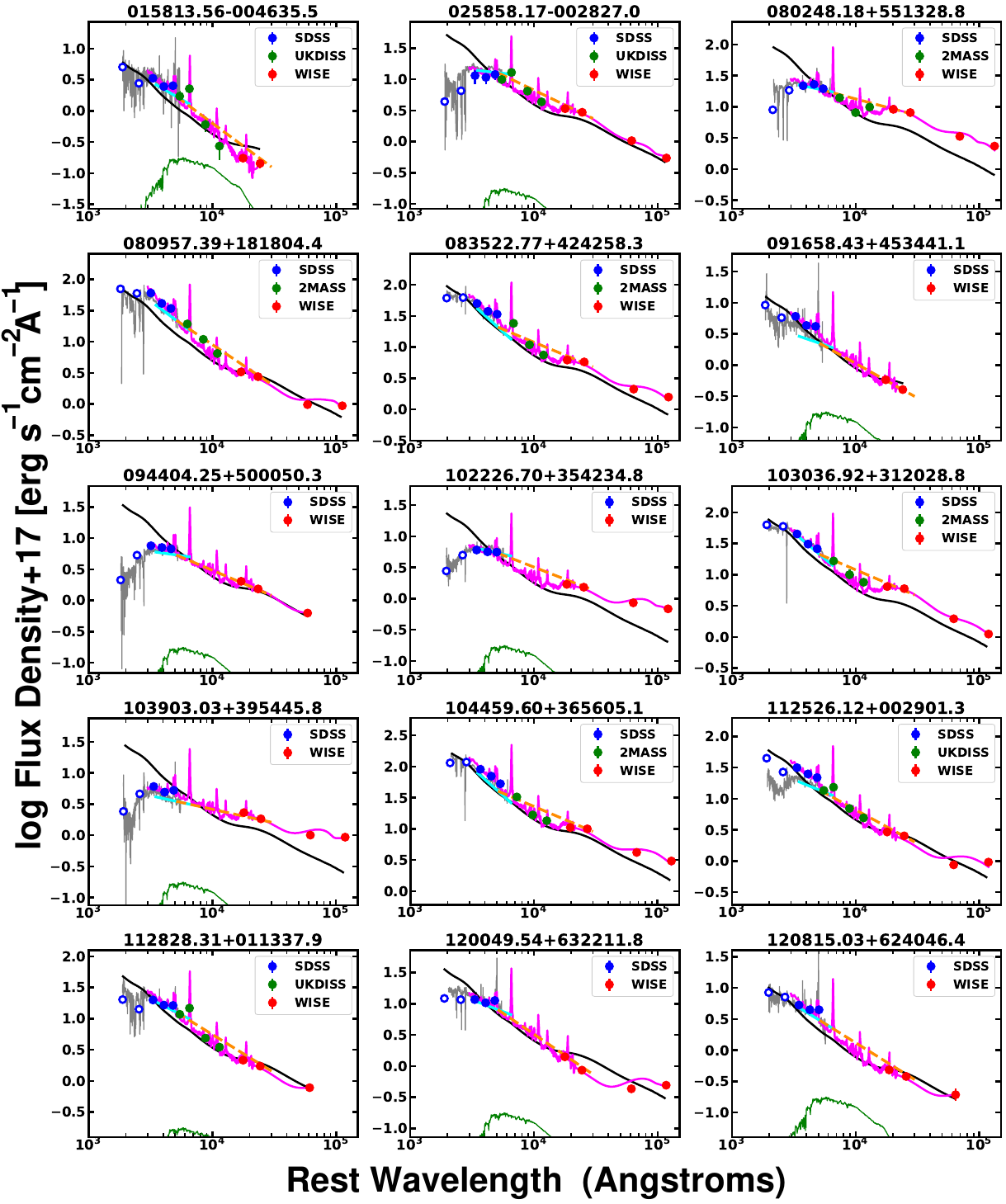}
\caption{Photometry spectral energy distributions of the FeLoBALQ
  sample.   The magenta line shows the empirical model
  (\S\ref{sed_composites}), while the 
  black line shows the \citet{krawczyk13} continuum normalized to the
  model continuum at 9730\AA\/, and the grey line shows the heavily
  rebinned spectrum.  The green line shows the estimated contribution
  of the host galaxy, which is weak or absent in most cases.  Open
  markers denote points not used 
  in the SED fitting.  The orange dashed (cyan solid) lines show
  $\alpha_{oi}$ (power law index) defined in \S~\ref{sed_modeling}. The
  SED shapes range from a simple SMC-reddened Seyfert (e.g.,
  SDSS~J080248.18$+$551328.8) to objects that appear to lack a
  near-infrared bump (e.g., SDSS~J080957.39$+$181804.4) to objects
  with very red optical spectra (e.g.,  SDSS~124014.04$+$444353.4).  
  \label{phot1}}
\end{center}
\end{figure*}

\addtocounter{figure}{-1} 
\begin{figure*}[!t]
\epsscale{1.0}
\begin{center}
\includegraphics[width=5.5truein]{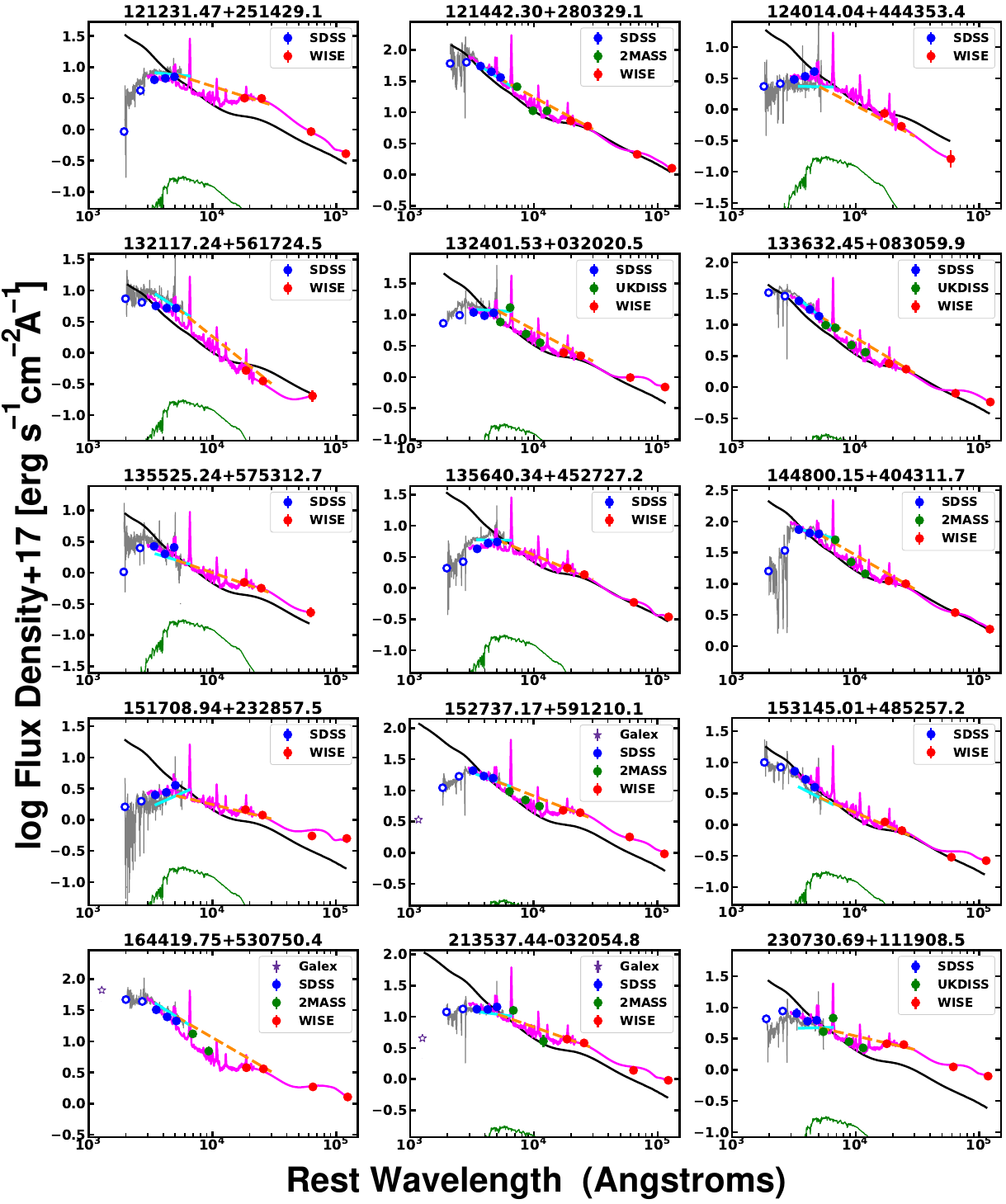}
\caption{Continued.  \label{phot2}} 
\end{center}
\end{figure*}

{ At low redshift and low luminosity, the host galaxy may
  contribute significantly to the spectral energy distribution.  To
  determine whether galaxy subtraction is necessary, we} made a
zeroth-order estimate of the contribution of the host galaxy to 
the photometry using Equation 1 of \citet{richards06}.  This
relationship, adapted from \citet{vandenberk06}, relates the
luminosity density of the quasar at 6156\AA\/ to the luminosity
density of the host galaxy at the same wavelength after adjusting for
the Eddington 
ratio.  We estimated the luminosity density of the quasar from the
photometry fits described below.  We used the Eddington ratios from
\citet{leighly22}. A 5-Gyr old elliptical template from the SWIRE
template library \citep{polletta07} was scaled to the galaxy
flux-density estimates, and these are shown in Fig.~\ref{phot1}.  With
the exception of one or two cases where the  luminosity and acccretion
rate are low, the quasar is very much brighter than the host
galaxy.  Because one of the criteria defining the 
unabsorbed comparison sample was the luminosity, which was based on
the 3-micron flux density \citep{leighly22}, we expect very similar
properties for the comparison sample.  We therefore do not correct the
photometry for the host galaxy emission.

The SEDs of the FeLoBALQs show a wide range of shapes.  There are some
objects that appear indistinguishable from a slightly reddened average
quasar (e.g, SDSS~J104459.60$+$365605.1, SDSS~J103036.92$+$312028.8)
or a somewhat more reddened average quasar  (e.g.,
SDSS~J080248.18$+$551328.8). Others have more peculiar SEDs.  Several
Seyfert 1.8 objects show an abruptly flatter optical spectrum
shortward of $\sim 6000$\AA\/ than the extrapolated
optical-to-infrared slope would predict (e.g.,
SDSS~J120049.54$+$632211.8, SDSS~J124014.04$+$444353.4).  Finally,
some objects lack the upturn toward longer wavelengths expected from
the torus (e.g., SDSS~J080957.39$+$181804.4,
SDSS~J121442.30$+$280329.1). 

\subsection{{\it WISE} Photometry Variability}\label{wise_variability}

We used {\it WISE} photometry to investigate the near-IR variability
of the FeLoBALQs and the comparison sample between 2010 and 2020.  We
used {\it unTimely}, the time-domain catalog of 
detections from the Wide-field Infrared Survey Explorer ({\it WISE}) and
{\it NEOWISE} missions.  The catalog was constructed from
the time-resolved unWISE coadded images, i.e., the stacked images from
each biannual pass   \citep{meisner22}.   The catalog yields
mid-infrared light curves from the W1 and W2 {\it WISE} filters (3.4 and 4.6
$\mu$m) respectively.  The sampling rate and duration are very
uniform: every 6 months between 2010 and 2020, with the exception of a
3-year gap between the {\it WISE} and {\it NEOWISE} missions; each light curve
includes 16 or 17 points.  For the $0.8 \lessapprox z < 1.0$
objects investigated in this paper, the 10-year time span is equivalent
to about 5 years at $\sim 1.8$ and $\sim 2.5\rm \, \mu m$ in the rest
frame.  The lightcurves sample principally the hot-dust
component of the torus and secondarily the accretion disk 
emission extending into the near infrared.

We downloaded the data using the {\tt unTimelyCatalogExplorer}
\citep{kiwy22}.  We investigated the variability properties by
selecting star candidates ($-0.2 < W1-W2 <0.2$) in a 500 arcsecond
region around the targets with magnitudes similar to those of the
targets.  Most classes of stars are not substantially variable.
However, analysis of the star properties found that the light curves
were significantly variable.  This result implies that the errors in
the catalog are underestimated.  We devised a correction of the
errors.  This analysis and the revised errors are given in
Appendix~\ref{varerr}. 

\subsection{A Sample of LoBAL Quasars}\label{lobals}

In \citet{leighly22}, we analyzed the optical properties of a sample
of FeLoBAL quasars with redshifts less than 1.0.  At these redshifts, the
rest-frame emission lines between $\sim 3500$ and $\sim 5500$\AA\/ may
be modeled if the bandpass extends to sufficiently long 
wavelengths and if the sky emission, which dominates observed-frame
emission longward of $\sim 8000$\AA\/, is cleanly subtracted.  While
this redshift range is far too low to identify high-ionization BAL
quasars, it is low enough that LoBAL quasars can be identified, and
the same analyses performed on them.

We used the SDSS DR12 BAL quasar catalog \citep{paris17} to build a
sample of LoBAL quasars.  To match the comparison sample used in
\citet{leighly22}, we chose the redshift range $0.75 < z < 1.0$
\citep[][Fig.\ 2]{leighly22}.  We found 176 objects in that redshift
range.  The 29 FeLoBAL quasars were excluded.  Another 57 were
rejected because they did not show evidence for \ion{Mg}{2}
absorption.  In many cases, the objects were extreme iron emitters,
and the strong upturn near 2600\AA\/ was apparently identified as
recovery from a longer-wavelength absorption line
\citep[e.g.,][Fig.\ 6]{lm06}.  An additional 26 were excluded because
the long wavelength region was too noisy to analyze.

\subsection{IRTF Observation of SDSS~J1448$+$4043}\label{irtf}

SDSS~J144800.15$+$404311.7 was observed using  IRTF {\it SpeX}
\citep{rayner03} as part of program 2015A041 on May 1, 2015
for a total exposure time of 40 minutes using a 0.8 arc sec slit and
the short wavelength cross-dispersing grating. The near-infrared
observation was done to observe the predicted metastable
\ion{He}{1}*$\lambda 10830$ absorption line.   A standard ABBA
integration scheme was used. The A0 star HD 128039 was used for flux
and telluric corrections.   The spectra were reduced 
and the telluric correction applied  in the standard manner using
{\tt Spextool} and accompanying software \citep{cushing04, vacca03}.
The spectrum was corrected for reddening using $E(B-V)=0.011$ \citep{sf11}
and redshift using $z=0.805$ \citep{paris18}.  

\begin{figure*}[!t]
\epsscale{1.0}
\begin{center}
\includegraphics[width=5.0truein]{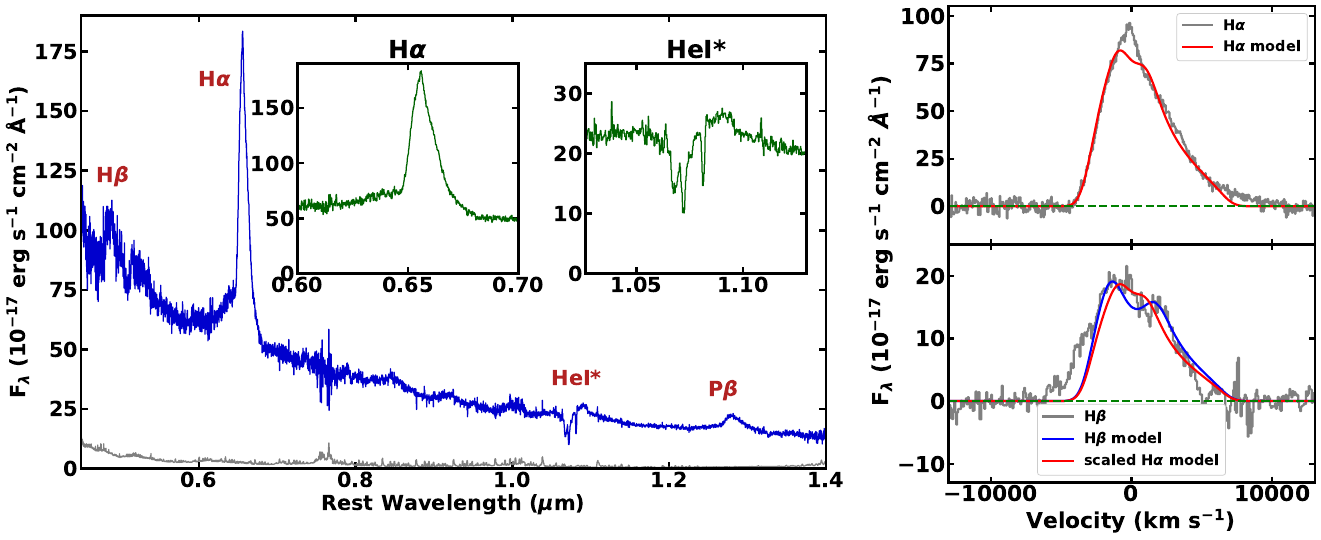}
\caption{{\it Left:} The IRTF {\it SpeX} spectrum of
  SDSS~J144800.15+404311.7. Principal emission lines are labeled.  The
  inset panels highlight the unusually-shaped H$\alpha$ line and the
  \ion{He}{1}*$\lambda 10830$ absorption line. {\it Right:}  The
  H$\alpha$  emission line from an IRTF SpeX observation of
  SDSS~J1448$+$4043 is fit well by a diskline profile plus a very
  broad and blue-shifted pedestal (not shown). The H$\beta$ line can
  also be modeled with the diskline profile.
  \label{irtf_spectrum}} 
\end{center}
\end{figure*}

The full spectrum is shown in Fig.~\ref{irtf_spectrum}.  The left
inset panel shows the H$\alpha$ line, which has a unusual shape in
this object.  It is not symmetric; it has an extended blue wing and a
steeply rising blue side, with a gradual decline toward longer
wavelengths.    While red wings have been
observed in Balmer lines from AGNs previously
\citep[e.g.,][]{lamura09}, the steep blue side is rare.  This profile
is an example of line emission from an accretion disk viewed at a low
angle from the normal \citep[e.g., Fig.\ 2 of][]{ch89}.
The right inset panel shows \ion{He}{1}*$\lambda 10830 $
absorption line.  The shape matches that of the \ion{He}{1}*$\lambda 3889$
absorption line, and very roughly corresponds to the {\it SimBAL}
inferred opacity for the \ion{Mg}{2} absorption.

\section{Preliminary Analysis}\label{preliminary}

We introduced the data used in this paper in \S\ref{data}.  In this
section, we extract parameter measurements that characterize the data,
and that can be used to compare among samples and with parameters
characterizing the accretion rate and outflow properties.  

\subsection{Spectral Energy
  Distributions: Parameters}\label{sed_parameters}

\begin{figure*}[!t]
\epsscale{1.0}
\begin{center}
\includegraphics[width=6.5truein]{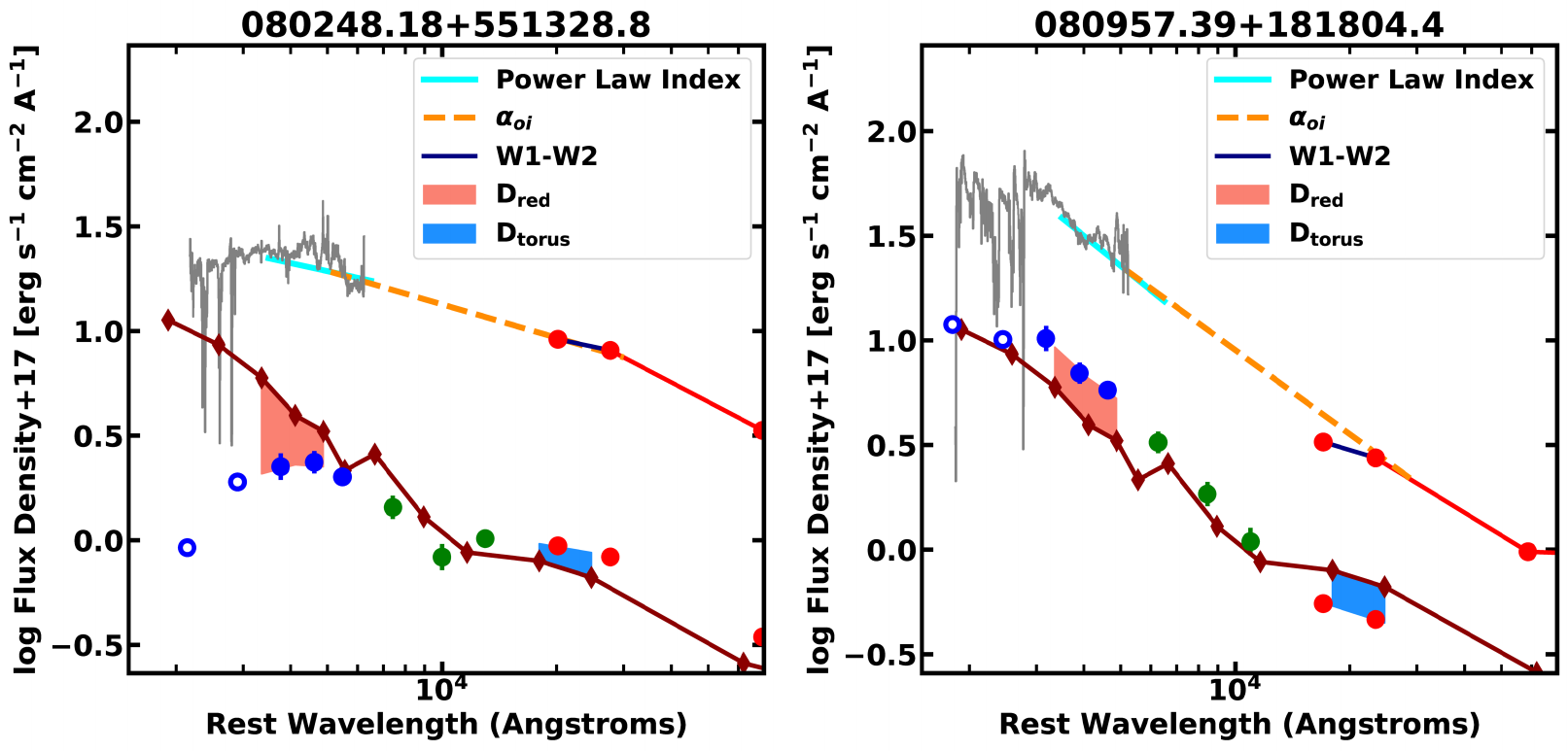}
\caption{Illustrative example of defined photometry parameters.  The
  solid circular points show the flux densities derived from the
  photometry, the grey line shows 
  the heavily rebinned spectrum, and the dark red line { and red
    diamonds} shows the
  composite photometry spectrum from the unabsorbed objects.  { The
    upper plot shows parameters defined in the rest frame for each
    object individually (the power law
    index, $\alpha_{oi}$, and W1-W2), while the lower one shows the
    parameters defined from the normalized spectra ($D_\mathrm{red}$
    and $D_\mathrm{torus}$)}.  The colored patches show the regions 
  of the spectra where  the  D$_\mathrm{red}$ and D$_\mathrm{torus}$
  parameters   are 
  defined; see text for details.    The left panel shows an example of
  an overall   flat   spectrum with negative $D_\mathrm{red}$ due to
  significant   reddening and positive $D_\mathrm{torus}$ due to a
  prominent   torus.  The right panels   shows an overall steep
  (blue)   spectrum with positive  $D_\mathrm{red}$ and negative
  $D_\mathrm{torus}$.     
  \label{phot_params}}
\end{center}
\end{figure*}

We compiled five empirical measurements to describe the shape of the
continuum; { these are decribed in Table~\ref{definitions} and}
illustrated in Fig.~\ref{phot_params}.   The 
effective power law index between 4500 and 5500\AA\/ was constructed
from the continuum component of the spectral fitting model presented
in \citet{leighly22}.  We also defined $\alpha_{oi}$, the
point-to-point flux density slope between 5100\AA\/, also extracted 
from the spectral fitting model, and the 3 micron flux density
interpolated from the {\it WISE} photometry.  Both of these parameters are
potentially sensitive to both reddening and intrinsic differences in
spectral shape.  The third parameter is {\it WISE} W1-W2 in magnitude
units.  Most of our 
objects have redshifts between 0.8 and 1.0, which means that W1 and W2
measure the flux densities near 1.7--1.86 and 2.3--2.56 $\mu$m
respectively. The values fall on the upturn from the 1-micron
dust-sublimation break toward the torus infrared bump; see, for
example, the \citet[][]{krawczyk13} composite spectrum. Therefore, the
W1-W2 color is a measurement of the prominence of the hot dust portion
torus in the SED; a larger value of W1-W2 implies that the spectrum is
flatter (redder) and the torus is more prominent.  The final two
parameters are defined in \S\ref{sed_composites}.  

\movetabledown=0.5in
\begin{deluxetable*}{lcl}
\tablecaption{Continuum Parameter Definitions\label{definitions}}
\tablehead{
\\
\colhead{Parameter Name} & \colhead{Computational Defintion} &
\colhead{Utility and Interpretation} \\
}
\startdata
Power Law Index &
$\frac{\log_{10}(F_\lambda(5500))-\log_{10}(F_\lambda(4500))}{\log_{10}(5500)-\log_{10}(4500)}$
  & Local optical continuum index\tablenotemark{a} \\
$\alpha_{oi}$ &
$\frac{\log_{10}(F_\lambda(30000))-\log_{10}(F_\lambda(5100))}{\log_{10}(30000)-\log_{10}(5100)}$
& Overall optical-NIR continuum shape\tablenotemark{b} \\
W1-W2 & $W1-W2$ & Color that measures the strength of 1-micron upturn  \\
$D_\mathrm{red}$ &
$\frac{1}{3}\sum_{r,i,z}(\log_{10}F_\mathrm{obs}-\log_{10}F_\mathrm{composite})$ &
Isolates the relative optical band reddening\tablenotemark{c} \\
$D_\mathrm{torus}$ & $\frac{1}{2}\sum_{W1,W2}
(\log_{10}F_\mathrm{obs}-\log_{10}F_\mathrm{composite})$ & Isolates the relative
torus strength\tablenotemark{c} \\
\enddata
\tablenotetext{a}{Computed from the continuum portion of the optical
  emission line models   presented in \citet{leighly22}.}
\tablenotetext{b}{Computed from interpolations of the log flux density
  from the photometry.}
\tablenotetext{c}{The composite SED constructed from the unabsorbed
  sample, as described in \S\ref{sed_composites}.}
\end{deluxetable*}

\subsection{Spectral Energy
  Distributions: Composites}\label{sed_composites}

The empirical parameters have the advantage that they can be robustly
measured from the SEDs.  However, their interpretions may be
ambiguous.  For example, the $\alpha_{oi}$ parameter may be larger
(flatter or redder) either because of reddening or because of strong
hot dust and torus  emission.  To address this limitation, we also
perform photometry SED  
fitting to the rest-frame optical through near IR bands, and use the
results, as described below and as illustrated in
Fig.~\ref{phot_params}, to separate the contributions of the rest-frame
infrared and optical bands to $\alpha_{oi}$.  

We used the simplified and parameterized model described by
\citet{temple21} on the $F_\lambda$ SED spectra, { modified
  appropriately for our
  BAL quasar data.}   Their model consisted of a broken power law with  
fixed break wavelength at 2820\AA\/ and indices of $-1.52$ and $-1.84$
for the short and long wavelengths respectively.  To that they added
line emission extracted from the UV-optical template from
\citet{francis91} and the NIR template from \citet{glikman06} and
convolved with the relevant filter curves.  They 
added a blackbody component with fixed temperature $T=1280\rm \, K$.
The result was convolved with an SMC reddening curve.   { The torus
  emits over a broad wavelength range, and it cannot be
  modeled using a single blackbody when the {\it WISE} W3 and W4 points
are present. We included a second blackbody component with fixed
temperature $T=350\rm \, K$ when necessary to model the {\it WISE} W3
and W4 points.}  The free
parameters were therefore the normalizations of the fixed-shape
optical continuum and the two blackbodies, and the SMC $E(B-V)$, which
was allowed to take negative as well as positive values to model
spectra both intrinsically steeper (bluer) and flatter (redder) than
the fixed broken power-law values.   Following \citet{temple21}, the
photometry error minimum is assigned to be 0.05 magnitudes.

Although we used the same model as \citet{temple21}, our analysis 
differed somewhat because our data are more limited.  All of the
objects in the \citet{temple21} sample were required to have UKIDSS
photometry, a fact that meant that the models were well constrained
near the crossover point between the  broken power law modeling the
accretion disk emission at shorter wavelengths and the blackbody
modeling the torus emission at longer wavelengths.  In contrast, only
7 of our 30 low-redshift FeLoBAL quasars had UKIDSS 
photometry, and 13 lacked any near-infrared photomtry. Also, we
ignored {\it WISE} photometry points with signal-to-noise ratios less than
3.  Finally, we only fit longward of 3000\AA\/ to avoid the BAL
absorption.  The result is that there are objects that have as few as 
five photometry points, a fact that might result in model dependence.
We therefore adopted the following approach to 
parameterize our data: we used the model fit parameters to provide
the normalization at 1 micron so that we could create composite
spectra, and then compared the model fit results of the individual
objects to the composite spectra. 

For objects in our redshift range, the $J$, $H$, $K$ bands correspond
to approximately rest frame 6500, 8750, and 11300 \AA\/ (computed for
$z=0.9$), i.e., spanning our 1-micron normalization point.  
Since many of our objects lack near infrared photometry, we first
investigated whether the model fits provided sufficiently robust
normalizations at 1 micron.  We tested this by fitting all of the
comparison sample objects that have NIR photometry (62 objects), and then
repeating the model fitting with the NIR points removed.  We found
that the mean of the ratio of the no IR photometry normalization factor 
to the nominal one was 1.008, with a standard deviation of 0.09, i.e.,
about 10\%.    Considering that we
performed our subsequent analysis on the log of the flux densities,
this small uncertainty was deemed acceptable.  We speculate that the
highly  
constrained nature of the model and the limited redshift range of the
targets is the origin of the small uncertainty.  Specifically, the
fixed-temperature blackbody normalizations were well constrained by
the {\it WISE} data, while the normalization and reddening of the fixed-shape
continuum component were well constrained by the SDSS photometry.

\begin{figure*}[!t]
\epsscale{1.0}
\begin{center}
\includegraphics[width=5.0truein]{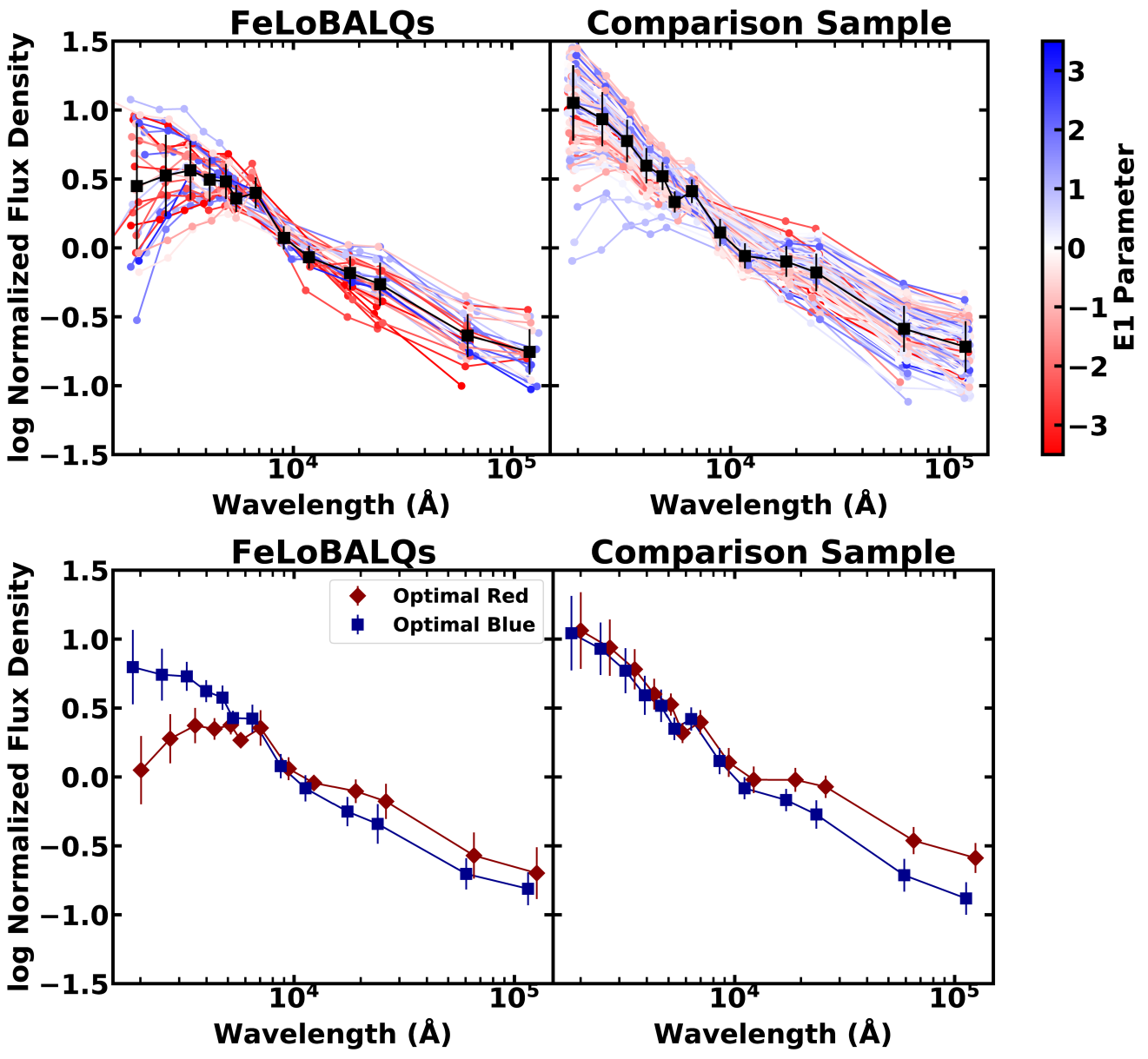}
\caption{{\it Top:} The normalized photometry for the FeLoBAL quasars
  (left) and   the comparison spectrum (right) colored according to
  the $E1$   parameter.  Overlaid are the mean normalized spectra
  binned by   filter, where the error bars show the standard deviation
  of the   points.  {\it Bottom:}  The optimal red and optimal blue
  spectra constructed from the FeLoBALQs and comparison sample
  spectra as described in the text.  The points are slightly shifted
  in wavelength for plot clarity. The peak near 6500\AA\/ is
  H$\alpha$. Both sets of objects show 
  significant variance in their near-IR components.  The FeLoBALQs
  clearly show deficits at the shorter wavelengths consistent with
  reddening.   \label{photometry_sed}} 
\end{center}
\end{figure*}

Armed with the normalized photometry, we proceeded to make composite
spectra.   Because of the limited redshift range, we could average
filter by filter without any K-correction.  We used a straight average
rather than a weighted average to avoid  bias towards brighter
objects.   Fig.~\ref{photometry_sed} shows all of the normalized
photometry and the mean spectra for both the comparison sample data
and the FeLoBAL quasar spectra.   

The photometry SEDs show a range of shapes and it appears that these
are not obviously correlated with the $E1$ parameter. In the optical
band, the spectral shape range seen in Fig.~\ref{photometry_sed} could
be due to differences in reddening or the variations in the intrinsic
slope of the power law.  In the near infrared, the relative strength
of the hot dust component may change the spectral shape
\citep[e.g.,][]{temple21}.  We defined two deviation parameters, one
in the optical band called $D_\mathrm{red}$, and another in the
near-infrared band called $D_\mathrm{torus}$.  For $D_\mathrm{red}$,
we used the SDSS $r$, $i$, and $z$ measurements; the shorter bands
were ignored because they may be contaminated by the BALs. For
$D_\mathrm{torus}$, we used {\it WISE} W1 and W2 because those data
exist for the entire sample; some objects are not detected in {\it
  WISE} W3 and W4.   To create these deviation parameters, we first 
performed a linear interpolation on the log  flux densities from each
object onto the wavelengths from the composite spectrum of the
comparison sample.   The deviations were computed as the mean
difference between the log of the normalized flux densities inferred
from the photometry  and the log of the corresponding flux densities
of the composite spectrum (i.e., from the three values in the optical
band and from the two values in the near infrared). These parameters
are illustrated in Fig.~\ref{phot_params}, and their definitions are
given in Table~\ref{definitions}.

The sign of the deviation indicates whether the photometry lies above
the composite model (positive) or below (negative).  By definition,
the means of $D_\mathrm{red}$ and $D_\mathrm{torus}$ are very close to
zero for the comparison sample.  The standard deviations for the
comparison sample are 0.13 and 0.12 for $D_\mathrm{red}$ and
$D_\mathrm{torus}$ respectively.  In contrast, the mean of
$D_\mathrm{red}$ for the FeLoBALQs is $-0.12$, implying that
FeLoBALQs are generally more reddened than unabsorbed objects.
Reddening is known to be more prevalent among BAL quasars compared
with unabsorbed objects \citep[e.g.,][and references
  therein]{krawczyk15}, so this result is 
no surprise.   The standard deviation is 0.17 for the 
FeLoBALs, implying there is also more scatter than among the
comparision objects. Indeed, 11 objects (1/3 of the sample) have
$D_\mathrm{red} < 0$.  Likewise, the mean of $D_\mathrm{torus}$ for
the FeLoBALQs is $-0.08$, which implies that the torus emission in the
FeLoBALQs is weaker than among the unabsorbed quasars.  The standard
deviation of $D_\mathrm{torus}$ for the FeLoBALQs is 0.14, comparable
to that of the unabsorbed quasars, and 11 objects (1/3 of the sample)
show positive values of $D_\mathrm{torus}$.

\begin{figure*}[!t]
\epsscale{1.0}
\begin{center}
\includegraphics[width=5.0truein]{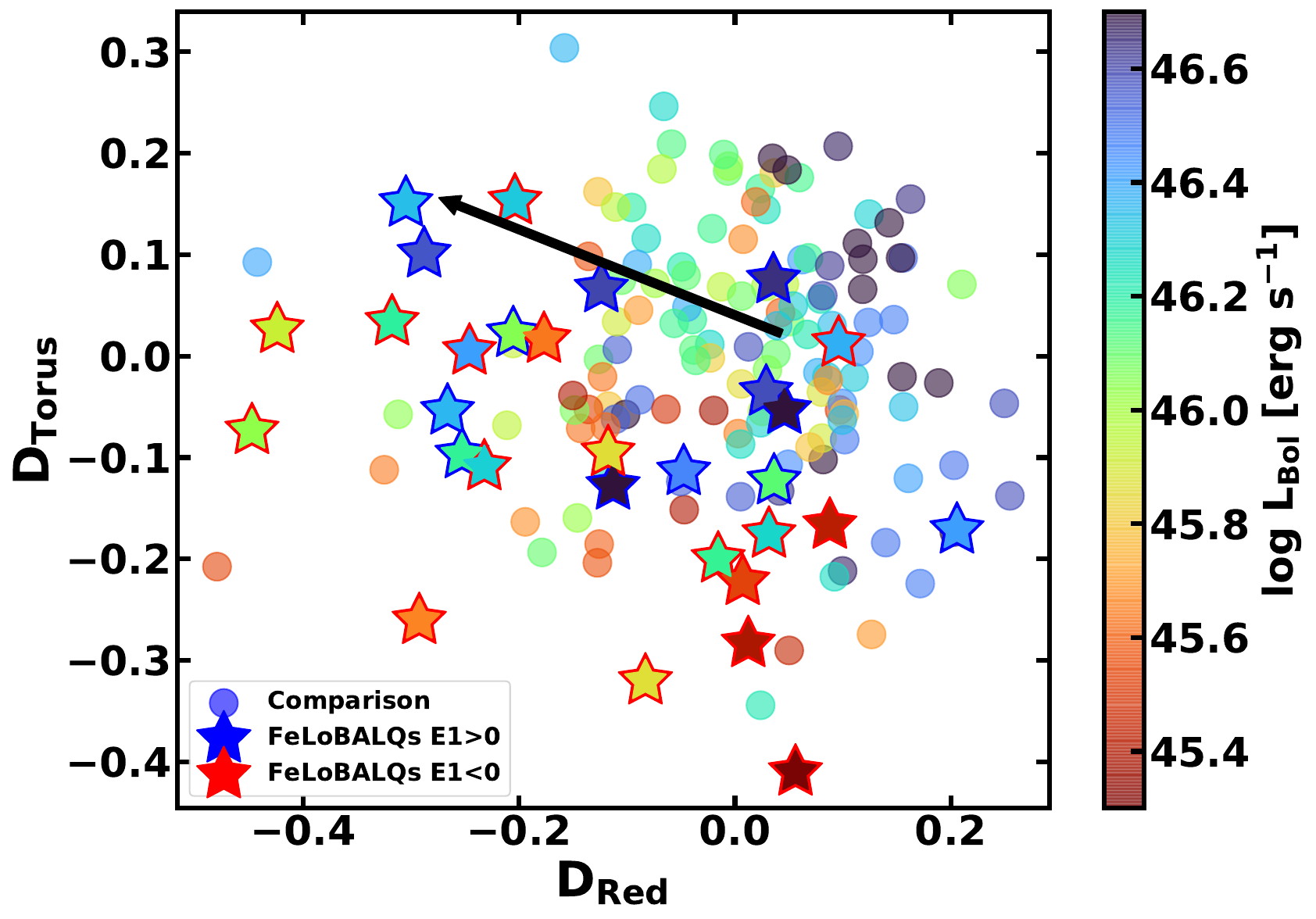}
\caption{The $D_\mathrm{torus}$ paramemeter as a function of the
  $D_\mathrm{red}$ parameter.  The black arrow shows the prediction
  for the \citet{krawczyk13} composite spectrum for $A_V=0$ to
  $A_V=1$.  The normalization of the SEDs at 1
  micron predicts that, if the SED shape differences are dominated by
  differences  in reddening, then these two parameters should be
  tightly   anticorrelated.  No correlation or trend is present among
  these two   parameters, indicating that other factors must
  contribute to the   range in SED shapes.
  \label{dred_dtor}}
\end{center}
\end{figure*}

Normalization at 1 micron might lead to an apparently stronger torus
contribution in reddened objects compared with, for example, a
normalization at longer wavelengths.  That is, reddening attenuates
the spectra at all wavelengths, but more toward shorter wavelengths.
Tilting a spectrum down at all wavelengths, then normalizing at 1
micron effectively tilts the near-IR spectrum up a small amount.
Thus, $D_\mathrm{red}$ and $D_\mathrm{torus}$ should be anticorrelated
if reddening dominates the spectral variability.  However, there is no  
evidence for an anticorrelation (Fig.~\ref{dred_dtor},
\S~\ref{correlation_cont}, Fig.~\ref{correlation}).  So while
reddening must influence the SED shapes to some degree, it does not
dominate the spectral variance in the sample.   

We were able to divide the redder and blue spectra into two groups
from which we made composite spectra.   This was done using an
expectation/maximization algorithm\footnote{E.g.,
https://en.wikipedia.org/wiki/Expectation–maximization\_algorithm}.
First, seed red and blue composites were made based on the
$\alpha_{oi}$ values.  We then computed the mean squared distance of
the photometry points from the seed spectra (expectation step). The
results were divided into two groups of objects: those that were
closer to the red spectrum and those that were closer to the blue
spectrum according to our metric (maximization step).   New composite
spectra were computed from those groupings, and the process iterated
until it converged.  The process was performed 
separately for the FeLoBALQs and the comparison sample quasars.  The
results are termed optimal red and optimal blue composite spectra;
they are shown in the lower panel of Fig.~\ref{photometry_sed}.  As
before, points shortward of 3000\AA\/ were ignored for all the spectra
to avoid the contamination of the broad absorption lines.  

We found that the optimal red and blue spectra from the comparison
sample differed only at long wavelengths, implying that the variance
among the SEDs from unabsorbed objects is principally due to variance
in the prominence of the hot dust emission.  In contrast, the
FeLoBALQs differed both at short wavelengths and at long wavelengths
suggesting that both reddening and prominence of hot dust emission are
important.

\subsection{{\it WISE} Variability Analysis}\label{wise_analysis}

\begin{figure*}[!t]
\epsscale{1.0}
\begin{center}
\includegraphics[width=6.0truein]{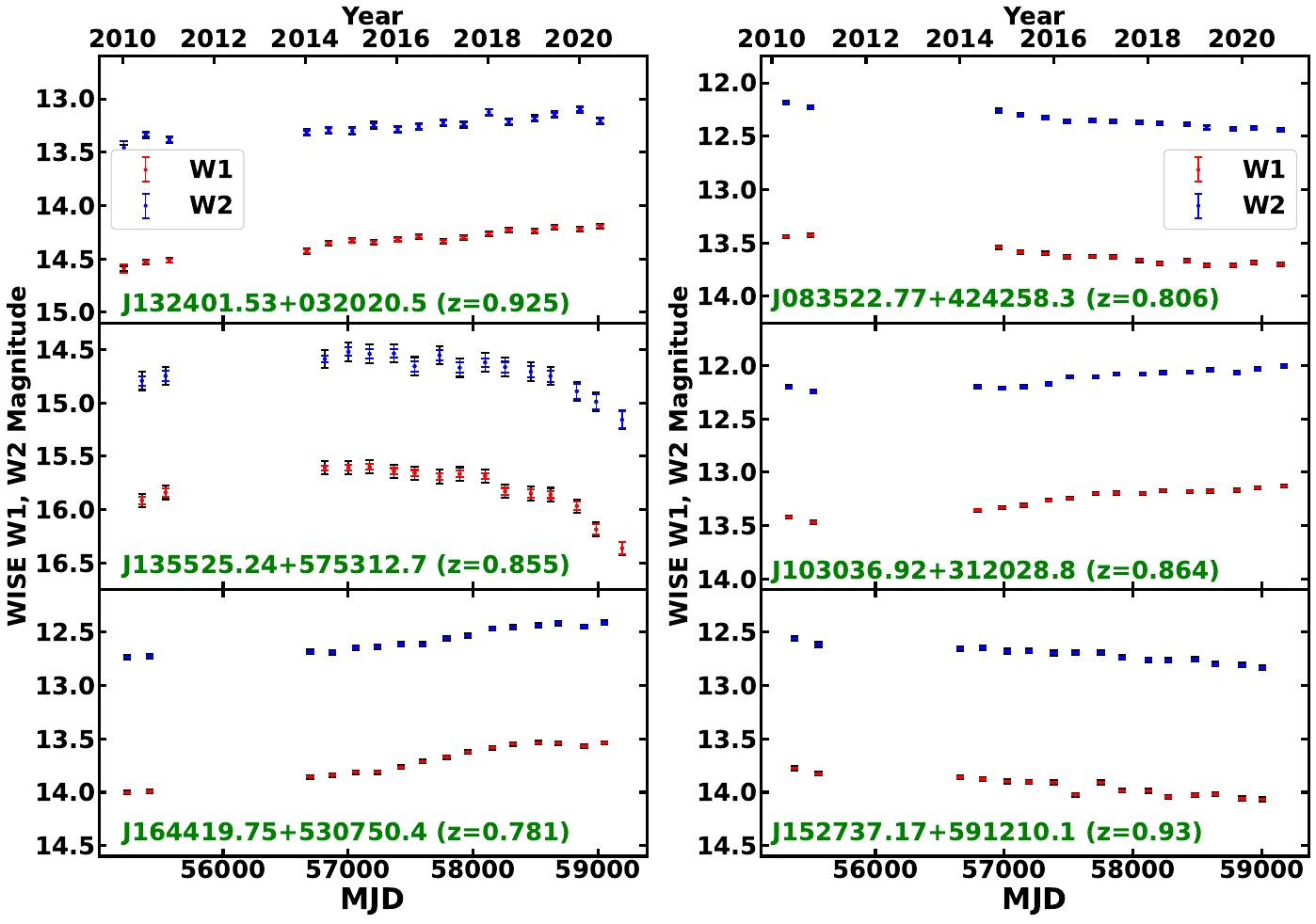}
\caption{{\it WISE} W1 and W2 lightcurves of six of the more variable
  FeLoBALQs.  {\it Left:} Objects with $E1<0$; {\it right:} objects with
  $E1>0$. The colored error bars show the photometry errors obtained
  from the   unTimely catalog, and the black error bars show the
  systematic errors derived using the scheme presented in
  Appendix~\ref{varerr}.  The light curves are given in the observed
  frame spanning about 10.5 years.  The redshifts of these objects are
  mostly between 0.8 and 1, yielding rest-frame light curves
  spanning about 5 years.
  \label{wiselc}}
\end{center}
\end{figure*}

We constructed four parameters to describe the variability properties.
The principal measures were the excess variance of the W1 and W2 light
curves.  The excess variance is the variance of the magnitude light
curve corrected for the variance due to noise by subtracting the mean
error squared. 
{ The equation for excess variance is given as
$$ EV = Variance(Magnitude)-Error(Magnitude)^2$$ where
$Variance(Magnitude)$ is the measured variance in the
magnitudes, and $Error(Magnitude)^2$ represents the variance due to
corrected statistical uncertainty.}
 The excess variance has a long history in the studies
of AGN X-ray variability \citep[e.g.,][]{nandra97, leighly99a}, and has
more recently been used to characterize optical variability
\citep[e.g.,][]{lopez-navas23}.  The errors on the excess variance
were taken to be the 1$\sigma$ bounds of the systematic error
described in Appendix~\ref{varerr}. We found that most of the objects were
variable.  Of the FeLoBALQs, 83\% (70\%) showed excess variance
$EV_\mathrm{W1}$ ($EV_\mathrm{W2}$) more than 3$\sigma$ larger than zero, while 85\%
(64\%) of the comparison sample showed $EV_\mathrm{W1}$ ($EV_\mathrm{W2}$) more than 
3$\sigma$ larger than zero.  Lightcurves of the six FeLoBALQs with the 
highest signal-to-noise ratio excess variance measurements are shown in
Fig.~\ref{wiselc}.  The variability is characterized by smooth
increases or decreases by a few tenths of a magnitude.

The third parameter was the color excess variance, i.e, the
difference between EV$_\mathrm{W1}$ and EV$_\mathrm{W2}$.  The errors
were propagated using the Monte Carlo scheme described in
\citet{leighly22}. Specifically, 10,000 normally distributed values
scaled to the positive and negative uncertainties were created for
each measurement. The difference between each pair of values was
computed, and the error was taken to be the 16 and 84\% cumulative
bounds on the resulting distribution.  A positive (negative) value of
the color excess variance indicates that the amplitude of variability
is larger (smaller) in W1 than in W2.  The final parameter measured
the correlated varibility between the W1 and W2 light curves, and was
defined as the { probability that a correlation would be
  detected in variations solely due to statistical fluctuations.}

\subsection{Restframe Optical Band Spectral Fitting of a Sample of
  LoBAL   quasars}\label{lobal_model} 

 The sample of 62 LoBAL quasr spectra was analyzed following
\citet{leighly22}. Specifically, the optical emission line region was
analyzed using {\it Sherpa} \citep{freeman01} using a model including
a power-law continuum, Balmer emission lines, [\ion{O}{3}] emission
lines, and a template model for \ion{Fe}{2}; see \citet{leighly22} \S
2.2 for details. The spectra were also analyzed using the PCA
eigenvectors developed in \citet{leighly22}.  Because eigenvector
model fitting uses correlated features over a broad band pass, this
type of analysis can be less dependent on individual features and
therefore can be more robust when the signal-to-noise ratio is low.

\subsection{Accretion Disk Models of
  SDSS~J1448$+$4043}\label{sdss1448_models}

We fitted the profile of the H$\alpha$ line of SDSS
J1448+4043 with a model that attributes the line emission to an
annulus in the accretion disk. The model, developed by \citet{chen89} and
\citet{ch89}, includes relativistic effects, such as light
bending, Doppler boosting, and transverse and gravitational
redshift. The disk is circular and inclined so that its axis makes an
angle $i$ with the line of sight. The line-emitting annulus is bound
between inner and outer radii $\xi_1$ and $\xi_2$, expressed in units
of the gravitational radius, $r_{\rm g}=GM_{\mathrm BH}/c^2$, where
$M_{\mathrm BH}$ is the mass of the black hole. The lines originate in a
photoionized surface layer in the disk with an emissivity that varies
as a power-law with radius, $\epsilon(\xi)\propto\xi^{-q}$. The
combination of all local broadening mechanisms, such as turbulence in
the emission layer and scattering, is captured by a Gaussian local
line profile of velocity dispersion $\sigma$ (referred to as the
broadening parameter). Thus the model has five free parameters: the
inclination angle of the disk, $i$, the inner and outer radii of the
line-emitting region, $\xi_1$ and $\xi_2$, the emissivity power-law
index, $q$, and the broadening parameter, $\sigma$. The inclusion of
relativistic effects, especially the transverse and gravitational
redshifts, makes the line profile asymmetric and breaks the classical
degeneracy between the radius and the inclination angle. As a result,
fitting the model to an observed line profile allows for the
determination of $\xi_1$ and $\xi_2$ separately from $i$.  

The best-fitting model is shown in the right panel of
Fig.~\ref{irtf_spectrum} as a magenta line superposed on the observed
profile of the H$\alpha$ line. The model was scaled to match
the wings of the line while 
missing the narrow H$\alpha$+[N~II] complex, compared by eye to the
observed line H$\alpha$ profile, and the parameters adjusted
iteratively until a good fit was achieved. Uncertainties on the model
parameters were determined by progressively perturbing their values
about the best fit and adjusting other parameters to compensate. Thus,
we obtained an inclination angle of $i=15^\circ\pm2^\circ$, inner and
outer radii of the line-emitting portion of the disk of
$(\xi_1,\xi_2)=(240,9500)\pm 10$\%, an emissivity power-law index of
$q=2.4\pm0.1$, and a broadening parameter of $\sigma=600 \pm
200~\rm{km~s}^{-1}$. The disk is viewed close to face-on in the
context of this model. The pronounced red wing is the result of
substantial transverse and gravitational redshifts of photons emitted
near the inner radius.

The lower panel of the right side of Fig.~\ref{irtf_spectrum} shows
the profile of the H$\beta$ line after subtraction of the Fe~II
complex and plotted on 
the same velocity scale as the H$\alpha$ line. The magenta line shows
the model that fits the H$\alpha$ line on the same velocity scale as
H$\beta$. Although the H$\beta$ profile is noisier, the H$\alpha$
model does not describe it very well; the H$\beta$ profile is somewhat
broader than that of H$\alpha$, which is a well known trend \citep[see][and
  references therein]{stirpe91}. The agreement between the model
and the H$\beta$ profile improves if we assume that the line emitting
region is more centrally concentrated by reducing the outer radius to
$\xi_2=3200$ and changing the emissivity power-law index to
$q=2.3$. The improved model is shown as a blue line in the lower
panel of Fig.~\ref{irtf_spectrum}. Such a modification to the model is
consistent with theoretical expectations for the line emissivity of
photoionized accretion disks \citep{dumont90}.

\section{Distributions of the SED and {\it WISE} Variability Parameters}\label{dist_comp}

In this section, we compare the SED and {\it WISE} variability properties
between the FeLoBALQ and unabsorbed comparison samples.  Following
\citet{leighly22}, we present the comparison for the full sample, and
also for the $E1>0$ and $E1<0$ subsamples.  As in \citet{leighly22}
and \citet{choi22}, we used the two-sample Kolmogorov-Smirnov (KS)
test and the two-sample Anderson-Darling (AD) test.  The KS test
reliably tests the difference between two distributions when the
difference is large at the median values, while the AD test is more
reliable if the differences lie toward the maximum or minimum values
(i.e., the median can be the same, and the distributions different at
larger and smaller values)\footnote{E.g.,
https://asaip.psu.edu/articles/beware-the-kolmogorov-smirnov-test/}.
The parameter value and false-alarm probabilities for these tests are
included in Table \ref{distributions}.

\movetabledown=0.5in
\begin{longrotatetable}
\begin{deluxetable*}{lCCCCCCCCCCC}
\tabletypesize{\scriptsize}
\tablecaption{Parameter Distributions Comparison\label{distributions}}
\tablehead{
\colhead{Parameter Name} & \multicolumn{6}{c}{FeLoBALQs vs Comparison}
& \multicolumn{2}{c}{FeLoBALQs} \\
\colhead{} & \multicolumn{2}{c}{All (30/132)} &
\multicolumn{2}{c}{E1$<0$ (17/61)} &
 \multicolumn{2}{c}{E1$>0$ (13/71)}  &  \multicolumn{2}{c}{E1$<0$ vs
   E1$>0$\tablenotemark{a}}  \\
\colhead{} & \colhead{KS\tablenotemark{b}} & \colhead{AD\tablenotemark{c}} & \colhead{KS\tablenotemark{b}} & \colhead{AD\tablenotemark{c}} & \colhead{KS\tablenotemark{b}} & \colhead{AD\tablenotemark{c}} 
& \colhead{KS\tablenotemark{b}} & \colhead{AD\tablenotemark{c}} \\
}
\startdata
4500-5500\AA\/ Slope & {\bf 0.48 / 1.3\times 10^{-5}} & {\bf 17.6 /  <0.001} & 
{\bf 0.56 / 2.3\times 10^{-4}} & {\bf 9.2 / <0.001} & 
{\bf 0.48 / 6.9\times 10^{-3}} & {\bf 7.1 / <0.001} & 
0.46 / 0.055 & 1.9 / 0.055 \\
$\alpha_{oi}$ (5100\AA -- 3$\mathrm{\mu m}$ slope) & 
0.21 / 0.20 & 0.41 / 0.23 & 
0.29 / 0.16 & 0.67 / 0.17 & 
0.23 / 0.55 & -0.78 / >0.25 & 
0.28 / 0.51 & -0.06 / >0.25 \\
W1-W2 & 0.11 / 0.91 & -0.86 / >0.25 & 
0.26 / 0.26 & 0.006 / >0.25 & 
0.24 / 0.46 & 0.054 / >0.25 & 
{\bf 0.52 / 0.025} & {\bf 3.8 / 9.3\times 10^{-3}} \\
$D_\mathrm{red}$ & 
{\bf 0.37 / 1.5\times 10^{-3}} & {\bf 9.3 / <0.001} & 
{\bf 0.42 / 1.1 \times 10^{-2}} & {\bf 5.4 / 2.6\times 10^{-3}} & 
0.39 / 0.05 & {\bf 3.7 / 1.1 \times 10^{-2}} & 
0.18 / 0.93 & -0.82 / >0.25 \\
$D_\mathrm{torus}$ & 
{\bf 0.31 / 1.4 \times 10^{-2}} & {\bf 4.9 / 3.7\times 10^{-3}} & 
{\bf 0.42 / 1.3 \times 10^{-2}} & {\bf 5.7 / 2.0 \times 10^{-3}} & 
0.25  / 0.42 & -0.002 / > 0.25 & 
0.42 / 0.12 & 1.4 / 0.08 \\
\hline
W1 Excess Variance & 0.14/0.69 & -0.80/0.25 & 
0.20 / 0.59 & -0.28 / 0.25 & 
0.31 / 0.21 & 0.64 / 0.18 & 
0.42 / 0.11 & {\bf 2.2 /0.04 } \\
W2 Excess Variance & 0.23 / 0.13 & 0.98 / 0.13 & 
0.28 / 0.20 & 1.6 / 0.07  &
0.30 / 0.21 & 0.66 / 0.18 & 
0.47 / 0.05 & {\bf 2.4 / 0.035} \\
Color Excess Variance & {\bf 0.29 / 0.026} & {\bf 2.0 / 0.048} & 
0.35 / 0.056 & 1.72 / 0.06 & 
0.35 /0.11 & 0.88 / 0.14 & 
{\bf 0.57 / 0.01} & {\bf 3.2 / 0.016} \\
W1-W2 Correlation Prob. & 0.20 / 0.22 & 1.03 / 0.12 & 
0.23 / 0.41 & 0.62 / 0.18 & 
0.24 / 0.44 & 0.37 /0.24 & 
0.35 / 0.25 & 0.02 / $<0.25$ \\
\enddata
\tablenotetext{a}{The optical data from the $E<0$ ($E>0$) FeLoBALQ
  subsamples  include 17 (13) objects. }
\tablenotetext{b}{The Kolmogorov-Smirnov Two-sample test.  Each entry
  has two numbers:  the first is the value of the statistic, and the
  second is the probability that the two samples arise from the same
  parent sample.  Bold type indicates entries that yield $p<0.05$,
  i.e., statistically significant.}.  
\tablenotetext{c}{The Anderson-Darling Two-sample test.  Each entry
  has two numbers:  the first is the value of the statistic, and the
  second is the probability that the two samples arise from the same
  parent sample.  Note that the implementation
  used does not compute a probably   larger than 0.25 or smaller than
  0.001. Bold type indicates entries that yield $p<0.05$, i.e.,
  statistically significant.}
\end{deluxetable*}
\end{longrotatetable}

\subsection{SED Parameters}\label{sed_params}

\begin{figure*}[!t]
\epsscale{1.0}
\begin{center}
\includegraphics[width=6.5truein]{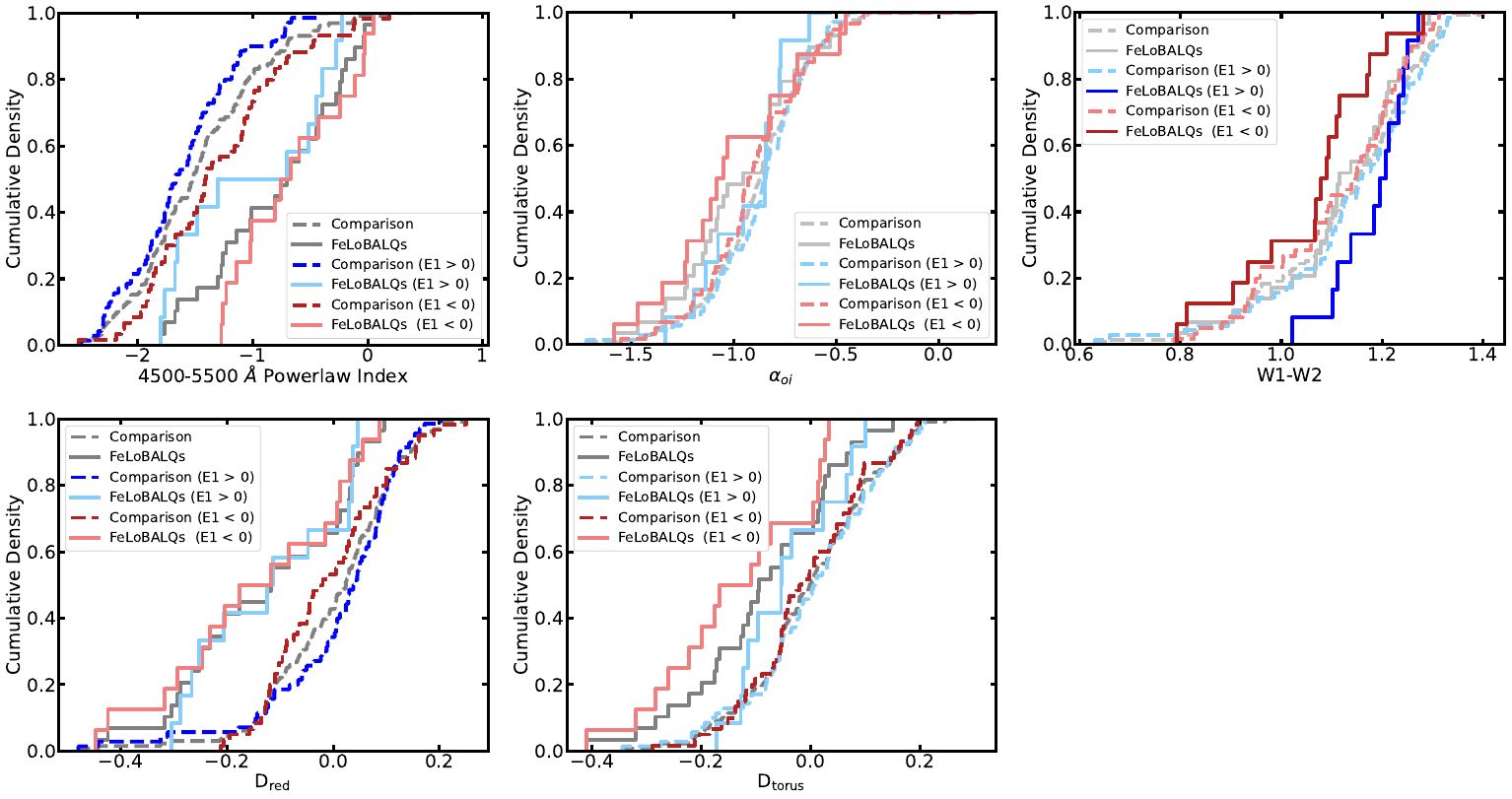}
\caption{Cumulative distribution plots of the continuum shape
  parameters.  The data are sampled in three ways: the grey lines show
  the full FeLoBALQ and comparison samples, and the blue (red) lines
  show the FeLoBALQ and comparison samples for E1 parameter $E1 > 0$
  ($E1<0$).  The Kolmogorov-Smirnov and Anderson-Darling statistics
  for four different comparisons are given in
  Table~\ref{distributions}.  Distributions that are significantly
  different ($p<0.05$) are shown in dark red, dark blue, or dark grey,
  while distributions that are not significantly different are shown
  in pale colors.  The full sample distributions are statistically
  significantly different for the 4500-5500 \AA\ power law slope, the
  reddening excess parameter $D_\mathrm{red}$, and the torus excess
  parameter $D_\mathrm{torus}$.  The FeLoBALQs partitioned by the E1
  parameter are statistically significantly different for only W1$-$W2.
   \label{plot_comp_4}} 
\end{center}
\end{figure*}

The cumulative probability distributions of the SED parameters are
shown in Fig.~\ref{plot_comp_4}.  We found that the power law index
between 4500 and 5500 \AA\/ measured from the continuum model fits
under the H$\beta$ / [\ion{O}{3}] / \ion{Fe}{2} model is significantly
flatter (redder) for all of the FeLoBAL groups compared with the
unabsorbed comparison sample groups.  The samples are not disjoint,
however; for example, the median slope of the $E1>0$ FeLoBALQs is only
slightly flatter than the unabsorbed objects.  The  $D_\mathrm{red}$
parameter shows very similar behavior indicating that these two
parameters are to some extent degenerate; a flatter power law is
equivalent to a larger difference from the composite spectrum compiled
from the unabsorbed objects.  

There are no significant differences in $\alpha_{oi}$ between any of
the groups of objects, although the $E1<0$ FeLoBALQs show the steepest
values of any of the groups.  This implies that this group of objects
tends to be bluer than other objects.  Recalling that $\alpha_{oi}$
measures the optical to NIR slope, a partial explanation for this
may be found in the $D_\mathrm{torus}$ distributions.  The FeLoBAL
quasars uniformly show a deficit in hot dust emission compared with the
unabsorbed quasars as a whole. As discussed in \S\ref{sed_composites},
if reddening is the principal origin of SED differences, the
normalization at 1 micron would tend to enhance torus emission in
objects inferred to be reddened.   The fact that
the FeLoBALQs are both reddened and have overall weaker torus emission
means that the deficit in hot dust emission is slightly stronger than
it appears.  The group with the greatest  $D_\mathrm{torus}$, i.e., the
least evidence for hot dust emission, are the $E1<0$ FeLoBALQs. 

Finally, W1-W2 is consistent between the FeLoBALQs and the unabsorbed
comparison sample overall, but there is a significant difference among
the $E1<0$ and $E1>0$ FeLoBAL quasars.  Since W1-W2 measures the
upturn past one micron towards the torus for these redshifts, this
parameter can be interpreted as the prominence of hot dust emission.
The $E1<0$ objects have lower values of W1-W2, i.e., evidence for a
lack of hot dust emission, and thus this result echoes the
$D_\mathrm{torus}$ result. 

Overall, these distributions reveal that FeLoBAL quasars have redder 
optical spectra than unabsorbed objects, and the SEDs suggest that
this may in part be due to reddening.  In addition, FeLoBAL quasars
with $E1<0$ ($E1>0$) as a group show bluer (redder) infrared spectra
suggesting weaker (stronger) hot dust signatures than the unabsorbed
comparison sample quasars.

\subsection{{\it WISE} Variability Parameters}\label{dist_var}

\begin{figure*}[!t]
\epsscale{1.0}
\begin{center}
\includegraphics[width=6.5truein]{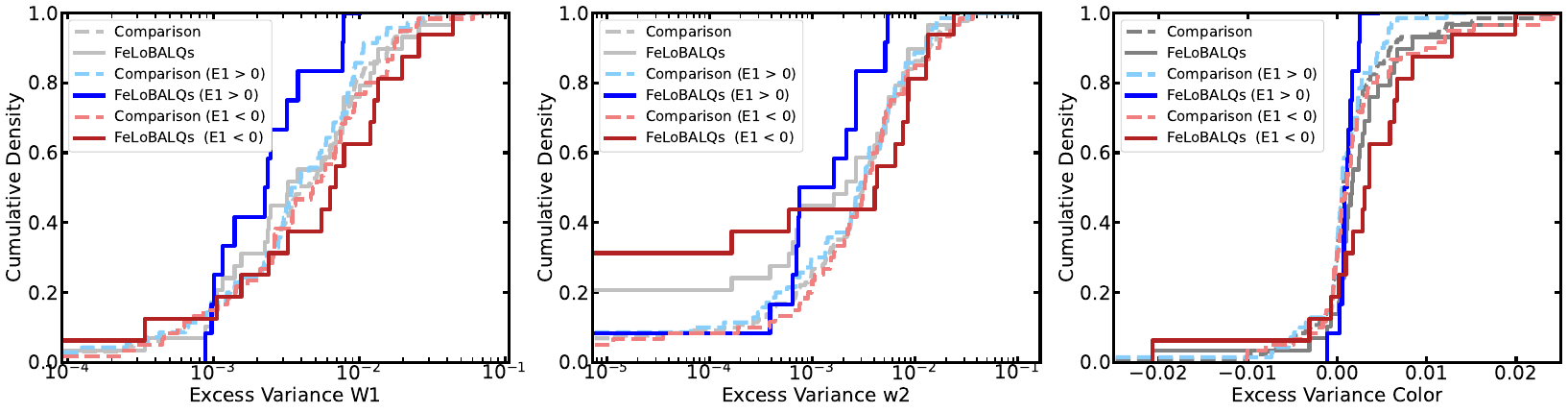}
\caption{The distributions of three of the four variability parameters
  including   excess variance for W1 (left) and W2 (center) and the
  color excess variance (right) where a positive value indicates that W1 varies more than
  W2.  In most cases, distributions of pairs of properties were
  indistinguishable; the exceptions were the $E<1$ versus the $E>1$
  FeLoBALQs for all parameters, and the color excess variance for the
  FeLoBALQs versus the comparison objects. The fourth parameter, the
  correlation $p$-value between W1 and   W2 light curves, showed no
  differences between any of the samples   and is not shown.
   \label{varcumdist}} 
\end{center}
\end{figure*}

We found that the FeLoBALQ and unabsorbed comparison sample {\it WISE}
excess variance distributions are indistiguishable in most cases
(Table~\ref{distributions}).  The most significant distribution
difference was found between the $E1$-divided FeLoBALQs.  The $E1>0$
objects, i.e., the ones with larger $L_\mathrm{bol}/L_\mathrm{Edd}$, 
were less variable than the $E1<0$ objects.  In addition, there is a
difference in excess variance color whereby the $E1<0$ objects varied
more in W1 than in W2.  The distributions of the three parameters
that have statistically significant differences are shown in
Fig.~\ref{varcumdist}.  

The objects in this sample are relatively nearby and have moderate
bolometric luminosities.  In \S\ref{sed_modeling} we showed 
that in most objects, the host galaxy contamination in  the
near-infrared band is expected to be negligible.  We confirm that
result here, because if galaxy contamination were significant, one would
expect that the lower luminosity objects would { appear to have}
a lower variability amplitude.  As shown in
\citet[][Fig.\ 8]{leighly22}, the $E1<0$ objects have a mean inferred
bolometric luminosity about 0.5 dex lower than the $E1>0$ objects, and
yet we found that they are more variable in both W1 and W2. Therefore,
while there could be some  supression of variability due to galaxy
contamination in some objects, it does not dominate the variability
properties.

\section{Correlations with the SED and {\it WISE} Variability Parameters}\label{correlations}

\begin{sidewaysfigure}
\epsscale{1.0}
\begin{center}
\includegraphics[width=9.2truein]{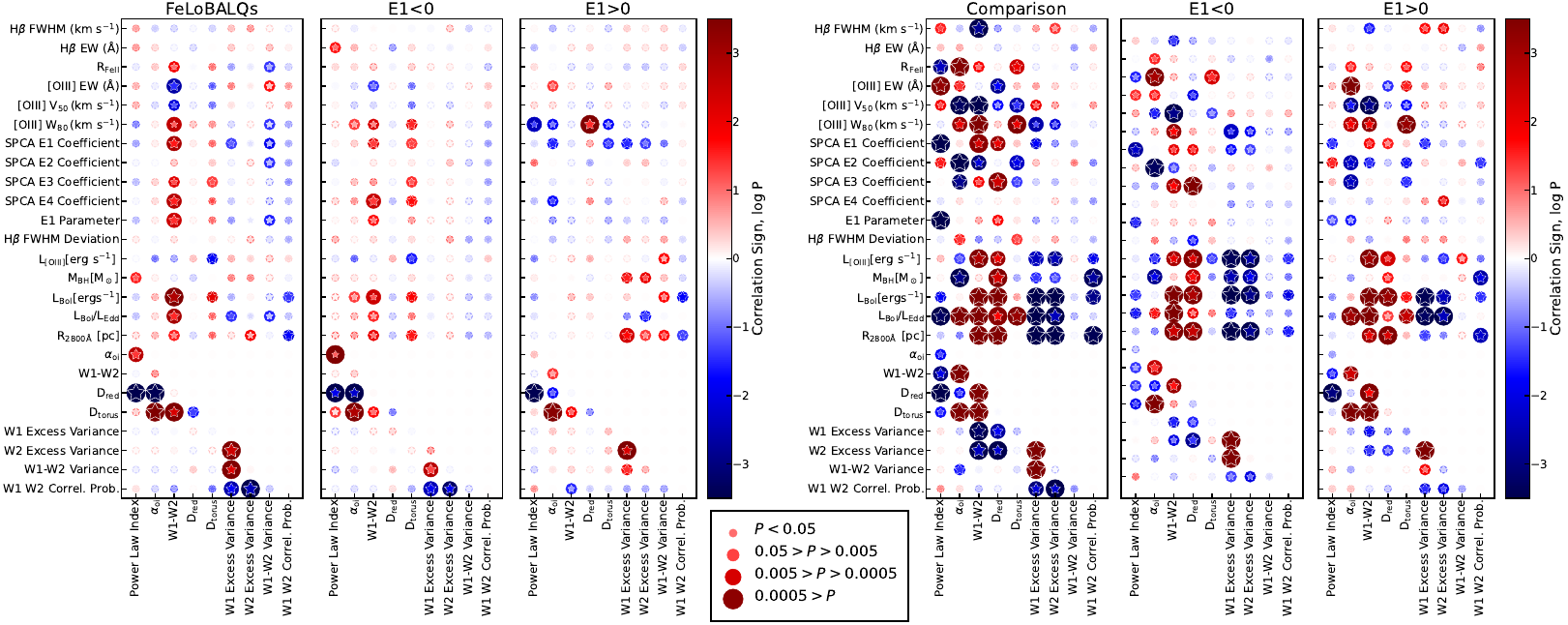}
\caption{The results of the Spearman rank correlation analysis for the
  FeLoBAL quasars (left) and the comparison sample (right), and
  divided according to the $E1$ parameter.
  The size  and color of the marker indicates the sign and $p$ value
  of the correlation.  Anticorrelations are shown in blue, and
  correlations are shown in red.  The shade of color of the points
indicates the significance of the correlation as a continuous
variable, while discrete sizes of the points characterizes a range of
$p$ values: $p<0.05$, $0.05 > p > 0.005$, $0.005 > p> 0.0005$, and
$0.0005 > p$.  The circular markers show the results for
  parameter values.  The stars show the results for a Monte Carlo
  scheme to estimate the effects of the errors (see text for
  details).  Many significant correlations are found among the 
  comparison sample properties, specially with Eddington ratio, that
  are not mirrored among the FeLoBALQs.  \label{correlation}}  
\end{center}
\end{sidewaysfigure}

\begin{figure*}[!t]
\epsscale{1.0}
\begin{center}
\includegraphics[width=6.5in]{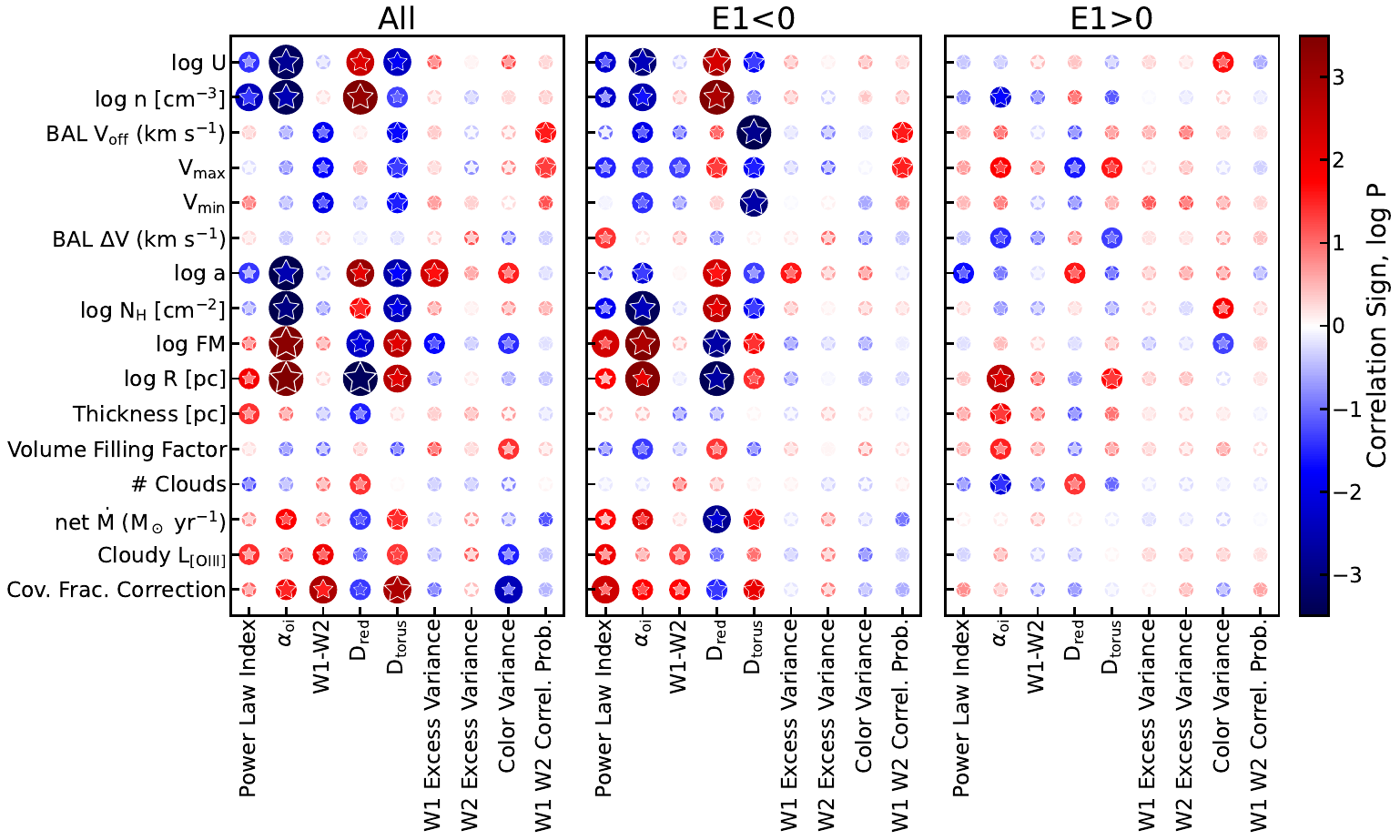}
\caption{The results of the Spearman rank correlation analysis for
  five 
  continuum and four variability parameters and the 16 {\it SimBAL}
  parameters.  The   symbols have the same meaning as in
  Fig.~\ref{correlation}.  The left
  plot shows the results for   the   whole sample, while the middle
  and right plots show the   results for   $E1<0$ and $E1>0$
  respectively.  Among the continuum parameters, $\alpha_{oi}$,
  $D_\mathrm{red}$, and $D_\mathrm{torus}$ are the
  most   strongly correlated with the   {\it SimBAL} parameters.
No significant patterns were found among the variability
parameters.  \label{corr_optical_simbal}}    
\end{center}
\end{figure*}

We examine the relationships between the five SED and
four variability parameters and the emission-line and global
parameters presented in \citet{leighly22} and the {\it SimBAL}
parameters presented in \citet{choi22b}.  First
we present Spearman-rank correlations, and then we discuss the
relationships among various parameters.  The results are presented in
Fig.~\ref{correlation} and \ref{corr_optical_simbal} which represent
the statistical properties of the correlations graphically.  { We plot
the false-alarm probabilities rather than the correlation coefficients
because we compare samples of different sizes.}

Parameter uncertainties were propagated through the correlations
using a Monte Carlo scheme; see also \citet{leighly22}.  We made
10,000 normally distributed draws 
of each parameter, where the distribution was stretched to the size of
error bar.  Asymmetrical errors were accounted for by using a
split-normal distribution (i.e., stretching  the positive draws
according to the positive error, and the negative draws according to
the negative error).  We chose $p<0.05$ as our threshold for
significance.  The overplotted stars in Fig.~\ref{correlation} and
\ref{corr_optical_simbal} show the fraction of draws that yield $p$
values greater than our threshold value; these are clearly seen only
when accounting for the uncertainty dramatically changes the
significance of the correlation.  Generally, taking the errors into
account did not affect the significance of a correlation, if present.  

As discussed in \citet{leighly22},  correlations among quasar
properties associated with the \citet{bg92} Eigenvector 1, here
parameterized using the $E1$ parameter, may dominate and potentially 
obscure other correlations.  We would like to determine whether there
were correlations independent of the $E1$ parameter.  It is also
possible that the two classes of FeLoBALQs show different correlation
behaviors  that would be washed out in the 
whole-sample correlations.  Therefore, we also computed the
Spearman-rank correlation coefficient between the parameters divided
by $E1<0$ and $E1>0$ (Fig.~\ref{correlation} and
Fig.~\ref{corr_optical_simbal}, middle and right panels).

\begin{figure*}[!t]
\epsscale{1.0}
\begin{center}
\includegraphics[width=5.0in]{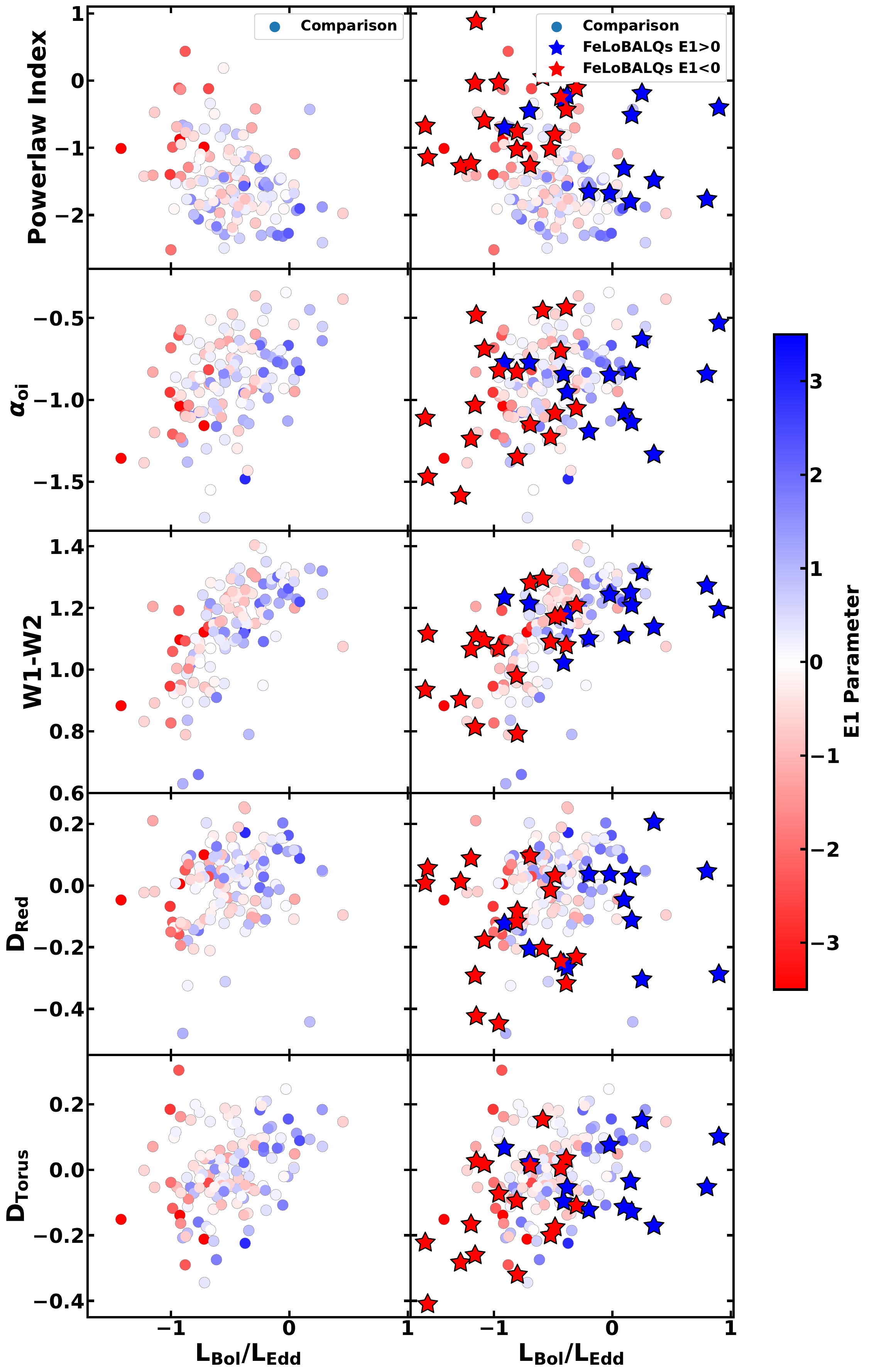}
\caption{The parameters used to describe the optical-near IR continuum 
  properties as a function of Eddington ratio. The unbsorbed
  comparison sample is shown in the left column, and those plus the
  FeLoBAL quasars are shown in the right column.  Significant
  correlations are seen between all parameters and the Eddington
  ratio among the unabsorbed quasars (also Fig.~\ref{correlation}). Among
  the FeLoBALQs, only the correlation with W1-W2 is significant.  The
  other  parameters are more scattered, perhaps as a consequence of 
  different contributions of reddening.   \label{cont_eddrat}}   
\end{center}
\end{figure*}

\subsection{Continuum Parameters}\label{correlation_cont}

\subsubsection{The Power Law Index}\label{corr_plaw}

We first considered the power law index, which was defined as the slope
between 4500 and 5500\AA\/ measured from the continuum portion of the
spectroscopy model from \citet{leighly22} { (Fig.~\ref{phot_params},
Table~\ref{definitions}). This parameter measures 
the shape of the optical continuum arising from the accretion disk.
This parameter may be intrinsically different from object to object,
and it also may be flattened by reddening if present.}  The
distribution analysis (\S\ref{sed_params})  showed that the power law
index is significantly flatter in FeLoBALQs than in unabsorbed quasars
(Fig.~\ref{plot_comp_4}, Table~\ref{distributions}), indicating that
FeLoBALQs have redder spectra.   

Among the unabsorbed comparison objects, the power law index is
strongly anti-correlated with with the Eddington ratio ($p=9\times
10^{-5}$; Fig.~\ref{correlation}, Fig~\ref{cont_eddrat}); the origin
of this correlation is discussed in \S\ref{disc_continuum}.  It is
similarly correlated with other parameters related to  the Eddington
ratio, including $R_\mathrm{FeII}$, the Spectral Principal Component
Analysis (SPCA) E1 coefficient \citep{leighly22}, and the $E1$
parameter.  Although the power law index for the FeLoBALQs is not
correlated with the Eddington ratio, Fig.~\ref{cont_eddrat} shows that
a rough  relationship is present, with the higher accretion rate
($E1>0$) objects having overall steeper spectra than the lower
accretion rate ($E1<0$) objects.  There is a large scatter that may be
due to a range of reddening among the FeLoBAL quasars.

\subsubsection{$\alpha_{oi}$}\label{corr_alphaoi}

{ We next consider $\alpha_{oi}$, which is defined as the
  point-to-point slope between 5100\AA\/ and 3$\mu \rm m$
  (Fig.~\ref{phot_params}, Table~\ref{definitions}).  This
  parameter measures the overall steepness of the optical through near
  IR spectrum.  It may be steep (blue) because the torus contribution
  is weak.  It may be flat (red) because the torus contribution is
  strong, or because there is significant reddening.}

Among the unabsorbed comparison objects, $\alpha_{oi}$ is correlated
with the Eddington ratio as the power law index  is ($p=7\times
10^{-6}$; Fig.~\ref{correlations}, Fig.~\ref{cont_eddrat}), but with
the opposite dependence: objects with steeper (bluer) spectra have
lower Eddington ratios. $\alpha_{oi}$ is correlated with the other
four SED parameters, as expected, but it is most strongly correlated
with $D_\mathrm{torus}$ ($p=2\times 10^{-19}$), suggesting that the
$\alpha_{oi}$ is dominated by the range of hot dust emission rather
than the range of optical deficits. This correlation may be expected
based on the optimal red and blue comparison sample composites shown
in Fig.~\ref{photometry_sed}, where the blue composites show stronger
torus emission.   

\begin{figure*}[!t]
\epsscale{1.0}
\begin{center}
\includegraphics[width=4.5truein]{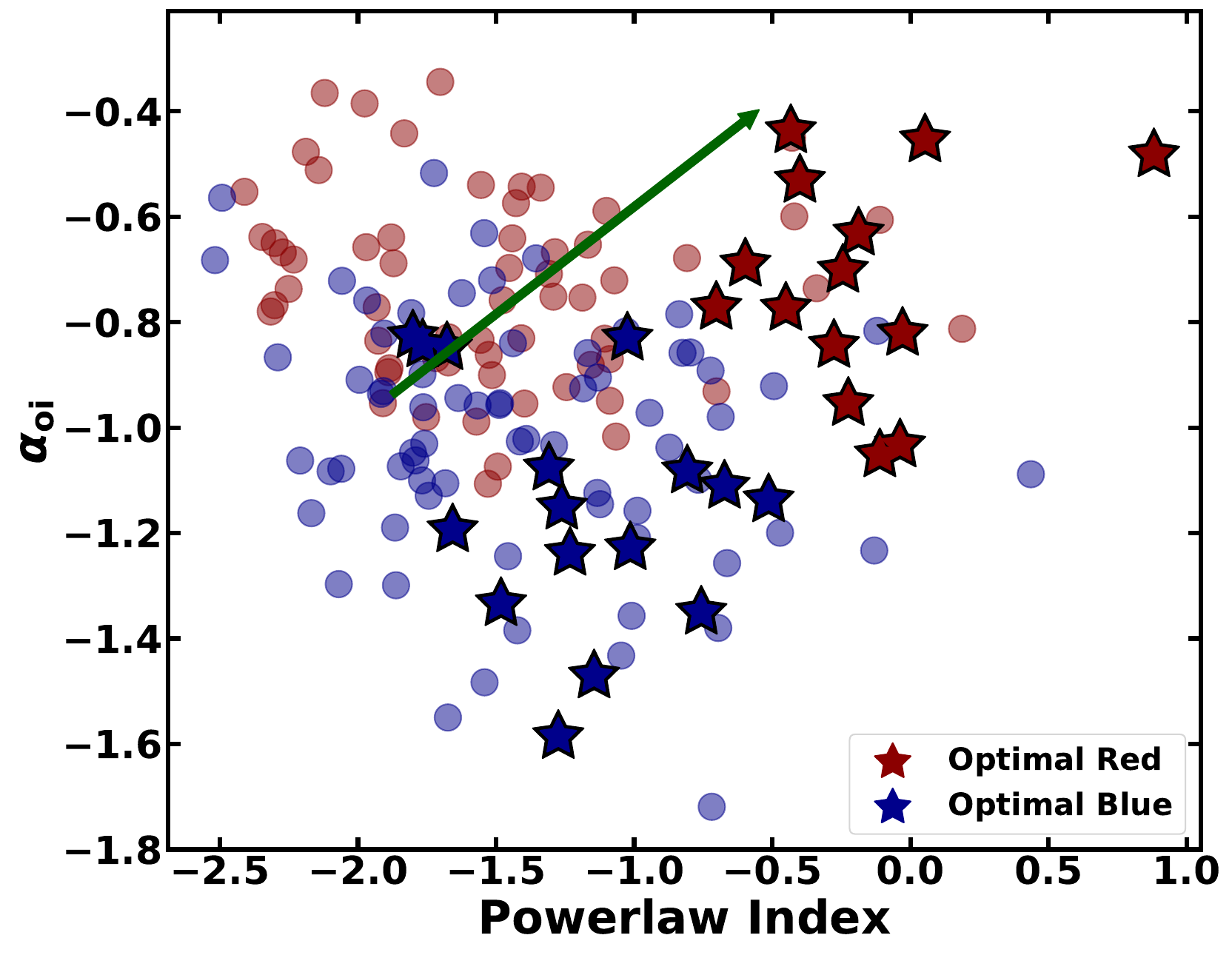}
\caption{The relationship between  $\alpha_{oi}$ and the local power-law
  index measured under the H$\beta$/[\ion{O}{3}]/\ion{Fe}{2}
  emission.  An  anticorrelation is found for the unabsorbed objects.
  The arrow shows $A_V=1$ for the \citet{krawczyk13} composite
  continuum.  If the FeLoBALQs were simply reddened versions of 
  unabsorbed quasars, they would also show similar anticorrelations,
  but broaden and shifted upward and to the right. Instead, they show
  a correlation.     \label{reddening}}        
\end{center}
\end{figure*}

The FeLoBALQs and unabsorbed comparison sample objects show opposite
relationships between $\alpha_{oi}$ and the power law index
(Fig.~\ref{reddening}), where we color code the markers using the
  optimal red and optimal blue classification discussed in
  \S\ref{sed_composites}.  While the FeLoBALQs show a correlation
between these two parameters ($p=0.002$), the unabsorbed objects show
an anticorrelation ($p=0.016$).  Also, the anti-correlation between
the Eddington ratio and the power law index in the unabsorbed objects
is mirrored by a correlation with  $\alpha_{oi}$; objects with steep
(blue) power law spectra have flat (red) values of $\alpha_{oi}$ and
vice versa.    

To help interpret this behavior, we overplotted the inferred
$\alpha_{oi}$ and 4500--5500\AA\/ power-law from the
\citet{krawczyk13} composite spectrum subject to SMC reddening with a
range of $A_V$ between 0 and 1.  This result
shows that reddening moves objects upward and to the right
(Fig.~\ref{reddening}).  
{ This plot and superimposed arrow shows
that the continua of FeLoBALQs are not {\it simply}  reddened
versions of unabsorbed quasar continua.  Specifically, the unabsorbed
quasars show an anticorrelation between power law index and
$\alpha_{oi}$,  stretching from upper left to lower middle.  If
FeLoBALQs were all reddened versions of the unabsorbed quasars, they
would show a similar anticorrelation, with the whole pattern shifted
upward and to the right along the arrow. Instead, they show a
correlation.  In fact, the optimal red FeLoBAL quasars are consistent
with being reddened versions of the optimal blue FeLoBAL quasars,
since the FeLoBAL correlation in this plot has roughly the same slope
as the arrow.   

The key to interpreting this plot lies in an examination of the
optimal red and blue spectra shown in Fig.~\ref{photometry_sed}.  The
unabsorbed quasars differ in the strength of the torus, while
reddening is much less important.  An interesting feature is that
among the unabsorbed quasars, stronger torus emission is associated
with a steeper (bluer) power law, yielding the points in the upper
left corner of the plot; the optimal red unabsorbed quasars comprise
these points.  That is, a very steep (blue) optical power law is
associated with a very flat (red) near-infrared spectrum in the
unaborbed objects (Fig.~\ref{correlation}).  That the power law and
the prominence of the torus are correlated with the
Eddington ratio (Fig.~\ref{cont_eddrat}) might be a clue about the
origin of the SED.

FeLoBAL quasars are noticably absent from the upper-left corner of the 
plot.  This means that FeLoBAL quasars either have intrinsically
flatter power law spectra, or that all FeLoBAL quasars have at least
some reddening (or both). FeLoBAL quasars that have relatively steeper
(bluer) 
optical spectra (power law indices less than $\sim -0.5$) all have
weak torus emission (the optimal blue FeLoBALQs), while the FeLOBAL
quasars that have flat optical spectra (suggesting more reddening)
also have relatively stronger torus emission than the optimal blue
FeLoBAL quasars (the optimal red FeLoBALQs).  This is the opposite
behavior from the unabsorbed quasars.  This behavior supports 
the idea that at least for the optimal blue FeLoBAL quasars, the 
flatter power law index compared with unabsorbed quasars is, in fact,
intrinsic. Otherwise, the optimal red and optimal 
blue FeLoBAL quasar spectra would have the same torus emission
(Fig.~\ref{photometry_sed}); there should be no reason for
strong-torus objects to have more reddening and vice versa.  The
principal group that is missing from this plot are reddened unabsorbed
quasars.}  

\subsubsection{W1-W2}\label{corr_w1w2}

{ At the low redshifts considered in this paper, W1-W2 measures the
  rest near-infrared color (Fig.~\ref{phot_params},
  Table~\ref{definitions}).  It quantifies the upturn toward
  the torus and the presence of hot dust emission.  A small value of
  W1-W2 indicates weak hot dust emission (blue near-infrared
  continuum), while a large value points to strong hot dust emission
  and a prominent torus (red near-infrared continuum).}

W1-W2 shows some of the stronger correlations with the optical
parameters, and unlike the other continuum parameters, correlations
among the FeLoBALQs are present.  Fig.~\ref{cont_eddrat} 
shows the correlations with Eddington ratio; $p$ values are $1.0\times
10^{-3}$ ($3.5\times 10^{-20}$) for the FeLoBALQs (unabsorbed objects)
respectively.  

As noted in \S\ref{sed_modeling}, most of our objects have redshifts
between 0.8 and 1.0, which means that W1 and W2 describe the flux
densities near 1.7--1.86 and 2.3--2.56 microns respectively. The
values fall on the upturn from the 1-micron dust-sublimation break
toward the torus infrared bump.  Thus, the correlations that we find
mean that at low accretion rates, the spectrum lacks this upturn, a
property that could be interpreted as a lack of hot dust emission.
The lack of the hot dust component can be seen in some of the SED
photometry plots (Fig.~\ref{phot1}) and in the optimal blue FeLoBAL
composite (Fig.~\ref{photometry_sed}).  

Previous investigations have associated a lack of hot dust emission
with a low accretion rate.  Objects that lack hot dust are termed hot
dust deficient (HDD), and in the 87 best studied Palomar-Green quasars
they are linked to relatively lower accretion rates \citep{lyu17}. A
lack of torus emission in low accretion rate objects is supported by a
number of other investigations \citep[e.g.,][]{wa04, trump11,
  vanderwolk11,   izumi17, gonzalez-martin17, ricci17, temple21}. 

\subsubsection{$D_\mathrm{red}$ and  $D_\mathrm{torus}$}\label{dparams}

{ As discussed in \S\ref{sed_composites}, the $D_\mathrm{red}$ and
$D_\mathrm{torus}$ parameters were devised to break the degeneracy
inherent in $\alpha_{oi}$. Specifically,  $\alpha_{oi}$ may be
flat (red) because of reddening in the optical band or strong hot dust 
emission in the near-infrared band.  These parameters were defined
relative to the composite created from the unabsorbed objects.  
$D_\mathrm{red}$ was defined in the rest optical band; it has a low value
if the optical spectrum of an object is flat (red) compared with the
composite.  $D_\mathrm{torus}$ was defined in the rest near-IR band;
it has a low value when the torus / hot dust emission is weak compared
with the composite.}

In the correlation anlysis, $D_\mathrm{red}$ and $D_\mathrm{torus}$
are somewhat redundant with 
the other parameters; their chief utility is to distinguish the
influences on $\alpha_{oi}$.  As noted above, they are not correlated
with one another (Fig.~\ref{dred_dtor}), showing  that reddening
alone can not be responsible for the range of continuum properties
found.  We found that $D_\mathrm{red}$ and $\alpha_{oi}$ are
correlated for the FeLoBALQs but not for the unabsorbed objects,
indicating that reddening is more important for shaping the continuum
among the FeLoBALQs.  In contrast,  $\alpha_{oi}$ is correlated with
$D_\mathrm{torus}$ for both samples.   This result was
also obtained from the  optimal red and optimal blue composites
shown in Fig.~\ref{photometry_sed}: the optimal composites from the
unabsorbed comparison sample differed longward of 1 micron, indicating
variations in the hot dust component of the torus, while the optimal
composites of the FeLoBAL quasars also differed shortward of 1 micron,
indicating that both reddening and hot dust emission variations are
important.  

\subsection{SimBAL Parameters}\label{corr_simbal}

\citet{choi22b} reported the correlations between the optical
emission-line properties and the {\it SimBAL} parameters.  While they found
correlations with the outflow velocity and parameters related to the 
volume filling factor, no correlations were found with the physical
properties of the gas such as the ionization parameter $\log U$, the
gas density $\log n$, or the column density $\log N_H$.  These outflow
properties are associated with the location of the gas, i.e., the
distance of the absorbing gas from the central engine, $\log R$, which
is derived from $\log n$ and $\log U$ { using standard,
  well-established methods based on the theory of photoionized gas as
  outlined in \S\ref{review}.}

Surprisingly, we found very strong correlations between the continuum
parameters $\alpha_{oi}$, $D_\mathrm{red}$, and $D_\mathrm{torus}$ and
the physical properties of the outflowing gas
(Fig.~\ref{corr_optical_simbal}). The strongest correlations are with
$\log R$ ($p=2.2\times 10^{-6}$, $1.0\times 10^{-4}$, $1.9 \times
10^{-3}$ for $\alpha_{oi}$, $D_\mathrm{red}$, and $D_\mathrm{torus}$,
respectively), arising from an anticorrelation with both $\log U$ and
$\log n$. Notably, the $\alpha_{oi}$ correlation persists in both the
$E1<0$ and $E1>0$ subsamples  ($p=4.0\times 10^{-4}$ and $p=2.1\times
10^{-3}$, respectively), so it is not a consequence of accretion rate
trends.  

\begin{figure*}[!t]
\epsscale{1.0}
\begin{center}
\includegraphics[width=6.5truein]{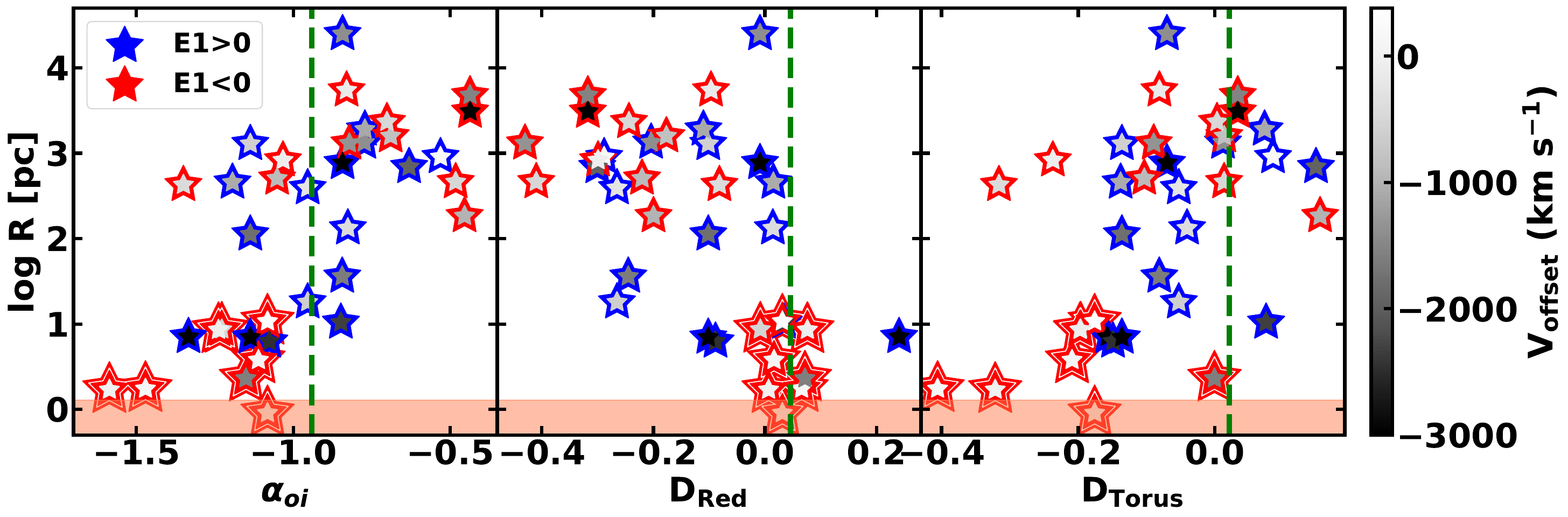}
\caption{Relationships between the location of the outflow, $\log R$,
  and several of the continuum properties shaded by the outflow
  velocity.  The vertical dashed lines 
  show the measurement from the \citet{krawczyk13} composite continuum for
  reference.  The coral shaded area shows the range of dust
  sublimation radii, i.e., the inner edge of the torus, among the
  FeLoBALQs. The double stars show the loitering outflow 
  objects, defined by \citet{choi22} to have $\log R < 1.2$ [pc] and
  $v_\mathrm{offset} > -2000\rm  \, km\,s^{-1}$. The location of the outflow
  is a function of the   
 ionization parameter $\log U$ and the density $\log n$ measured using
 {\it SimBAL}. Similar correlations are found with the covering
 fraction parameter $\log a$, the column density $\log N_\mathrm{H}$,
 and the force multiplier (Fig.~\ref{corr_optical_simbal}).  Outflows
 close to the central engine show  overall steep (blue) optical/IR
 continuum  spectra (more negative values of $\alpha_{oi}$), while
 those located  far from the central engine have flat (red)
 optical/IR continuum  spectra (less negative values of
 $\alpha_{oi}$).   \label{cont_corr_plot}}
\end{center}
\end{figure*}

The $\alpha_\mathrm{oi}$ parameter is not
easily attributable to a single physical origin.  The fact that both
$D_\mathrm{red}$ and $D_\mathrm{torus}$ are also correlated with the
{\it SimBAL} parameters suggests
that both the optical and the infrared regions are participating.  A
clue may be found in the properties of the
loitering outflows.   Loitering outflows were
identified as a new class of FeLoBAL outflow by \citet{choi22}.
Despite the location of the outflow near the central engine, they
showed very low outflow velocities and relatively narrow absorption
lines, yet an overall high absorption opacity due to a large number of
high excitation \ion{Fe}{2} lines \citep[the so-called 'iron
  curtain',][]{lucy18}. \citet{leighly22} found that they were
uniformly low accretion rate objects, and \citet{choi22b} showed that
their outflows had high volume filling factors approaching one.  In
the middle panel of Fig.~\ref{cont_corr_plot}, the loitering outflow
objects that have  $0 < \log R < 1$ are notably clustered
around $D_\mathrm{red}=0$ indicating that 
they have as blue optical photometry as the comparison composite
spectra.  At the same time, these objects have low accretion rates,
and therefore show the lack of hot dust emission indicated by the
correlation between accretion rate and  W1-W2 noted in
\S\ref{corr_w1w2}.  In contrast, the objects with larger $\log R$ have
evidence for a deficit in the optical band ($D_\mathrm{red}<0$) and a
surplus in the infrared band ($D_\mathrm{torus} > 0$).   

Among the objects with outflows located close to the central engine
($\log R < 1.2$ [pc]), low accretion rate objects (8) outnumber the
high accretion rate  ones (4).  More high accretion rate objects, in
particular, high accretion rate overlapping trough objects, are found
to have  outflows located close to the central engine in the
\citet{choi22} sample of 50 FeLoBAL quasars
\citep[e.g.,][Fig.\ 6]{choi22}; however, their SDSS spectra did not have
the H$\beta$ region at the long wavelength end, principally because
their redshifts were larger than 1.0.    

Regardless, we interpret these correlations as a general lack of dust
signatures near the central engine.   Specifically, we suggest that the
increase in continuum reddening with outflow location 
suggested by the $\log R$ / $D_\mathrm{red}$ anticorrelation 
is evidence for an increased prevalence of SMC-style dust attenuation
farther from the central engine; this dust may or may not be directly
associated with the outflow.  Thus, it may be that intrinsic reddening 
increases as a function of radius in unabsorbed quasars as well, but
since constraining the location of reddening in unabsorbed quasars is
difficult, that correlation may be correspondingly difficult to
detect.   

Similarly, the increase in torus emission with more distant outflow
location suggested by the $\log R$ - $D_\mathrm{torus}$ correlation
could be evidence for a  lack of hot dust at locations close to the
central engine, particularly 
among the low accretion rate objects.  Examination of the right panel
in Fig.~\ref{cont_corr_plot}  indicates that low accretion
rate objects dominate the objects with $D_\mathrm{torus}$ less than
$\sim -0.2$, and if they were removed, it appears likely that the
correlation would not be significant.  Indeed, while the p-value for 
the full correlation is $2\times 10^{-3}$, the p-values are 0.037 and
0.021 for the $E1<0$ and $E1>0$ objects, respectively, i.e., still
significant but not strong.  

We note finally that the lack of evidence for dust attenuation near the
central engine  extends only to dust with significant $E(B-V)$, i.e.,
attenuation in the optical band; the photometry data with wavelengths
less than 3000\AA\/ 
were not used in the spectra fitting due to the contamination by
BALs.  At shorter wavelengths, we have found evidence for so-called
anomalous reddening \citep[e.g.,][\S 4.2]{choi20} in several objects
with outflows close to the central engine.  Anomalous 
reddening is characterized by little attenuation in the optical band,
transitioning fairly abruptly to strong attenuation shortward of the
near-UV \citep[e.g.,][]{hall02,leighly09,fynbo13,jiang13,meusinger16}.
Such reddening may be produced by a dust distribution 
dominated by small grains \citep[e.g.,][]{jiang13}.   In particular,
FeLoBAL quasars with high velocity outflows and overlapping troughs
frequently showed this type of reddening, and this correspondence may
be the origin of the correlation between reddening and outflow
velocity found by \citet{choi22}.  

\subsection{Correlations with {\it WISE} Variability
  Parameters}\label{correlation_var}

\begin{figure*}[!t]
\epsscale{1.0}
\begin{center}
\includegraphics[width=6.5truein]{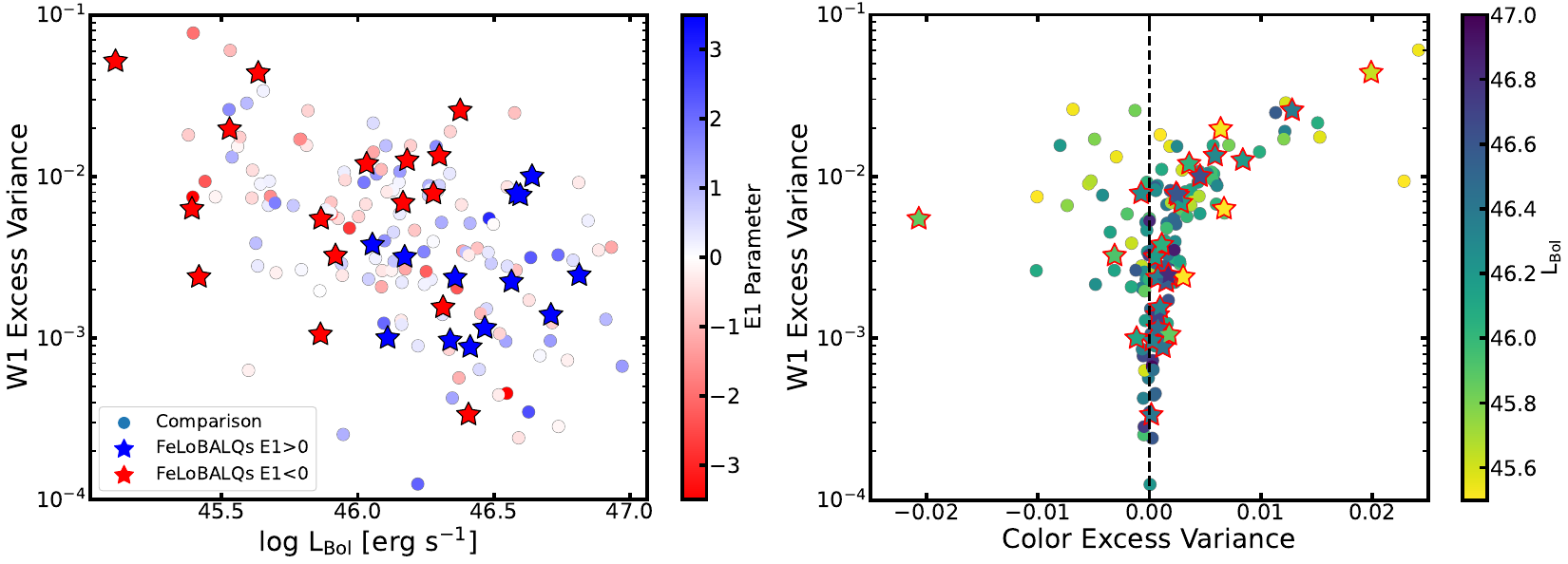}
\caption{{\it Left:}  The W1 excess variance is significantly
  statistically anticorrelated with   the bolometric luminosity for 
  the comparison sample (circles). While there is a tendency for the
  FeLoBAL quasars (stars) to be more variable for lower luminosities,
  the  relationship is very scattered and no significant correlation is
 present.  {\it Right:}  The W1 excess variance as a function of the
 difference between the excess variances for W1 and W2 (color excess
 variance).  The FeLoBALQs  show
 stronger variability in W1 than in W2 (positive color excess
 variance); this behavior is dominated by the low accretion rate
 objects (Fig.~\ref{varcumdist}).  
   \label{var_results}} 
\end{center}
\end{figure*}

Among the comparison sample objects, the variability amplitude
parameters are principally anticorrelated with parameters that depend
upon the luminosity and Eddington ratio (Fig. \ref{correlation}). This
pattern of behavior is commonly found  \citep[e.g.,][]{vandenberk04,
wilhite08, kelly09, macleod10, zuo12, gallastegui14, simm16,
caplar17, sun18, laurenti20, suberlak21,  decicco22, yu22}. 
Fig.~\ref{var_results} shows the relationship between W1 excess
variance and luminosity.  The excess variance values for the FeLoBAL
quasars are very scattered and there are no strong correlations with
any of the optical or continuum parameters.  This
phenomenon could point to some alterative parameter dependence, 
but it is also possible that the stocastic nature of quasar
variability is the origin of the scatter and lack of correlation.

The variability parameters for the comparison objects are 
strongly correlated with the Eddington ratio,
$L_\mathrm{bol}/L_\mathrm{Edd}$ and with $R_\mathrm{2800}$, the
characteristic radius for 2800\AA\/ continuum emission \citep[defined
  in][]{leighly19}.  Both of these parameters depend on 
$L_\mathrm{bol}$. The Eddington ratio dependence is expected to be 
$L_\mathrm{bol}^{1/2}$ (i.e., through $L_\mathrm{bol}$ itself and
through the computation of the radius of the broad line region
necessary to compute the black hole mass), and a numerical experiment
shows that the $R_\mathrm{2800}$ dependence should be
$L_\mathrm{bol}^{0.32}$ for a constant black hole mass. 

The variability parameters show few correlations with any of the
{\it SimBAL} parameters (Fig.\ref{corr_optical_simbal}).  In
\citet{choi22b}, we found that the outflow velocity was very strongly
correlated with the luminosity, so it might be expected that the
commonly known inverse relationship between luminosity and excess
variance would spawn a correlation with outflow velocity.  None is
observed.  

We plot color excess variance as a function of the W1 excess variance
in Fig.~\ref{var_results}.  Recalling that the color excess variance
is defined as the difference between the excess variance of W1 and the
excess variance of W2, the excess of points right of the color excess
variance $=0$ line means that objects are more variable at shorter
wavelengths.  The color excess variance appears more
scattered for the comparison sample objects than for the FeLoBALQs.
Specifically, 87\% of the FeLoBAL quasars  have color excess variance
greater than zero, while only 69\% of the comparison objects show this
property.  Statistically, this situation constitutes the comparison of
two binomial distributions.  We can  determine whether these
distributions are statistically different using the Fisher Exact 
 test\footnote{https://docs.scipy.org/doc/scipy/reference/generated/scipy.stats.fisher\_exact.html}.
 The hypothesis is that the FeLoBAL  quasars show greater variance in
 W1 than in W2 than the comparison  sample as measured by color
 variance greater than zero.  The $p$ value for this test is 0.037,
 and therefore there is evidence that the FeLoBAL quasars are more
 variable in W1 than in W2 than the comparison objects.  Because the
 color excess variance of the $E1>0$ (high accretion rate) objects is
 uniformly close to 
 zero (Fig.~\ref{varcumdist}), this difference is driven by the
 properties of the $E1<0$ (low accretion rate) objects.

\section{A Comparison with LoBAL Quasars}\label{comp_lobal}

In \citet{leighly22}, we presented evidence that FeLoBAL quasars have
significantly different H$\beta$ / \ion{O}{3} / \ion{Fe}{2} properties
than a luminosity, redshift, and signal-to-noise ratio matched sample
of unabsorbed quasars.  \citet{leighly22} found that the distribution
of the accretion rate parameter $E1$ was significantly different
between the FeLoBAL quasars and the comparison sample.  Likewise, the
distributions of $R_\mathrm{FeII}$ and [\ion{O}{3}] were also
significantly different. The FeLoBAL quasars were characterized by
optical emission line properties that were interpreted as evidence for
a low accretion rate (broad Balmer lines, weak \ion{Fe}{2}, strong
\ion{O}{3}, and $E1<0$) or a high  accretion rate (narrow Balmer
lines, strong \ion{Fe}{2}, weak \ion{O}{3}, and $E1>0$).  

\begin{figure*}[!t]
\epsscale{1.0}
\begin{center}
\includegraphics[width=6.5truein]{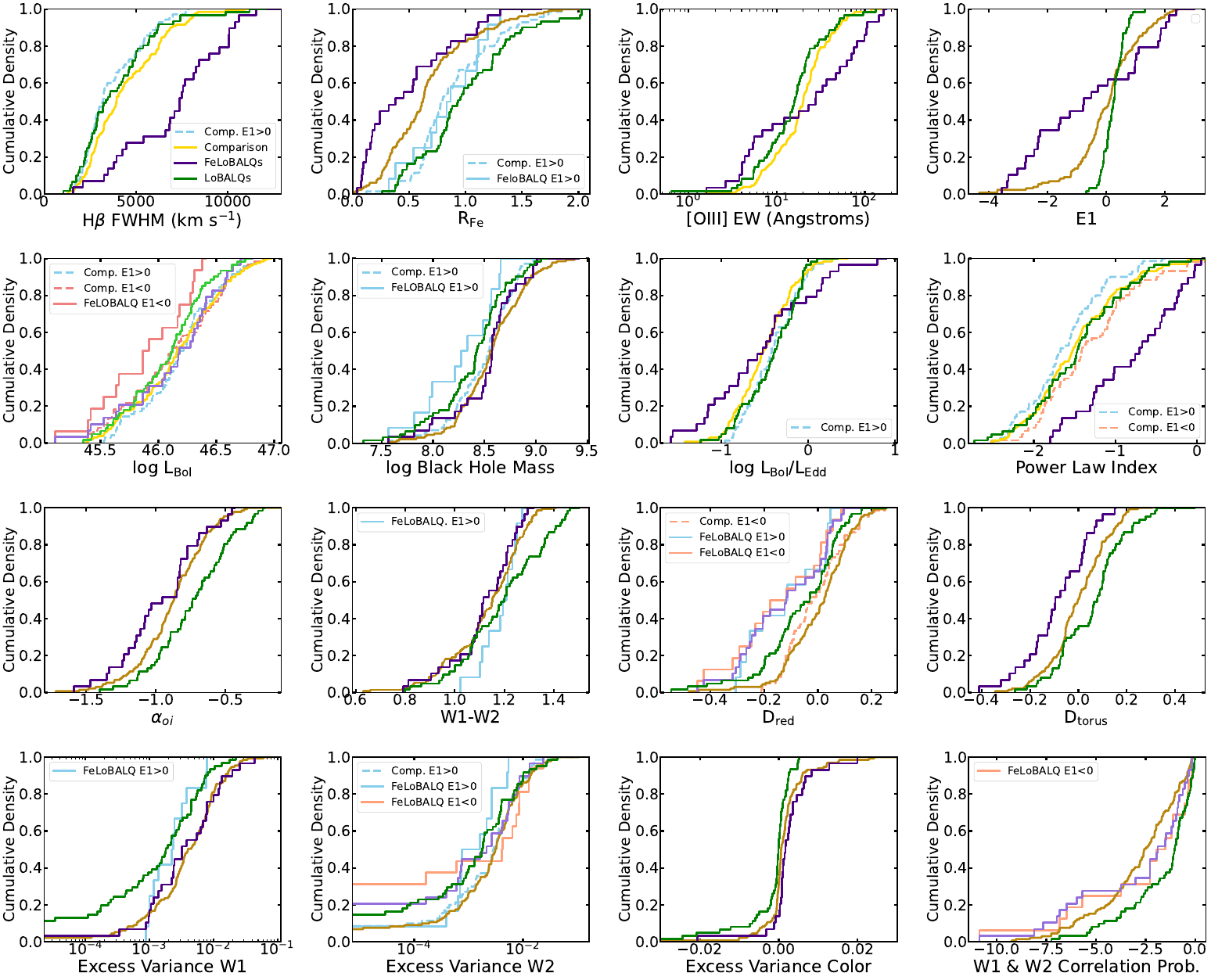}
\caption{Distribution comparison of optical emission line, global,
  continuum, and variability properties between the sample of 62
  LoBALQs (green) presented   in this paper and the 30 FeLoBALQs
  (purple) and 132  unabsorbed comparison   sample objects  (mustard)
  presented in   \citet{leighly22}. Darker (paler) colors indicate significant
  (insignificant) statistical differences between the distributions.
  The legends and lines are included in a panel when an
  $E1$-partitioned parameter is statistically consistent with the
  LoBAL parameter  distribution.  Overall, the subsample that is the most
  similar to the LoBAL quasars   is the  $E1>0$ comparison 
  sample objects.  \label{lobaldists}}   
\end{center}
\end{figure*}

An obvious next question is whether these trends are shared by
other types of broad absorption line quasars, i.e., the LoBAL quasars
and the HiBAL quasars.  In this paper, we analyzed the sample of LoBAL 
quasars described in \S\ref{lobals}, comparing them with both the
FeLoBAL quasars and the comparison sample described in
\citet{leighly22}. Key cumulative distribution functions are shown in
Fig.~\ref{lobaldists} and the results of the KS and AD tests are given
in Table~\ref{lobaltab}.  

\begin{deluxetable*}{lCCCCCC}
\tabletypesize{\scriptsize}
\tablecaption{LoBAL Parameter Distribution Comparison\label{lobaltab}}
\tablehead{
  \\
\colhead{Parameter Name} & \multicolumn{2}{c}{FeLoBALQs vs Comparison}
& \multicolumn{2}{c}{LoBALQs vs Comparison} &
\multicolumn{2}{c}{FeLoBALQs vs LoBALQs} \\ 
\colhead{} & \colhead{KS\tablenotemark{a}} & \colhead{AD\tablenotemark{b}} & \colhead{KS\tablenotemark{a}} & \colhead{AD\tablenotemark{b}} & \colhead{KS\tablenotemark{a}} & \colhead{AD\tablenotemark{b}}  \\
}
\startdata
Redshift & 0.14 / 0.65 & -0.03 / $>0.25$ & 0.13 / 0.47 & 0.13 /
$>0.25$ & 0.20 / 0.35 & 0.54 / 0.20 \\
W1 Magnitude & 0.08 / 0.99 & -0.82 / $>0.25$ & 0.19 / 0.09 & {\bf 3.0
  / 0.020} & 0.25 / 0.14 & {\bf 2.09 / 0.045} \\
W2 Magnitude & 0.10 / 0.93 & -0.74 / $>0.25$ & 0.16 / 0.23 & 1.43 /
0.08 & 0.21 / 0.29 & 0.93 / 0.14 \\
\hline
$H\beta$ FWHM & {\bf 0.56 / $1.5 \times 10^{-7}$} & {\bf 18.0 / $<
  0.001$} & 0.14 / 0.32 & 0.98 / 0.13 & {\bf 0.60 / $2.3 \times
  10^{-7}$} & {\bf 15.4 / $<0.001$} \\
$R_\mathrm{FeII}$ & {\bf 0.35 / $3.6 \times 10^{-3}$} & {\bf 4.5 /
  $5.4 \times 10^{-3}$} & {\bf 0.45 / $3.7 \times 10^{-8}$} & {\bf
  17.0 / $< 0.001$} & {\bf 0.51 / $2.9 \times 10^{-5}$} & {\bf 14.3 /
  $<0.001$} \\
$[$\ion{O}{3}$]$ EW  & {\bf 0.28 / 0.034} & {\bf 6.5 / $1.1\times 10^{-3}$}
& 0.18 / 0.10 & 1.6 / 0.07 & {\bf 0.32 / 0.022} & {\bf 3.7 / 0.010}  \\
E1 Parameter & {\bf 0.34/$4.7\times 10^{-3}$} & {\bf 5.7 / $1.9\times
  10^{-3}$} & {\bf 0.52 / $1.6 \times 10^{-5}$} & {\bf 12.7
  /$<0.001$} & {\bf 0.35 / $3.3 \times 10^{-5}$} & {\bf 8.5 /
  $<0.001$} \\
SPCA $E1$ & {\bf 0.33 / $7.2 \times 10^{-3}$} & {\bf 5.0 / $3.5 \times
  10^{-3}$}  & {\bf 0.31 / $5.0 \times 10^{-4}$} & {\bf 9.0 /
  $<0.001$} & {\bf 0.44 / $5.3 \times 10^{-4}$} & {\bf 8.0 / $< 0.001$}\\
SPCA $E2$ & {\bf 0.40 / $4.7 \times 10^{-4}$} & {\bf 5.1 /
  $3.1\times 10^{-3}$} & {\bf 0.22 / 0.03} & {1.1 / 0.12} & {0.22 /
  0.26} & {0.37 / 0.24}\\
SPCA $E3$ & {\bf 0.37 / $1.5\times 10^{-3}$} & {\bf 10.3 / $<0.001$} &
{\bf 0.44 / $7.5 \times 10^{-8}$} & {\bf 17.2 / $<0.001$} & {0.25 /
  0.12} & {1.7 / 0.07} \\
\hline
$L_\mathrm{bol}$ & 0.11 / 0.91 & -0.72 / $>0.25$ & 0.15 / 0.27 & 0.53 /
0.20  & 0.21 / 0.29 & 0.26 / $>0.25$ \\
$M_\mathrm{BH}$ & 0.13 / 0.79 & $-0.67$ / $>0.25$ & {\bf 0.23 / 0.021}
& {\bf 4.8 / 4.1$\times 10^{-3}$} & {\bf 0.31 / 0.029} &  {\bf 2.1 /
  0.044} \\
$L_\mathrm{bol}/L_\mathrm{Edd}$ &  0.20 / 0.23 & {\bf 3.1 / 0.018} &
0.18 / 0.11 & 1.74 / 0.063 & 0.22 / 0.22 & {\bf $3.2 / 1.7\times 10^{-2}$}\\
\hline
4500-5500\AA\/ Slope & {\bf 0.48 / $1.3 \times 10^{-5}$} & {\bf 17.6 /
  $<0.001$} & 0.10 / 0.79 & $-0.77$ / $>0.25$ & {\bf 0.48 / $9.1
  \times 10^{-5}$} &  {\bf 12.8 / $<0.001$}  \\
$\alpha_{oi}$ (5100\AA\/ - 3$\mu$m slope) & 0.21 / 0.20 & 0.41 / 0.23
& {\bf 0.26 / $5.2\times 10^{-3}$} & {\bf 7.8 / $<0.001$} & {\bf 0.35
  / 0.012} & {\bf 5.8 / $1.9 \times 10^{-3}$} \\
W1-W2 & 0.11 / 0.91 & -0.86 / $>0.25$ & 0.27 / 0.08 & {\bf 2.1 /
  0.046} & {\bf 0.25 / $7.6 \times 10^{-3}$} & {\bf 5.0 / $3.5 \times
  10^{-3}$} \\
$D_\mathrm{red}$ & {\bf 0.37 / 1.5 \times 10^{-3}} & {\bf 9.3 /
  $<0.001$} & 0.19 / 0.069 & {\bf 4.4 / $5.6 \times 10^{-3}$} & 0.27 /
0.08 & 1.32 / 0.09 \\
$D_\mathrm{torus}$ & {\bf 0.31 / 0.014} & {\bf 4.9 / $3.7 \times
  10^{-3}$} & {\bf 0.27 / $2.8 \times 10^{-3}$} & {\bf 5.0 / $3.6
  \times 10^{-3}$} & {\bf 0.48 / $9.1 \times 10^{-5}$} & {\bf 10.5 /
  $<0.001$} \\
\hline
W1 Excess Variance & 0.14 / 0.69 & -0.08 / $>0.25$ &  {\bf 0.30 /
  $7.5\times 10^{-4}$} & 
{\bf 10.2 / $<0.001$ } & 0.29 / 0.054 & {\bf 3.67 / 0.011} \\
W2 Excess Variance & 0.23 / 0.13 & 0.98 / 0.13 & {\bf 0.23 / 0.017} &
{\bf 2.3 / 0.035} & 0.18 / 0.49 & -0.60 / $>0.25$ \\
Color Excess Variance & {\bf 0.29 / 0.026} & {\bf 2.0 / 0.048 } & {\bf
    0.22 / 0.027} & {\bf 5.4 / $2.6 \times 10^{-3}$} & {\bf 0.46 / 
    $2.1 \times 10^{-4}$} & {\bf 9.4 / $<0.001$} \\
W1 \& W2 Correlation Prob. & 0.20 / 0.22 & 1.03 / 0.12 & {\bf 0.36 / $2.4
  \times 10^{-5}$} & {\bf 11.2 / $<0.001$} & 0.23 / 0.18 & 1.8 / 0.058
\\
\enddata
\tablecomments{The FeLoBAL quasar sample includes 30 objects.  The
  unabsorbed comparison sample includes 132 objects.  The LoBAL quasar
  sample includes 62 objects.}
\tablenotetext{a}{The Kolmogorov-Smirnov Two-sample test.  Each entry
  has two numbers:  the first is the value of the statistic, and the
  second is the probability that the two samples arise from the same
  parent sample.  Bold type indicates entries that yield $p<0.05$.}
\tablenotetext{b}{The Anderson-Darling Two-sample test.  Each entry
  has two numbers:  the first is the value of the statistic, and the
  second is the probability that the two samples arise from the same
  parent sample.  Note that the implementation
  used does not compute a probably   larger than 0.25 or smaller than
  0.001. Bold type indicates entries that yield $p<0.05$.}
\end{deluxetable*}

The top section of Table~\ref{lobaldists} shows that the redshift and
{\it WISE} W1 and W2 magnitude distributions are consistent among 
the FeLoBAL quasar sample, the unabsorbed quasar sample, and the LoBAL
quasar sample. Likewise, the bolometric luminosity distribution 
is also consistent.  This result is not surprising given the selection
criteria used to construct the comparison sample, namely the matching
in redshift, signal-to-noise ratio, and bolometric luminosity which
was estimated using the 3-micron bolometric correction from
\citet{gallagher07}.

Beyond these properties, there is abundant evidence that the LoBAL
quasars are neither similar to the FeLoBAL quasars nor to the unabsorbed
comparison sample objects (Fig.~\ref{lobaldists}).  For example, the
LoBAL quasars are characterized by stronger \ion{Fe}{2} emission than
either the FeLoBAL quasars or the comparison sample as whole, yet
their $E1$ values show a very small spread and are very close to
zero.   

To help interpret these results, we also compared the LoBAL
properties as a whole to the FeLoBALQ and comparison samples divided
by $E1$ parameter.  Those combinations that were consistent with the
LoBALQ property distributions are shown in Fig.~\ref{lobaldists}.
These comparisons show that the LoBAL quasars are more frequently
similar to the comparison objects with $E1>0$ among properties related
to accretion rate, such as H$\beta$ FWHM, $R_\mathrm{FeII}$, and
$L_\mathrm{bol}/L_\mathrm{Edd}$.   The LoBAL quasars are similar to
the FeLoBAL quasars in $D_\mathrm{red}$, but more similar to the
unabsorbed objects in power law index, perhaps indicating that the
LoBALs are more subject to reddening that includes attenuation in the
rest optical band rather than being possibly intrinsically flat as are
the FeLoBAL quasars.  

Our goal for this analysis was to determine whether the LoBAL quasars
are more similar to the FeLoBAL quasars or the unabsorbed comparison
quasars.  These results indicate that they are different from both
samples, but suggest that LoBAL quasars are characterized overall by a
high accretion rate relative to Eddington.   Perhaps they comprise the high
accretion rate BAL quasars that  \citet{boroson02} found \citep[see
  also][]{turnshek97}.  This 
superficial analysis also suggests that LoBAL quasars may be
interesting objects in their own right.  For this paper, the most
important result is that the accretion properties of the LoBAL quasars 
are more similar to those of the unabsorbed comparison sample, and
they do not show the apparent bimodality that the FeLoBAL quasars do.  

It is possible that selection effects influence the results.  As
stated in \S\ref{lobals}, the LoBALQ sample was taken from the DR12
BAL quasar catalog \citep{paris17}.  As discussed in \citet{leighly22},
the FeLoBAL quasars came from a variety of sources including the
literature, the results of our convolutional neural net classifier
FeLoNET \citep{dabbieri20}, and visual examination.  We partially
tested the possibility of selection effects by comparing the
aforementioned 27 FeLoBALQs found among the $0.75 < z < 1.0$ DR12 BAL
objects with the 30 objects discussed here.  After removing the
three objects in our FeLoBALQ sample that had $z<0.75$, we found that
12 objects were in both samples.  Among those twelve, 7 (5) have
$E1>0$ ($E1 <0$), so there is no evidence that either accretion state was
favored among the FeLoBALs in the DR12Q BAL catalog.  In addition, the
average $E1$ values of these 12 objects were 1.5 ($-1.7$) for $E1>0$
($E1<0$), i.e., much larger in amplitude than the typical values found
for the LoBALQs (Fig.~\ref{lobaldists}; $-0.22$ ($0.39$) for $E1<0$
and $E1>0$).  Thus, there is no strong evidence that our results are
dominated by selection. 

\section{Discussion}\label{discussion}

In this paper, we continued our exploration of a sample of 30
low-redshift FeLoBAL quasars. Briefly (see
\S\ref{review} for a full review),  \citet{choi22} analyzed their
outflows using {\it SimBAL}.  \cite{leighly22} analyzed their
rest-frame optical spectra, focusing in particular on the rich
diagnostics available 
in the H$\beta$ / [\ion{O}{3}] / \ion{Fe}{2} region of the spectrum. 
\citet{choi22b} combined the results from the first two papers,
investigating how the outflow properties depend on accretion rate.  This
paper focuses on their optical-NIR continuum spectra extracted from
archived photometry observations, and their near-infrared variability 
obtained from the decade-long {\it WISE} observations.  Finally,
rest-frame optical spectra of a new sample of LoBAL quasars were
analyzed to determine whether LoBALs are similar to or different from
FeLoBALQs in their accretion properties. 

\subsection{Optical-Infrared Continuum Properties}\label{disc_continuum}

The fact that the continuum properties in the unabsorbed objects are
most significantly correlated with the Eddington ratio and similar
parameters (Fig.~\ref{correlation}, \S\ref{correlation_cont},
Fig.~\ref{cont_eddrat}) indicates that much of the variance in the SED
shapes is intrinsic to 
the quasar rather than an extrinsic property such as a reddening
\citep[e.g.,][]{davis07}, or inclination \citep[e.g.,][]{capellupo15}.

The origin of the optical-UV power law slope in quasars has long been
a mystery. It is believed to be the emission from the accretion disk.
However, the accretion disk spectrum should have a $F(\nu) \propto
\nu^{1/3}$, corresponding to $F(\lambda) \propto \lambda^{-2.33}$ 
\citep[e.g.,][]{kb99}.  This is much steeper (bluer) than observed in
most quasars, although a few objects in our sample are nearly as
steep (Fig.~\ref{plot_comp_4}). Considerable effort has been devoted
to explaining the flatter observed spectra, although these efforts
have focused principally on the UV and shorter wavelengths and not the
optical wavelengths investigated in this paper.   For example, a
physical accretion disk is expected to have an atmosphere which will
modify the emitted spectrum through atomic processes
\citep[e.g.,][]{dl11}.  Alternatively, an accretion disk wind can also
flatten the spectrum \citep{sn12, capellupo15}. However, one might
expect a stronger disk wind in higher accretion 
objects, and a correlation between power law index and Eddington ratio
would be predicted, rather than the anticorrelation that we found.

We suggest that the origin of the anticorrelation between the
4500-5500\AA\/ slope parameterized by the power law index and the
Eddington ratio and similar factors originates in the expected changes
in the accretion disk structure due to changes in the accretion
rate \citep[e.g.,][]{af13}.  As the accretion rate decreases, the
accretion disk is expected to switch from an optically thick
geometrically thin standard disk to an ADAF in the inner regions of
the quasar central engine, and the radius at which the transition
happens is expected to vary with accretion rate.  The expected changes
in the spectral energy distribution are shown in
\citet[][Fig.\ 2]{mitchell22}; the spectrum is steeper for higher
accretion rates, as we observe.  As before, these changes are expected
to dominate at short wavelengths, and it is not clear whether the
optical band would be affected.  However, those models are
semi-empirical, and some leeway may be expected.

Turning to the near-infrared region of the spectrum, we examined the
W1-W2 color, which at these redshifts measures the upturn toward the
torus longward of 1-micron and therefore measures the presence of hot
dust emission.  We found that while there was no statistical
difference between the FeLoBALQs and the unabsorbed comparison
objects, the color distributions 
of $E1<0$ and $E1>0$ FeLoBALs are statistically significantly
different (Fig.~\ref{plot_comp_4}, Table~\ref{distributions}).  W1-W2
was found to be correlated with the Eddington ratio and similar
parameters  for the unabsorbed comparison
sample, and  also for the FeLoBALQs (Fig.~\ref{correlation}). Low
values of W1-W2 correspond a weak or absent upturn toward long 
wavelengths, and quasars exhibiting these properties are called
Hot-Dust Deficient (HDD) quasars. As noted 
in the introduction, the among the well-studied Palomar-Green quasars,
the HDD objects were found to have relatively lower accretion rates 
\citep{lyu17}.

Physically, what might be the origin of the correlation between the
accretion rate and the presence of hot dust?  We interpret this result
in terms of dynamical outflow models of quasars and the expected
properties of a quasar as the accretion rate is dialed up or down.
The presence of large-scale dynamical outflows in quasars has been
proposed by many people.  The torus may be the portion of this outflow
optically thick enough to both block the view to the central engine
and to thermalize incident radiation sufficiently to emit as a near
black body in the infrared
\citep[e.g.,][]{bp82,kk94,gallagher15}. \citet{es06} built upon this
idea; they suggested that the torus should disappear at low enough
accretion rates that the optically thick wind cannot be sustained.

There is a substantial amount of evidence that low-accretion-rate
quasars lack a  torus.  For example, \citet{wa04} showed that M87, an
object that has an ADAF-like spectral energy distribution, does not
show thermal infrared emission.  \citet{izumi17} found that the
low-luminosity AGN NGC~1097 lacks a torus.  \citet{gonzalez-martin17}
found evidence for the lack of a torus in low-luminosity
AGN. \citet{lyu17} found that, among the 87 $z < 0.5$ Palomar-Green
quasars, objects that are deficient in hot-dust emission are
characterized by a low accretion rate.  \citet{ricci17} found a low
fraction of obscured AGN at very low Eddington ratios.
\citet{trump11} found that the IR torus signature became weaker at low
accretion rates in quasars from the COSMOS
survey. \citet{vanderwolk11} found evidence that low-accretion-rate
radio galaxies lack a torus. 

At low accretion rates, \citet{eh09} found evidence that the
broad-line region also disappears.  These observations are confirmed
in other studies.  \citet{trump11} found that lower-accretion-rate AGN
in the COSMOS survey were unobscured but also lacked a broad line
region.  \citet{constantin15} found that narrow-line LINERs have
weaker emission lines than broad-line LINERs, consistent with the idea
that a low accretion rate causes the disappearance of the broad-line
component. 

These ideas are combined in \citet{elitzur08}.  If we start with a
typical AGN with a well-developed broad line region and torus, and
dial the accretion rate down, we should find the infrared emission
from the torus disappearing first, since the reprocessing of the
incident continuum to thermal emission requires a large column density
of gas.   The outer broad 
line region disappears next, making the Balmer lines appear
broader. Ultimately, the broad line region emission becomes dominated
by the double-peaked disk lines seen in broad-line radio galaxies and
LINERs \citep[e.g.,][]{sb17}.   An example is NGC~3147, a candidate
true type 2 AGN which apparently lacks broad lines but also suffers no
X-ray absorption; an {\it HST} observation of that object revealed a
disk-line profile in H$\alpha$ \citep{bianchi19}.   

\subsection{Reddening}\label{reddening_interp}

Next, we consider the reddening properties.  There is evidence for
reddening in some but not all FeLoBALQs, recognized in the optimal red
and blue composite spectra (\S~\ref{sed_composites},
Fig.~\ref{photometry_sed}) and possibly indicated in the lack of
correlations between the continuum parameters and the other parameters
(\S~\ref{corr_alphaoi}, Fig.~\ref{reddening}).  Specifically, in the
unabsorbed comparison sample, the power law index was found to be
significantly anticorrelated with the Eddington ratio for the
unabsorbed comparison sample  (Fig.~\ref{cont_eddrat}).  Among the
FeLoBAL quasars, no statistically significant correlation was found,
but a general impression of anticorrelation with larger scatter toward
flatter indices was seen, as well as a deficit of objects with the
steepest optical power law indices (Fig.~\ref{plot_comp_4},
Fig.~\ref{reddening}).   Our results confirm  that it is likely that
there is at least a little bit of reddening in the optical band in
most FeLoBAL quasars. 

More significantly, we found that the presence of strong versus weak
optical reddening is not randomly distributed among the FeLoBAL
quasars.  The {\it SimBAL} correlations, shown in
Fig.~\ref{cont_corr_plot}, revealed that objects with more distant
outflows show more evidence for reddening than the ones with outflows
located close to  the central engine.  

What is the physical origin of this difference in reddening properties?
We speculate that the dust is associated with the outflows in some
way.  For example, distant outflows may be formed by a shocked ISM 
\citep[the ``cloud-crushing'' scenario;][]{fg12}; dust is expected to
be present in the interstellar medium.  Alternatively, dust that
causes significant reddening in the optical band, for example dust
with an SMC attenuation curve, may not survive near the central
engine, at least in FeLoBAL quasars. Finally, the anomalous reddening
that is found in some outflows close to the central engine
\citep{choi20, choi22} may be formed in the outflow itself
\citep{elvis02}; we speculate that such newly-formed dust might have
small dust grains, resulting in lack of rest-optical attenuation
\citep[e.g.,][]{jiang13}.   

\subsection{{\it WISE} Variability Properties}\label{disc_var}

We probed the variability properties of the FeLoBALQs and the
unabsorbed quasars using the {\it WISE} photometry which samples the
near infrared variability over five years in the rest frame
(\S\ref{wise_variability}).  We found that overall the variability properties were
consistent between the FeLoBALQs and unabsorbed quasars, but the variability
properties of the $E1<0$ and $E1>0$ subsamples were significantly different
(\S\ref{dist_var}, Fig.~\ref{varcumdist}).  The 
$E1<0$ objects were more variable in both the W1 and W2 filters.
This may be attributed to the well-known anticorrelation between
variability amplitude and luminosity.  Also, while the $E1>0$ objects
showed almost no color variability, the $E1<0$ objects showed a
greater amplitude of variability in W1 than in W2.  

Weak hot dust emission in the low accretion rate ($E1<0$) FeLoBAL
quasars may provide an explanation for their distinctly different
variability properties.  If the torus emission is weak in the near-IR,
the long wavelength tail of the accretion disk component makes up a
larger fraction of the near-IR flux.  The accretion disk size scales
are much smaller than the torus size scales, and therefore can be
expected to vary with larger amplitude on the sampled time interval,
producing the higher amplitude NIR variability observed in 
the $E1<0$ objects.   It is well known that the optical-UV emission in
quasars shows larger amplitude variability at shorter wavelengths \citep[i.e.,
  bluer   when brighter,
  e.g.,][]{vandenberk04,macleod10,zuo12,kokubo14,simm16} potentially
explaining the larger amplitude of variability in W1 compared with W2
among the $E1<0$ objects.    

Some of the correlations between the variability amplitudes and the
optical parameters are suggestive.  We found that in the comparison
sample, the {\it WISE}  excess variance is anti-correlated with the
luminosity, as well as other measurements that depend on the
luminosity, including the black hole mass, the Eddington ratio, and
$R_\mathrm{2800}$, the characteristic radius for the 2800\AA\/ 
continuum emission.   

It is well known that lower-luminosity objects have a shorter 
variability time scales than higher-luminosity objects
\citep[e.g.,][]{vandenberk04, wilhite08, kelly09, macleod10, zuo12,
  gallastegui14, simm16, caplar17, sun18, laurenti20, suberlak21,
  decicco22, yu22}.   The anticorrelation between the excess variance
and $R_\mathrm{2800}$ that we found may provide an explanation for
this behavior through a consideration of interpretations of X-ray
variability.   Seyfert galaxies are well known to show significant
X-ray variability on short time scales, and this variability is one of
the principal pieces of evidence that AGN are powered by accretion
onto a central supermassive black hole \citep[e.g.,][]{mchardy85}.
It has long been known that the time scale of X-ray variability is
anticorrelated with the X-ray luminosity
\citep[e.g.,][]{barr86,lp93,green93,nandra97}, a fact that has been
interpreted as an indication that lower luminosity objects have
smaller emission regions, i.e, black hole masses. For example, using
{\it ASCA} data, \citet{leighly99a} found that narrow-line Seyfert 1
galaxies  (NLS1s) showed systematically larger excess variances than
broad-line objects of the same X-ray luminosity. The larger excess
variance implies a smaller emission region size for the
$1/f$-noise typical of X-ray variability in Seyfert galaxies. Hence,
\citet{leighly99a} concluded that NLS1s have smaller black hole masses
for the same accretion rate than broad-line objects, i.e., a higher
Eddington ratio. This result formed a cornerstone for our
understanding of NLS1s.    

It therefore seems possible that the excess variance anticorrelation
with $R_\mathrm{2800}$ is simply an anticorrelation with the size of
the emission region.  Since $R_\mathrm{2800}$ is correlated with the
luminosity \citep[Fig.\ 12 in][]{leighly22}, the origin of the
commonly observed inverse correlation between variability amplitude
and luminosity may be simply be a consequence of a correlation between
luminosity and emission-region size.  

\subsection{The LoBAL Quasar Sample}\label{disc_lobal}

We compared the rest-frame optical emission line properties of a
sample of LoBAL quasars with similar redshifts and luminosities with
the FeLoBAL quasars and the unabsorbed comparison sample
(\S\ref{lobals}).  We also compared the global, continuum, and {\it WISE}
variability parameters.  Our goal was to determine whether the LoBAL
quasars split into two groups like the FeLoBAL quasars, or whether the
FeLoBAL quasars are a special group among BAL quasars.  We found that
many but not all of the properties are consistent with the unabsorbed
comparison sample (\S\ref{lobaldists}, Table~\ref{lobaltab}). The 
LoBAL quasars do not split into two groups in $E1$, 
Eddington ratio, or similar parameters.  We conclude that FeLoBAL
quasars are a special class of object, distinct from both unabsorbed
quasars and other broad absorption line quasars.

As discussed in \S\ref{comp_lobal}, it is possible that the LoBAL
sample is affected by selection effects.
Creating a selection-free sample of quasars is well known to be
difficult, and this difficulty is compounded if objects are reddened
and their emission lines absorbed. With {\it Gaia}, astrometric
selection provides a promising method that could yield a quasar sample
that, while not completely selection free, at least has simple
selection criteria that can be modeled.  \citet{krogager23}
describes the 4MOST-Gaia Purely Astrometric Quasar Survey (4G-PAQS), a
project that will carry out the first large-scale completely
color-independent quasar survey using a sample selected solely based
on astronometry from {\it Gaia}.  It is expected to obtain a sample of 
100,000 quasars over the five years of the 4MOST project, sufficient
to obtain a substantial sample of LoBAL and FeLoBAL quasars. 

The analysis presented in \S\ref{comp_lobal} does not support the
possibility that the differences between FeLoBALQs and LoBALQs are 
driven by selection effects.  Therefore, until the 4G-PAQS sample has
been analyzed, we conclude that FeLoBAL quasars are both different
from unabsorbed quasars, as shown in \citet{leighly22}, and different
from other BAL quasars. 

\subsection{The Accretion Properties of FeLoBAL
  Quasars}\label{accretion_properties} 

In \citet{leighly22}, we showed that FeLoBAL quasars are significantly
different in their accretion properties than the unabsorbed comparison
sample.   We established that while a luminosity, redshift,
and signal-to-noise ratio matched sample of unabsorbed quasars is
characterized by a single peaked distribution in accretion properties,
the FeLoBAL quasars more naturally divide into high accretion rate and
low accretion rate groups.   

\begin{figure*}[!t]
\epsscale{1.0}
\begin{center}
\includegraphics[width=6.5truein]{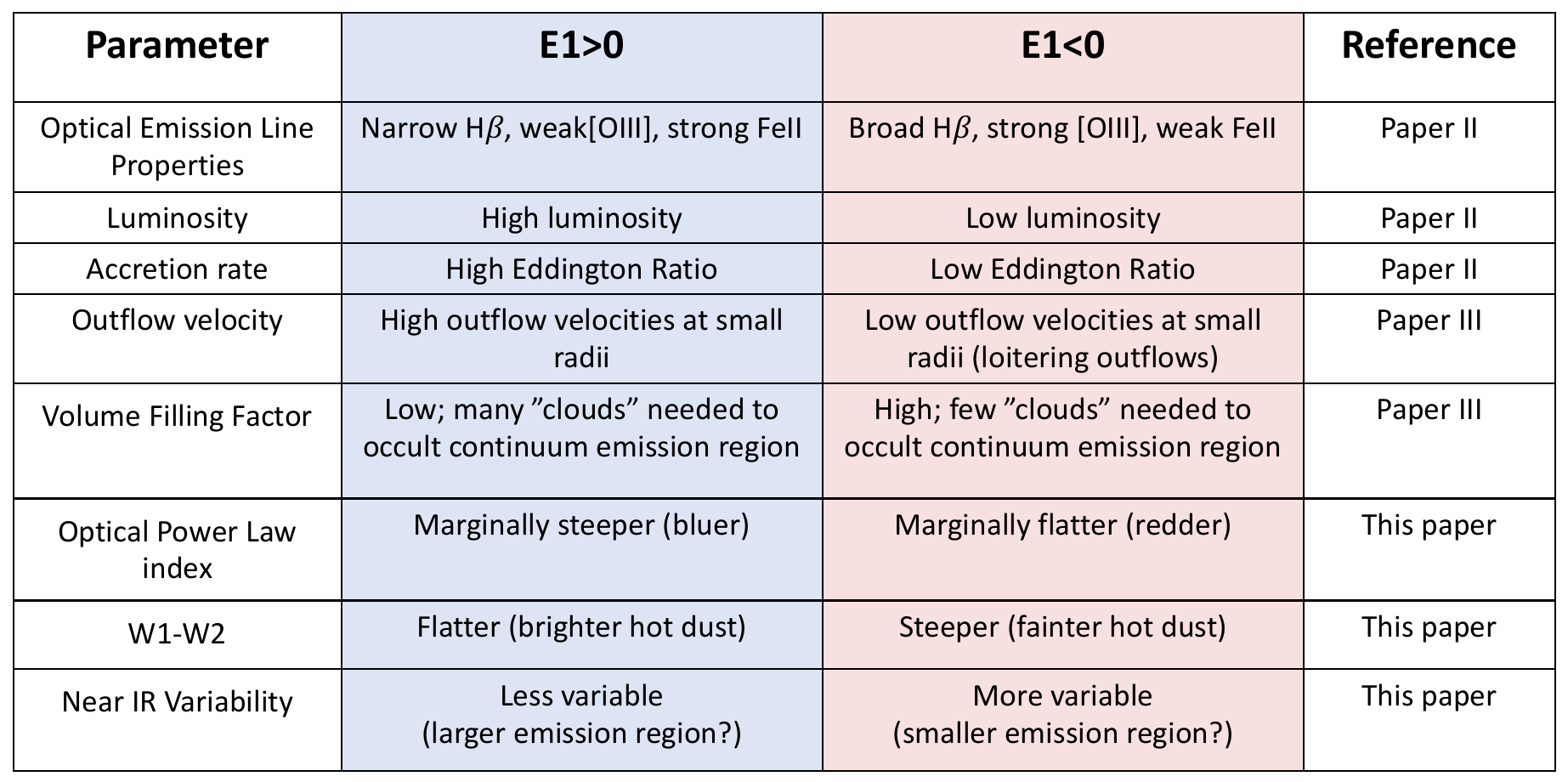}
\caption{A summary of the observed differences in emission line,
  continuum, variability, and outflow properties divided by accretion
  rate parameter $E1$ for the two FeLoBALQ classes.\label{prop_table}} 
\end{center}
\end{figure*}

In \citet{choi22b} and in this paper, we divided the FeLoBAL quasars
into two groups based on the $E1$ parameter, and bent to the task of
characterizing the differences between the two groups.  These
differences are summarized in Fig.~\ref{prop_table}. 

Using the ideas discussed in those papers and above in
\S\ref{disc_continuum}, we now present explanations for the origins of
the two populations of FeLoBAL quasars, followed by a description of
how FeLoBAL quasars fit into the quasar general population.   

\subsubsection{The Origin of  FeLoBALQs with $E1<0$}\label{e1lt0}

We have clearly established that the $E1<0$ FeLoBAL quasars are low
accretion rate objects.  As discussed in \citet{leighly22}, the
presence of such objects among a sample of BAL quasars is not
necessarily expected. \citet{boroson02}  suggested that BAL quasars
have high accretion rates and high  Eddington ratios.  All of the
$E1<0$ objects defy that conjecture; they fall among the radio-loud
objects in the \citet{boroson02} scenario
\citep[][Fig.~14]{leighly22}.  However, their radio properties have
not been systematically investigated yet.   

\ion{Fe}{2} absorption is not unknown in low accretion rate objects.
It has been documented in Arp~102B, the prototypical double-peaked
emission line AGN \citep{halpern96, eracleous03}.  The {\it HST}
spectrum of this object \citep[Fig.~1 in][]{halpern96} clearly 
shows high excitation \ion{Fe}{2} lines characteristic of a relatively
high density absorber.  \ion{Fe}{2} absorption has been found in a few
other broad-line radio galaxies \citep{eracleous02}, and in LINERs and
other low-accretion-rate objects \citep{shields02,sabra03}.  We
suggest that our $E1<0$ subsample are drawn from the higher accretion
rate portion of the same parent sample for the
broad-line radio galaxies and LINER objects reported in the literature
to exhibit \ion{Fe}{2} absorption.  

The key to understanding \ion{Fe}{2} absorption in low accretion-rate
objects may lie in the expected and observed properties of the quasar
central engine as the accretion rate is dialed down.   We propose that
the \ion{Fe}{2} absorption in the compact $E1<0$ objects (the
loitering outflows) marks a stage in this scenario, as follows. The
reprocessed continuum emission observed at infrared wavelengths that
characterize objects with a  moderate and high accretion rate 
requires the torus to be very optically thick.     \citet{elitzur08}
presented a scheme for the evolution of an AGN as the accretion rate
decreases  (see Fig.\ 12 in that publication).  Specifically, as the
accretion rate is dialed down, there is not enough energy to sustain
the wind that is the torus, and the torus disappears.   

We speculate that a  decreasing accretion rate does not
manifest as a sudden disappearance of the torus;  rather, it may
become optically thinner by degrees.  At first, the dust optical depth
drops to the point that it is no longer optically thick enough to
reprocess continuum light into the near infrared; the near- to mid-IR
torus emission would disappear, as we observe.  At some point, the
remaining  column density in the torus may become low enough that the
dust can be sublimated and the remaining gas photoionized.  If the
column density is still substantial enough to include the hydrogen ionization front
(Str\"omgren sphere) in the photoionized slab, Fe$^+$ ions may be
present, creating the signature of an FeLoBAL.   At this point, the
torus wind motion may be dominated by rotation rather than outflow,
explaining the low velocities characteristic of the $E1<0$ FeLoBAL
quasars and distinct from the $E1>0$ FeLoBALQs at with torus-scale
outflows \citep[Fig.\ 4,][]{choi22b}.  Likewise, the remaining torus gas may be
mostly uniformly distributed, explaining the large filling fraction
also characteristic of $E1<0$ FeLoBALQs \citep[Fig.\ 8,][]{choi22b}.
Finally, ablation of the outer broad line region may explain the
systematically broader H$\beta$ FWHM observed among the $E1<0$ objects
\citep[][\S3.4.1, Fig.~13]{leighly22}.   

\subsubsection{The Origin of FeLoBALQs with $E1>0$ }\label{e1gt0}

We have clearly established that the $E1>0$ FeLoBAL quasars are high
accretion-rate objects.   \citet{boroson02} suggested that all BAL
quasars have high accretion rates and high Eddington ratios, and the
$E1>0$ FeLoBAL quasars conform to that expectation
\citep[][Fig.~14]{leighly22}. 

To understand the $E1>0$ FeLoBALQs, we turn to their H$\beta$ FWHM
properties. \citet{leighly22} found H$\beta$ FWHM to be systematically
broader for the FeLoBALQs compared with the unabsorbed quasars
(\S~3.4.1, Fig.~13 of that paper).  As discussed in that paper, there
are two explanations for this result: 1.) $E1>0$ FeLoBAL quasars are
ordinary quasars viewed at a large angle from the normal to the system symmetry
axis; 2.) $E1>0$ FeLoBAL quasars lack emission in the core of the
line, making the line appear broader.  

The evidence that supports the larger inclination angle principally
comes from studies of LoBAL quasars, originally thought to be a
plausible parent sample of FeLoBALQs (although shown in this paper to
be different). LoBAL quasars are more likely to be  highly polarized
than HiBAL quasars \citep[e.g.,][]{dipompeo13}, suggesting supression
of the continuum along the direct line of sight and allowing the
polarized light from axial or high-latitude scattered emission to be
detected \citep[e.g.,][]{wills92, leighly97}. LoBAL quasars  can be
red or can be found preferentially in samples of red quasars
\citep[e.g.,][]{urrutia09}, and are generally extremely X-ray weak
\citep[e.g.,][]{green01, morabito11}, suggesting a line of sight to
the central engine through obscuration \citep[although the discovery of
radio-loud BAL quasars casts some doubt on the general applicability
of this scenario, e.g.,][]{dipompeo13}.  Thus, the broader
H$\beta$ lines found in our sample may be explained if they are viewed
at a larger angle with respect to the normal. \citet{leighly22} showed
that if the average inclination angle to unabsorbed quasars is 30
degrees to the normal \citet[e.g.,][]{shen_ho_14}, an inclination of
48 degrees is required for the FeLoBAL quasars. 

However, there are several $E1>0$ FeLoBAL quasars that have peculiar
properties that cannot be easily explained by a large inclination
angle.  We propose that the lines in these objects are
broad because there is less emission from slower-moving gas, i.e.,
diminished emission in the core of the line profile. This
interpretation is consistent with the $E1>0$ composite spectrum shown
\citep[][Fig.~9]{leighly22}, which shows that the \ion{Fe}{2} emission
is as strong as in the average composite spectrum, but the H$\beta$
line is weaker, yet no less broad at the base.  It may also be
consistent with the scenario proposed by \citet{sb17}, who suggested
that a disk line is a common component in the spectra of Seyfert 1
galaxies that is not generally recognized because of the presence of
additional lower-velocity non-disk clouds \citep[see
  also][]{bon09}. The additional lower-velocity component may be
comparable to the so-called intermediate line region
\citep[e.g.,][]{brotherton94}.  We suggest that in some of the $E1>0$
objects, the intermediate-line region emission is weak, producing a
broader H$\beta$ line with lower equivalent width, as well as a higher
R$_\mathrm{FeII}$ ratio.   

\begin{figure*}[!t]
\epsscale{1.0}
\begin{center}
\includegraphics[width=4.0truein]{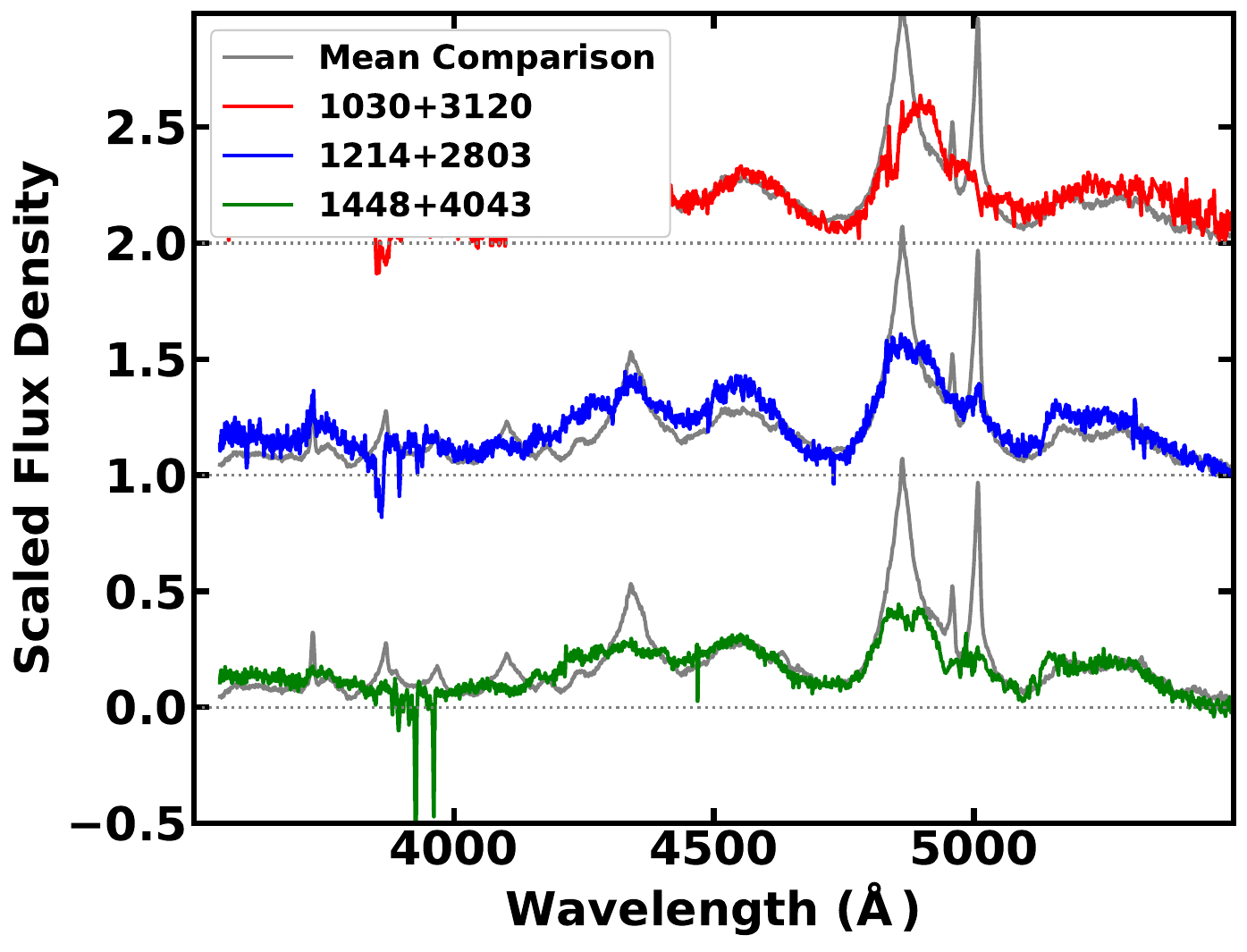}
\caption{    Unusual H$\beta$ line profiles among luminous $E1>0$
    FeLoBALQs. The lower H$\beta$ equivalent width than the comparison
    composite and boxy shape suggest that the broad-line region
    lacks low-velocity gas.
   \label{diskline_fig}}  
\end{center}
\end{figure*}

Fig.~\ref{diskline_fig} shows the H$\beta$ / [\ion{O}{3}] region of
the spectrum from three of our more luminous $E1>0$ FeLoBAL quasars.
The H$\beta$ line shows a stubby and boxy profile, and comparison with
the unabsorbed composite spectrum suggests that they are missing
the low-velocity gas. 

In \S\ref{irtf} we modeled the H$\alpha$ line from
SDSS~J144800.15$+$404311.7.  We found that it was adequately modeled
using a diskline with an outer radius of 9500 gravitational radii.
\citet{leighly22} estimated a log black hole mass of 8.55,
corresponding to an outer radius equal to 0.16 parsec.  In contrast,
single epoch reverberation estimation predicts an H$\beta$
reverberation radius of $\sim 0.3$ parsec.  Thus, the model fit
predicts that the emission line region has a significantly smaller
size than expected, i.e., intermediate-line emission gas is missing,
leading to overall broader lines.

Why would the intermediate-line region be missing in some objects? A
clue to this may come from the behavior of the H$\beta$ deviation
parameter discussed in \citet[][\S~3.4.1]{leighly22}.  The H$\beta$
deviation parameter was defined to measure of the difference between
H$\beta$ FWHM measurements and the average values of the unabsorbed
comparison sample as a function of $R_\mathrm{FeII}$.
\citet{leighly22} showed that the H$\beta$ deviation parameter is
significantly anticorrelated with the maximum outflow velocity and
correlated with the velocity widths: thus, objects with powerful
outflows have broader H$\beta$  lines for their $R_\mathrm{FeII}$
value.  The lower velocity gas producing the core of the emission line
lies farther from the central engine than the higher-velocity gas, and
it may be that the location of this gas is close to the torus.  We
speculate that there is a torus/wind connection {\it also} among the $E1>0$
objects; perhaps the outer broad line region where the line core would
be emitted is disrupted due to the outflow.  The nature of this
connection is not known, and is quite possibly indirect, as the
relationship again appears stochastic rather than deterministic, since
we lack evidence for a disk line in any other of our  $E1>0$
objects. However, the low redshift range of our sample means that
while we are rather more sensitive to low accretion rate objects, we
are less sensitive to very high accretion rate objects, since those
objects are rare and a large volume needs to be surveyed.   It may be
that a higher redshift sample will reveal more FeLoBAL quasars with
disk-line profile Balmer lines.  We are currently making observations
of a sample of higher-redshift FeLoBAL quasars, and we have not yet
found any with such an extreme H$\alpha$ profile.   

\subsection{Speculations on an Evolutionary Scenario} \label{evolution}

 The results presented in \citet{leighly22} and this paper link
  the properties of FeLoBAL quasars directly to differences in
  accretion rate.  As shown by \citet{leighly22}, $E1>0$ FeLoBAL
  quasars have high accretion rates, $E1<0$ FeLoBAL quasars have low
  accretion rates, and unabsorbed quasars have intermediate accretion
  rates.  If a quasar episode is characterized by an initially
  large accretion rate, perhaps originating from a merger, followed by
  a decreasing accretion rate as the fuel supply runs low, we can
  speculate that the direct link to accretion rate that we have
  uncovered implies an indirect link to quasar evolution. We explore
  this scenario in this section.

\citet{sanders88b, sanders88} proposed an evolutionary link between
infrared-luminous galaxies and quasars that follows a merger event.
Infrared galaxies at low redshift often show merger signatures.  A
merger may allow gas to be funneled to the nucleus, captured by the
black hole, and fuel the quasar during its lifetime.   Shrouded by
a thick layer of gas and dust, the energy provided by the black hole
is reprocessed into the infrared, creating a warm ultraluminous galaxy.
Eventually, the accreting black hole produces sufficient energy to
shrug off its cloak of gas and dust and a quasar is born.
Finally, the fuel source decreases producing a dormant early-type
galaxy \citep[e.g.,][]{klindt19}.  

LoBAL quasars have been proposed to play a role in this evolutionary
scenario.  At low redshifts, LoBAL  and FeLoBAL quasars are more
frequently found among  
infrared-selected quasars \citep{low89, boroson91} and red quasars
\citep{becker97, urrutia09, dai08}.  Many such reddened objects are
luminous \citep{dai12}, have high accretion rates
\citep{urrutia12}, and there is evidence they are in a blow-out phase
\citep{glikman17}.   The reddening and high accretion rates suggest
that the LoBAL and FeLoBAL stage should occur after the ULIRG stage 
and before the bare quasar stage.  These interpretations seem
plausible; because the BAL outflow in an FeLoBAL quasar must include 
the hydrogen ionization front to show absorption from Fe$^+$, the
ratio of column density to radiative-driving flux density must be
high, and so significant energy must be present to accelerate the gas.
A high accretion rate provides an environment in which these
conditions may be met. 

In this sequence of papers, we learned that not all FeLoBAL quasars
have high accretion rates.  Instead, we found a population of objects 
that have low accretion rates.  This set of objects is clearly
disjoint from the high accretion rate objects; the only property that
they share is the presence of Fe$^+$ ions in the line of sight. As
outlined in \S\ref{e1lt0}, we propose that in these objects, and
especially in the subclass of loitering outflow objects, the
\ion{Fe}{2} absorption  arises from the disintegrating torus.  

\begin{figure*}[!t]
\epsscale{1.0}
\begin{center}
\includegraphics[width=4.5truein]{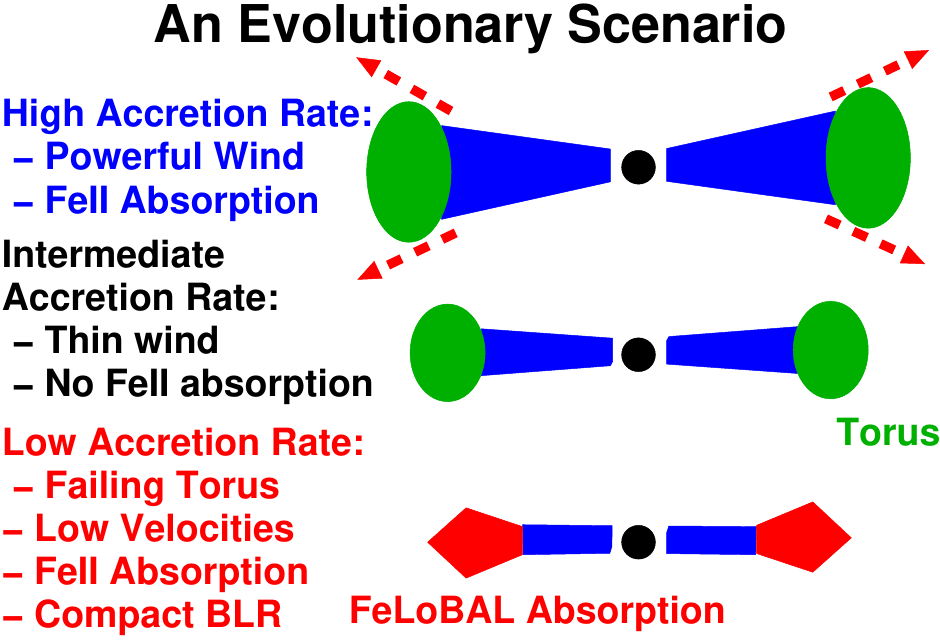}
\caption{A proposed evolutionary sequence for FeLoBAL quasars.  Black
  denotes the black hole, blue shows the accretion disk, green illustrates the
  infrared-emitting torus, and red displays the region producing
  Fe$^+$ ions. {\it
    Top:} The $E1>0$ objects are characterized by a high accretion
  rate relative to the Eddington value.  They emit sufficient energy
  to accelerate a thick wind from the vicinity of the torus or
  throughout the central engine.
  {\it Middle:}  The typical quasar may have a BAL wind, but it is
  generally not thick enough to include Fe$^+$ ions. {\it Bottom:}
  The low-accretion-rate quasar loses its ability to sustain an
  optically thick torus wind \citep{es06}, yet retains sufficient gas
  for   Fe$^+$ ions to be present in the torus
  remnant.    \label{scenario}} 
\end{center}
\end{figure*}

These results suggest that the FeLoBAL quasar phase may occur at two
points in a quasar's evolution: between the ULIRG and bare quasar
stage, while the object is accreting rapidly, and between the bare
quasar stage and the slide into senescence as the quasar runs out of
fuel and the black hole becomes quiescent.   This evolutionary
scenario is illustrated in Fig.~\ref{scenario}.  

Why was the low-accretion-rate branch of the FeLoBAL quasars not
recognized earlier?   There are several ways that our experiment
differs from the previous ones. First, our convolutional neural net
classifier discovered a substantial number of previously missed
FeLoBAL quasars in the SDSS quasar catalog.  Second, we were able to
compile a moderate-sized sample with uniform properties (e.g.,
luminosity) by using the broad bandpass available on the SDSS.
Previous experiements required followup observations in the
near-infrared \citep[e.g.,][]{schulze17, schulze18}, and so the
samples were smaller and the observational properties less uniform. 

Another reason that low accretion rate objects
are expected to be hard to find is because they are less luminous and 
therefore more likely to drop out of flux-limited samples
\citep{jester05, hopkins09}.  The $z<1$ selection of our sample
combined with the sensitivity of the SDSS and BOSS projects apparently
fortuitously probed the right region of parameter space to yield
about 17 $E1>0$ and 13  $E1<0$ objects.  In addition, 11 of
our thirty objects are newly identified FeLoBALQs
\citep[\S~2.1,][]{leighly22}.  The mean and median bolometric
luminosity  of the  previously discovered  objects is almost 0.4 dex
higher than the mean and median bolometric 
luminosity  of the eleven new objects, and 72\% of the newly
discovered objects are $E1<0$ objects.  

What, then, can we expect from the 4G-PAQS survey briefly described in
\S~\ref{disc_lobal}?  The selection cutoff in that survey is {\it
  Gaia} $G<20.5$ \citep{krogager23}.  It turns out that 27 of our 30
objects are brighter than that limit, so we should observe
sufficiently varied accretion rates among the new sample objects.  Not 
surprisingly, though, all three of the objects in our current sample
with $G>20.5$ are low 
accretion rate objects.  

\section{Summary, Conclusions, and Future Work} \label{future}

This is the fourth in a series of papers quantifying the outflow,
accretion, and variability properties of a sample of low-redshift
FeLoBAL quasars.  The first three papers are summarized in
\S\ref{review}.  In this paper, we investigated the optical to near-IR
continuum properties and the {\it WISE} variability properties.  We
also performed an analysis of a sample of LoBAL quasars to determine
whether the unusual set of accretion properties exhibited by FeLoBALQs
were shared with their lower outflow column density cousins, and
presented analysis of a near-IR spectrum of one of our objects.  We 
summarize the principal results below. 

\begin{itemize}

\item  Using the rest-frame optical to near-infrared
  continuum properties   (\S\ref{sed_modeling},
  \S\ref{sed_composites}) extracted from photometry measurements of
  the 
  30 FeLoBAL quasars and   132 unabsorbed quasars from a matched
  sample  introduced in
  \citet{choi22} and   \citet{leighly22}, we found 
  significant distribution differences   (\S\ref{sed_params}) and
  correlations   (\S\ref{correlation_cont}) among the 
  high- and   low-accretion rate FeLoBAL quasars, and the unabsorbed
  quasars.  Among the most statistically significant results was the
  discovery that the optical power law slope is overall steeper in the
  unabsorbed objects compared with the FeLoBAL quasars, likely
  pointing intrinsically flat spectra and/or at least a small amount
  of reddening in most FeLoBAL quasars.  We found that the low accretion rate
  FeLoBAL quasars lacked evidence for the near-IR upturn that
  indicates   the presence of  hot dust (Fig.~\ref{plot_comp_4}), a result that is
  explained by their low accretion rates (\S\ref{disc_continuum}).  We
  found that the continuum parameters were most strongly correlated
  with the Eddington ratio in the unabsorbed objects, arguing for an
  intrinsic origin, but that these patterns were muted among the
  FeLoBAL quasars, likely a consequence of a range of reddening
  (Fig.~\ref{cont_eddrat}, Fig.~\ref{reddening}).  

\item We examined correlations between the 
  continuum shape parameters and the {\it SimBAL} parameters
  compiled in \citet{choi22}  (\S\ref{corr_simbal}).  We found
  evidence for a lack of dust close to the central engine in the
  FeLoBAL quasars that is manifested by a lack of reddening in all
  objects, and weak hot dust emission in the low-accretion rate
  FeLobALQs. We attribute the distribution differences and correlations to
  structural differences in the  central engine driven by accretion
  rate (\S\ref{disc_continuum}), combined with a range of reddening. 

\item We investigated the near-infrared variability of the samples
  using {\it WISE} and {\it NEOWISE} data (\S\ref{wise_variability}).
  After   correcting a systematic underestimation in the photometry
  uncertainty   (\S\ref{varerr}), we found
  that while the variability amplitude did not differ between the FeLoBALQs
  and the comparison sample, the low accretion rate FeLoBALQs showed a
  higher amplitude of variability than the high accretion rate FeLOBALQs
  (\S\ref{dist_var}).  We examined correlations between the
  variability amplitude parameters and the   emission line and global
  properties compiled in \citet{leighly22}.   We recovered the typical
  anticorrelation between variability   amplitude   and luminosity,
  and found that evidence that W1 shows   stronger variability than W2
  in FeLoBALQs but not in the comparison   sample objects
  (\S\ref{correlation_var}).   We found evidence for an
  anticorrelation between the variability amplitude and R$_{2800}$,
  the predicted size of the continuum emission region at 2800\AA\/.
  We suggest that the commonly-found anticorrelation between
  variability amplitude and luminosity is a consequence of a 
  correlation between luminosity and emisssion region size
  (\S\ref{disc_var}).  

\item By analyzed the restframe optical spectra, the photometry, and
  the {\it WISE} variability of a redshift-matched
  sample of LoBAL quasars (\S\ref{lobals}), we concluded
  that FeLoBAL quasars, found to be different from unabsorbed objects, 
  are also different from other types of BAL quasars, and therefore
  they are truly special objects (\S\ref{disc_lobal}).  Specifically,
  the   LoBAL quasars were more similar to the unabsorbed comparison
  sample than to the FeLoBALQs in properties  related to accretion
  rate (\S\ref{comp_lobal}). 

\item  In \S\ref{e1lt0} and \S\ref{e1gt0}, we presented our explanations for
  FeLoBAL absorption in the low accretion rate and high accretion rate
  FeLoBAL quasars,  with particular focus on the  objects with
  outflows found by \citet{choi22} to be located  close   to the
  active nucleus.   
\end{itemize}

Based on the overwhelming evidence of the segregation of
FeLoBAL quasars into two groups, and the evidence that these two
groups represent high and low accretion rates \citep{leighly22}, we
presented an evolutionary scenario in \S\ref{evolution}.
Specifically, we imagined a scenario in which an event, perhaps a
merger, creates an initially high accretion rate that decreases over
time.  We suggested that \ion{Fe}{2} absorption characterizes the high
accretion rate stage when the infrared-luminous galaxy is shrugging
off its cloak of dust and gas in a massive, thick outflow to become an
ordinary quasar.  The intermediate accretion rate stage produces an
outflow that is no longer optically thick enough to include Fe$^+$
ions, keeping in mind that they are not found in the \ion{H}{2} region
of a photoionized gas, but rather at a larger depth, in the
partially-ionized zone.   
Finally, the Fe$^+$ absorption appears again when the quasar is
running out of fuel (the low accretion rate branch).  Specifically, we
suggested that among the FeLoBAL quasars with low accretion rates and 
outflows located close to the central engine, we are seeing absorption
from Fe$^+$ present in the remnants of the failing torus.  

While our speculative scenario involves the properties of the torus in
general, it is worth noting that the data in hand cannot constrain
torus models, as it is limited to the near infrared band and only
measures the hot dust emission likely from graphite grains at the very
inner edge of the torus.  Sophisticated torus models show that the hot
dust emission can be at least partially decoupled from the thermal
torus emission.  For example, \citet{stalevski12} present 3D radiative  
transfer models of a clumpy two-phase torus.  Their models reveal that  
the hot dust component may not be observed under various conditions.
For example, the hot dust emission may not be observed from large
angles from the normal, principally due to self absorption (their
Fig.\ 4); however, we don't believe this explanation applies to our
objects since they are all Type 1s.  But they also found weak hot dust
emission in the case of low contrast parameter, where the contrast 
parameterizes the density ratio between their high- and low-density
phases (their Fig.\ 9).  While our suggestions are bolstered by many
studies, as described in \S\ref{disc_continuum}, much more data,
especially at mid-IR wavelengths, would be necessary to test modern
torus models and our speculative scenario in these objects.  

Whether or not our speculative scenario has merit, there is no doubt
that these four papers have a substantial impact on our understanding
of FeLoBAL quasars.  For example, the global covering fraction
$\Omega$,  the fraction of 4$\pi$ sterradians that includes
outflow, is generally estimated based on the fraction of objects in a
sample.  This type of computation requires that the absorbed objects,
in this case, the FeLoBALQs, differ from the unabsorbed ones {\it
  only} in their line of sight viewing angle.  We have clearly shown
that this assumption is violated among the FeLoBAL quasars, implying
that other methods are needed to estimate the global covering
fraction.  For example, if the absorption lines are narrow, then a
comparison between observed emission lines and those predicted from
the absorbing gas could give an estimate of the global covering
fraction.  Even then, this method might only give an upper limit if
other sources of the line emission, i.e., from gas not associated with
the outflow, are present in the quasar.

This sequence of papers has revealed how primitive, phenomenological,
and incomplete some of our ideas about the quasar phenomenon are.  It
is startling that we could discover a new branch of  the BAL outflow
phenomenon using traditional analysis of archival ground-based
spectroscopic data, given the maturity of the field of study.  These
papers exemplify the power of {\it SimBAL} analysis.  The forward
modeling semi-empirical approach allows us to directly constrain the
physical parameters of the outflow as a function of velocity in the
most complicated spectra. The relative ease of use allows us to
compile detailed results from moderate-sized samples, which means that
we can not only look for trends, but also place our  analysis on a
statistical footing.   

At the same time, it is worth remembering that the quasar phenomenon
is stochastic rather than deterministic.  The stochasticity of the
quasar phenomenon was discussed in the context of Type 2 AGN by
\citet{en16} \citep[see also][]{elitzur12}.  They argued that the
luminosity cutoff consistent with the presence of a broad line region
depends on the black hole mass, but that the proportionality
coefficient is not a single value but has a range of values due to
factors that are not or cannot be modelled. Thus it is certain that
objects that do not conform to the scenario presented here can be
found.   While the cutoff for different accretion states in
X-ray binaries may be well constrained (e.g., the  $\dot
m_\mathrm{crit}=\dot M/\dot M_\mathrm{Edd}=0.01$--0.02 for the
transition from hard to soft state), it seems reasonable that the
different physics of accretion onto a million or billion times larger
black hole would shift this value.     A significant factor may be the
fact that the accretion disk around a large black hole should emit in
the UV where there is plenty of atomic opacity.  This idea has been
investigated by \citet{jiang16} who found that a modeled accretion
disk that included atomic transitions, in particular the iron opacity
that is also important in stellar winds, was more stable than one
without atomic transitions. 

Moving forward, we have embarked  on an analysis of higher-redshift
($1.3 < z < 3$) FeLoBALQs; preliminary results have been presented in 
\citet{voelker21}.  Because of the flux-limited nature of SDSS, the
higher-redshift objects are more luminous.  More luminous objects are
known to have faster outflows \citep[e.g.,][]{laor02, ganguly07}, and
we have already found evidence for a higher incidence of energetic
outflows among these more luminous objects.  An interesting question
is whether we will find many $E1<0$ FeLoBALQs, in particular, the
objects with loitering outflows.  In the sample presented here, we
found that about half of the objects have $E1<0$.  Low accretion rate
objects are predicted to fall out of flux-limited samples at higher
redshifts \citep{jester05}, so we expect to find fewer of them.  On
the other hand, the loitering outflow objects have very characteristic
FeLoBAL absorption spectra, and it may be that we can use those
spectra to predict which objects should have low accretion rates.  If
we can find such objects at higher redshifts, we may find very massive
black holes, and possibly be able to place constraints on the scaling
of the critical accretion rate $\dot m_\mathrm{crit}$ with black hole
mass and/or luminosity.   

\acknowledgements

KML acknowledges very useful conversations with Leah Morabito.
Support for {\it SimBAL} development and analysis is provided by NSF
Astronomy and Astrophysics Grants No.\ 1518382, 2006771, and
2007023. This work was performed in part at Aspen Center for Physics,
which is supported by National Science Foundation grant
PHY-1607611. SCG thanks the Natural Science and Engineering Research
Council of Canada. 

Long before the University of Oklahoma was established, the land on
which the University now resides was the traditional home of the
“Hasinais” Caddo Nation and “Kirikiris” Wichita \& Affiliated
Tribes. This land was also once part of the Muscogee Creek and
Seminole nations.

We acknowledge this territory once also served as a hunting ground,
trade exchange point, and migration route for the Apache, Comanche,
Kiowa and Osage nations. Today, 39 federally-recognized Tribal nations
dwell in what is now the State of Oklahoma as a result of settler
colonial policies designed to assimilate Indigenous peoples.

The University of Oklahoma recognizes the historical connection our
university has with its Indigenous community. We acknowledge, honor
and respect the diverse Indigenous peoples connected to this land. We
fully recognize, support and advocate for the sovereign rights of all
of Oklahoma’s 39 tribal nations.

This acknowledgement is aligned with our university’s core value of
creating a diverse and inclusive community. It is our institutional
responsibility to recognize and acknowledge the people, culture and
history that make up our entire OU Community.

Funding for the SDSS and SDSS-II has been provided by the Alfred
P. Sloan Foundation, the Participating Institutions, the National
Science Foundation, the U.S. Department of Energy, the National
Aeronautics and Space Administration, the Japanese Monbukagakusho, the
Max Planck Society, and the Higher Education Funding Council for
England. The SDSS Web Site is http://www.sdss.org/.

The SDSS is managed by the Astrophysical Research Consortium for the
Participating Institutions. The Participating Institutions are the
American Museum of Natural History, Astrophysical Institute Potsdam,
University of Basel, University of Cambridge, Case Western Reserve
University, University of Chicago, Drexel University, Fermilab, the
Institute for Advanced Study, the Japan Participation Group, Johns
Hopkins University, the Joint Institute for Nuclear Astrophysics, the
Kavli Institute for Particle Astrophysics and Cosmology, the Korean
Scientist Group, the Chinese Academy of Sciences (LAMOST), Los Alamos
National Laboratory, the Max-Planck-Institute for Astronomy (MPIA),
the Max-Planck-Institute for Astrophysics (MPA), New Mexico State
University, Ohio State University, University of Pittsburgh,
University of Portsmouth, Princeton University, the United States
Naval Observatory, and the University of Washington.

Funding for SDSS-III has been provided by the Alfred P. Sloan
Foundation, the Participating Institutions, the National Science
Foundation, and the U.S. Department of Energy Office of Science. The
SDSS-III web site is http://www.sdss3.org/.

SDSS-III is managed by the Astrophysical Research Consortium for the
Participating Institutions of the SDSS-III Collaboration including the
University of Arizona, the Brazilian Participation Group, Brookhaven
National Laboratory, Carnegie Mellon University, University of
Florida, the French Participation Group, the German Participation
Group, Harvard University, the Instituto de Astrofisica de Canarias,
the Michigan State/Notre Dame/JINA Participation Group, Johns Hopkins
University, Lawrence Berkeley National Laboratory, Max Planck
Institute for Astrophysics, Max Planck Institute for Extraterrestrial
Physics, New Mexico State University, New York University, Ohio State
University, Pennsylvania State University, University of Portsmouth,
Princeton University, the Spanish Participation Group, University of
Tokyo, University of Utah, Vanderbilt University, University of
Virginia, University of Washington, and Yale University.

\facility{IRTF (SpeX), Sloan, {\it WISE}, {\it NEOWISE}}

\software{Sherpa \citep{freeman01}, unTimelyCatalogExplorer \citep{kiwy22}}

\appendix

\section{Systematic Errors in the unTimely Catalog Photometry}\label{varerr}

The analysis was performed for each of the 30 FeLoBALQs and each of the
132 comparison objects.  For each object, star candidates were
identified from the AllWISE Source Catalog (TAP identifier {\tt
  allwise\_p3as\_psd}) as follows.  The potential star candidates were
objects lying within circular region with a radius of 500 arc seconds
around each quasar 
that also had W1$>9$, were less than 0.5 magnitudes dimmer than the
corresponding object, and with color range $-0.2< W1-W2 < 0.2$.  The
time series data was downloaded from the unTimely Catalog using the {\tt
  unTimely\_Catalog\_explorer} \citep{kiwy22} using a box size of  
1000 arc seconds.  All star candidates were required to have both W1 and 
W2 light curves.  

\begin{figure*}[!t]
\epsscale{1.0}
\begin{center}
\includegraphics[width=6.5truein]{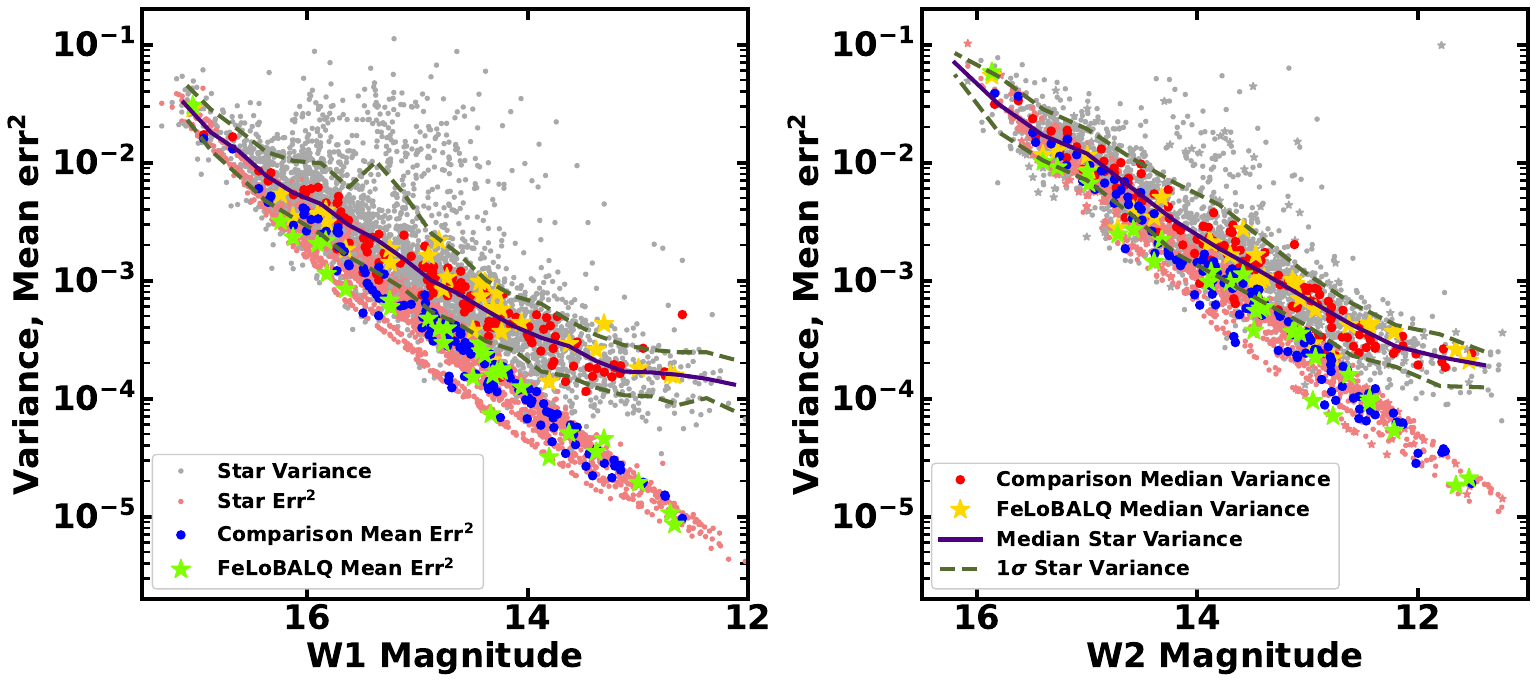}
\caption{A comparison of the stellar light curve variance and the
  mean catalog error squared for the star candidates for both the W1
  and W2 filters; legends apply to both panels.  Most stars do not
  vary significantly.  However, we 
  found that the measured light curve variance (grey) is larger than the
  statistical error squared (pink), especially at lower magnitudes.  We
  interpreted this as evidence for a systematic underestimation of the
  statistical errors.  We defined a modified error as a function of
  magnitude based on the observed median (black line) and 16 and 84\%
  cumulative distribution values (dashed grey lines).  
  \label{wise_var_plot}}
\end{center}
\end{figure*}

The stars were chosen from the star candidates by requiring that the
magnitude of the star lay within a range of the magnitude of the
object.  Because there are far more fainter stars than bright ones,
the range depended on the magnitude as $\Delta M=0.7 - 0.125
(W1_{object} - 12.5)$.  In addition, only lightcurves with 10 or more
points were considered.    The variance in the magnitude was computed for
each star candidate.  The mean of the catalog-obtained error squared
was also collected.  Fig.~\ref{wise_var_plot} shows the relationship
between these values for each star.   There were 3293 W1 comparison
stars and 661 W2 comparison stars.  The values are different because
quasars are relatively brighter in W2 compared with W1, and there is a
lower density of brighter stars in the sky. The observed variance was
found to be larger than the mean error squared, with the discrepancy
being larger for the brighter objects.  The origin of this discrepancy
is not known.   For each object, the median of the variance and the
error squared is also plotted.   Because of the outliers with unknown
origin seen in the plot, median was deemed more representative than
the mean. These values are also shown in the plot. 

As discussed in \S\ref{wise_analysis}, the principal metric used for
the variability was the excess variance, which is the difference
between the variance of the object and the contribution due to noise
in the data.  We determined the contribution due to the noise in the
data from the distribution of all of the stars shown in 
Fig.\ref{wise_var_plot}.  In bins of magnitude, the median, 16 and
84\% values of the distribution were computed.  The results are seen
in Fig.~\ref{wise_var_plot}.  For each object, the 
variance due to noise was interpolated from the median relationship
using the mean magnitude of the object.  The uncertainties used in
computing the excess variance in \S\ref{wise_analysis} were taken to
be the 1-sigma bounds shown.



\end{document}